%% file: quartzdet.tex
\documentclass[a4paper,12pt,twoside,english]{article}
\pdfoutput=1
\usepackage[latin1]{inputenc}
\usepackage{parskip}               
\usepackage{graphicx,xcolor}
\usepackage{wrapfig,rotating}
\usepackage{array}
\usepackage{hyperref}
\usepackage{amsfonts}       
\usepackage{amsmath}        
\usepackage{amssymb}        
\usepackage{hhline}
\usepackage{feynmp}
\usepackage{multirow}
\usepackage[,hang,labelfont=bf]{caption}
\usepackage{subcaption}
\usepackage{xspace}
\usepackage{textcomp}     
\usepackage{colonequals}  
\usepackage{cite}
 
\newlength{\dinwidth}
\newlength{\dinmargin}
\input{declarations.tex}

\begin{document}
\begin{titlepage}
  \begin{flushleft}
    {\tt DESY 15-028    \hfill    ISSN  1748-0221} \\
    {\tt  February 2015, update May 2015}                  \\
  \end{flushleft}

  \vspace{1.0cm}
  \begin{center}
    \begin{Large}
      {\bfseries \boldmath A Quartz Cherenkov Detector for Compton-Polarimetry at Future $e^+e^-$ Colliders}

      \vspace{1.5cm}
      Jenny List$^1$, Annika Vauth$^{1,2}$, and Benedikt Vormwald$^{1,2}$ 
    \end{Large}

    \vspace{.3cm}
    1- Deutsches Elektronen-Synchrotron DESY \\ 
    Notkestr. 85,  22607 Hamburg, Germany
    \vspace{.1cm} \\
    2- Universit\"at Hamburg, Institut f\"ur Experimentalphysik \\
    Luruper Chaussee 149,  22761 Hamburg, Germany
  \end{center}

  \vspace{1cm}

\begin{abstract}
\input{00_abstract.tex}

\end{abstract}

\end{titlepage}

\input{01_intro}

\input{02_concept}

\input{03_design}

\input{04_testbeam}

\input{05_conclusions}


\section*{Acknowledgement}
The construction of the prototype would not have been possible without the great technical 
support by Volker Prahl, Bernd Beyer and D\"orte David. In particular we thank Benno List
for fruitful discussions and the encouragement to pursue the idea of single-peak resolution.
The results presented here could not be achieved without the National Analysis Facility and
testbeam infrastructure at DESY and we thank the NAF and testbeam teams for their continuous
support. We thankfully acknowledge the financial support by the BMBF-Verbundforschung in
the context of the project ``Spin-Management polarisierter Leptonstrahlen an Beschleunigern''.

\appendix
\begin{footnotesize}
\input{bib.tex}

\end{footnotesize}

\end{document}

%% file: declarations.tex
\newcommand{\ex}[1]{\cdot 10^{#1}}

\newcommand{\degree}{\ensuremath{^\circ}\xspace}
\newcommand{\percent}{\,\%\xspace}

\newcommand{\NU}[2]{\ensuremath{#1\,\mathrm{#2}}\xspace}

\newcommand{\GF}{\textsc{Geant4}\xspace}
\newcommand{\Spec}{{Spectrosil\textsuperscript{\textregistered} 2000}\xspace}

\newcommand{\comment}[1]{}

\newcommand{\qqtext}[1]{``#1``\xspace}   
\newcommand{\reftitle}[1]{``#1``\xspace} 
\newcommand{\doi}[1]{\href{http://dx.doi.org/#1}{doi:#1}\xspace}
  
\newcommand{\deltaP}{\ensuremath{\delta \mathcal{P} / \mathcal{P}}}
\newcommand{\Pe}{\ensuremath{\mathcal{P}_{e^-}}}
\newcommand{\Pp}{\ensuremath{\mathcal{P}_{e^+}}}
\newcommand{\Pol}{\ensuremath{\mathcal{P}}}
\newcommand{\AP}{\ensuremath{\mathcal{AP}}}
\newcommand{\A}{\ensuremath{\mathcal{A}}}
\newcommand{\paveip}{\ensuremath{{\langle\Pol\rangle_{\text{lumi}}}}}

\makeatletter
\@ifundefined{subfloat}{\newcommand{\subfloat}}{}
\makeatother
\renewcommand{\subfloat}[3][.45\linewidth]{ \begin{subfigure}[b]{#1}
    \centering
    \includegraphics[width=0.95\textwidth]{#2}
    \caption{}
    \label{#3}
  \end{subfigure}}

\newcommand{\prosubref}[1]{(\protect\subref{#1})\xspace}

%% file: 00_abstract.tex
Precision polarimetry is essential for future $e^+e^-$ colliders and requires Compton polarimeters designed for negligible statistical uncertainties. 
In this paper, we discuss the design and construction of a quartz Cherenkov detector for such Compton polarimeters. The detector concept has been developed with regard to the main systematic uncertainties of the polarisation measurements, namely the linearity of the detector response and detector alignment.
Simulation studies presented here imply that the light yield reachable by using quartz as Cherenkov medium allows to resolve in the Cherenkov photon spectra individual peaks corresponding to different numbers of Compton electrons. The benefits of the application of a detector with such single-peak resolution to the polarisation measurement are shown for the example of the upstream polarimeters foreseen at the International Linear Collider. Results of a first testbeam campaign with a four-channel prototype confirming simulation predictions for single electrons are presented.

%% file: 01_intro.tex
\section{Introduction}        
\label{sec:intro}

Polarised beams are a key ingredient of the physics program of
future electron-positron colliders~\cite{power}. 
The precise knowledge of the beam polarisation is as important as the 
knowledge of the luminosity, since for electroweak processes,
the absolute normalisation of expected event rates depends on 
both quantities to the same order. The values relevant for the experiments,
namely the luminosity-weighted average polarisations at the interaction point \paveip, have to be
determined by combining fast measurements of Compton polarimeters with long-term scale calibration 
obtained from reference processes in collision data. 
 
In particular for the International Linear Collider~\cite{tdr}, where both beams are foreseen to be 
longitudinally polarised, it is required to control \paveip\ with permille-level precision. 
Two Compton polarimeters~\cite{boogert} per beam  aim to measure the longitudinal polarisation 
before and after the collision with a precision of $\deltaP \leq 0.25\%$\footnote{for typical 
ILC beam polarisation values of $\Pe \geq 80\%$ and $\Pp \geq 30\%$ or even $\geq 60\%$.}. 
It should be noted, though, that this goal is not driven by physics requirements, but by what 
used to be considered feasible experimentally. Thus, further improvements in polarimetry would 
still have direct benefits for the physics potential of the machine.
Spin tracking simulations are required to evaluate the effects on the mean polarisation vector 
caused by the beamline magnets between the polarimeters and the interaction point, 
by the detector magnets and by the beam-beam interaction. These effects 
have recently been studied in~\cite{spintracking}, concluding that a cross-calibration of the 
polarimeters to 0.1\percent is feasible, as well as individual extrapolations of upstream and downstream 
measurements to the $e^{+}e^{-}$  interaction point.

The long-term average of the polarisation at the interaction point can be determined from reference 
reactions, where in particular $W$ pair 
production~\cite{List:EPSproceedings, diss:marchesini, Rosca:2013lcnote } and single $W$ 
production~\cite{Graham} have been studied. Precisions of about $0.15\%$ can be achieved after 
several years of data taking. These results are quite robust with regard to systematic uncertainties: e.g.\ 
in the measurement using $W$ pair production, even for a conservative assumption of 0.5\percent uncertainty on 
the selection efficiency for the signal and 5\percent for the background, the impact on the uncertainty of the 
polarisation measurement was found to be below the statistical uncertainty for an integrated luminosity 
of $\NU{500}{fb^{-1}}$~\cite{diss:marchesini}.
Any imperfection in the beam helicity reversal, i.e.\ differences in the magnitude of the polarisation 
between measurements with left- and right-handed polarised beams, has to be corrected for based on the 
polarimeter measurements.

\begin{figure}[!htb]
\centering
  \includegraphics[width=0.95\linewidth]{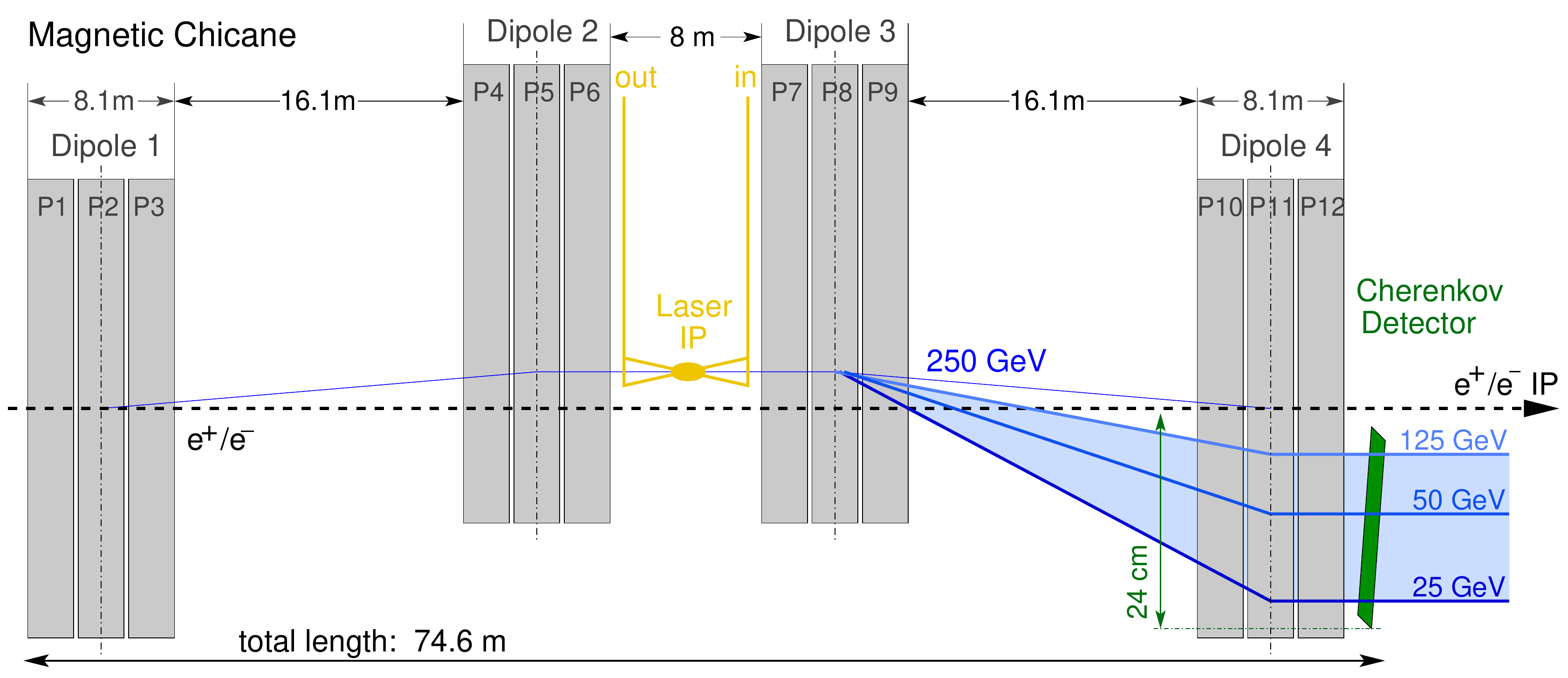}
  \caption{Sketch of the magnetic spectrometer for the ILC upstream polarimeter 
  (from~\cite{testbox_paper}). The Compton-scattered electrons with the 
  lowest energies are deflected most by the spectrometer.  }
  \label{fig:magnetic-chicane}
\end{figure}

The two polarimeters are located about $1.8$\,km upstream and $160$\,m downstream of the 
interaction point.
Both provide non-destructive measurements of the longitudinal beam polarisation based on the 
polarisation dependence of Compton scattering and have been designed for operation at beam 
energies between $45$\,GeV and $500$\,GeV. In the order of $10^3$ 
electrons\footnote{or positrons in case of the positron beam of the ILC which is equipped 
analogously.} per bunch undergo Compton scattering with
circularly polarised laser light which is shot under a small angle onto individual bunches.
The energy spectrum of these scattered particles depends on the product of laser and beam 
polarisations as shown in figure~\ref{fig:comptonXS}, and the differential rate asymmetry 
with respect to the sign of the circular laser polarisation $\lambda$ is directly proportional 
to the beam polarisation. The asymmetry expected for $\lambda\Pol = 100\%$ is
called analysing power \AP . The scattering angle in the laboratory frame is less than
$10\,\mu$rad, so that a magnetic chicane is employed to transform the energy spectrum into a spatial 
distribution. Such a spatial distribution is shown in figure~\ref{fig:lcpol80} for the example of 
the chicane foreseen for the ILC's upstream polarimeters, which is sketched in 
figure~\ref{fig:magnetic-chicane}.

\begin{figure}[tb]
  \centering
  \subfloat{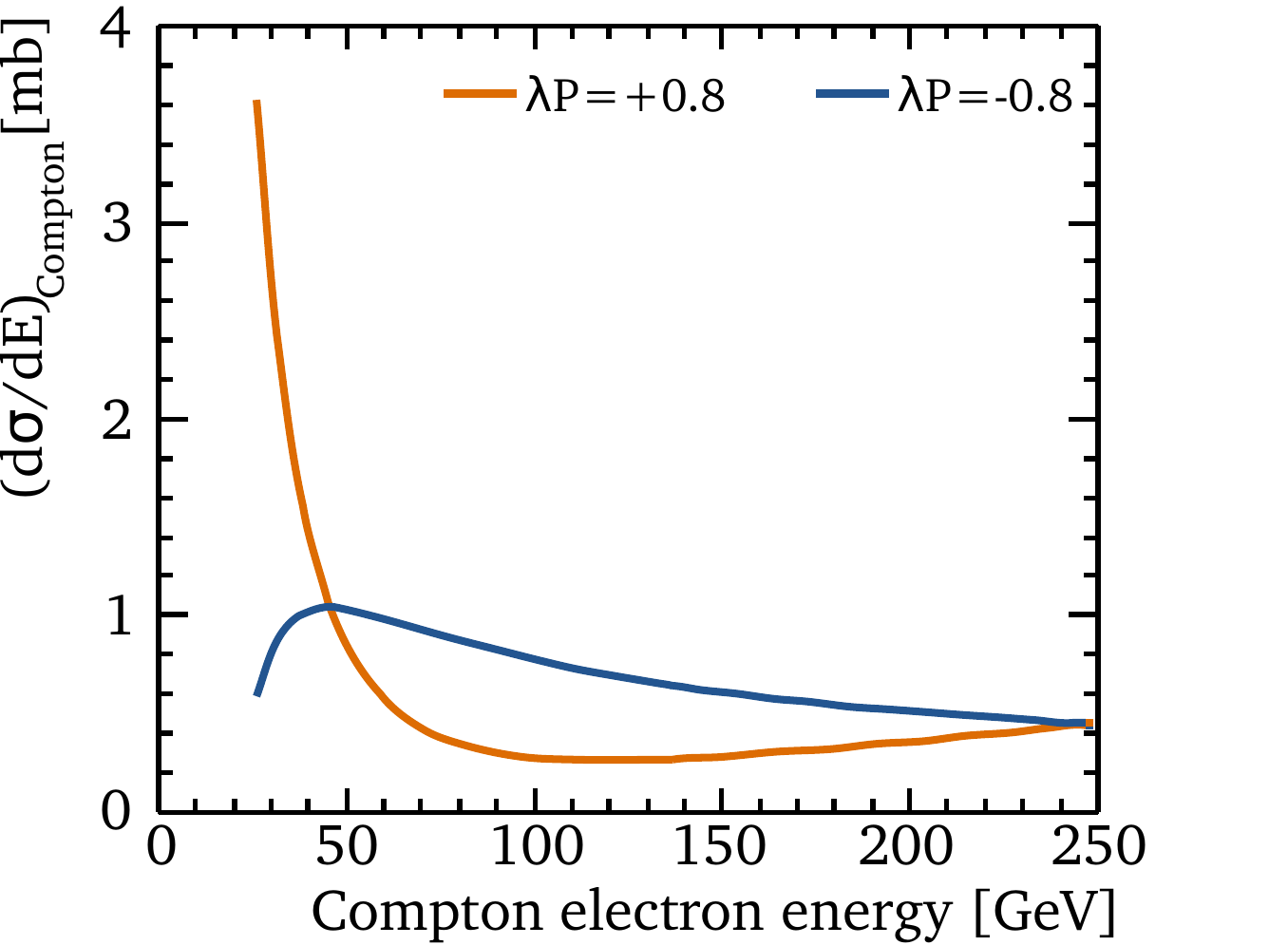}{fig:comptonXS}
  \hfill
  \subfloat{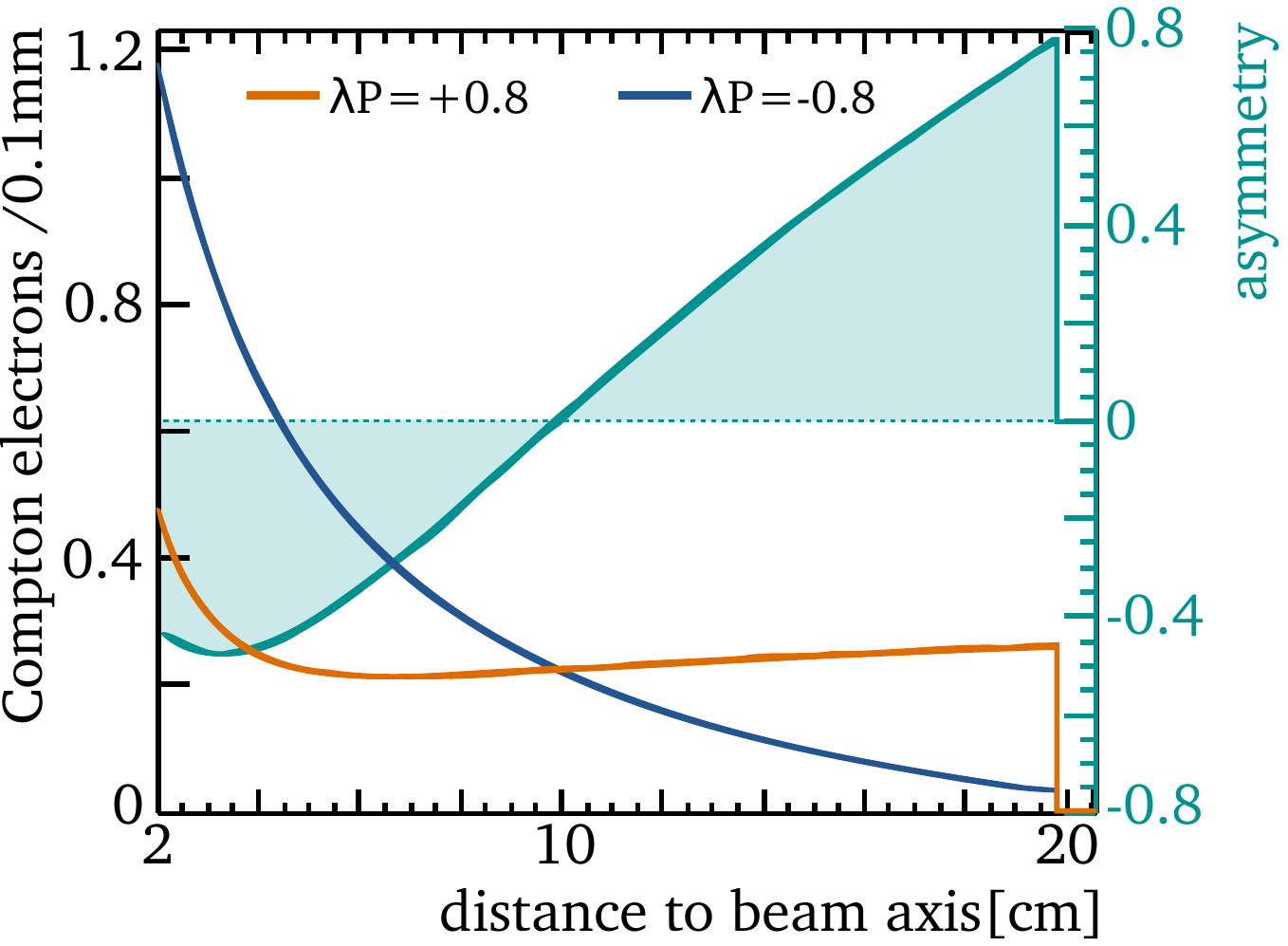}{fig:lcpol80}

  \subfloat{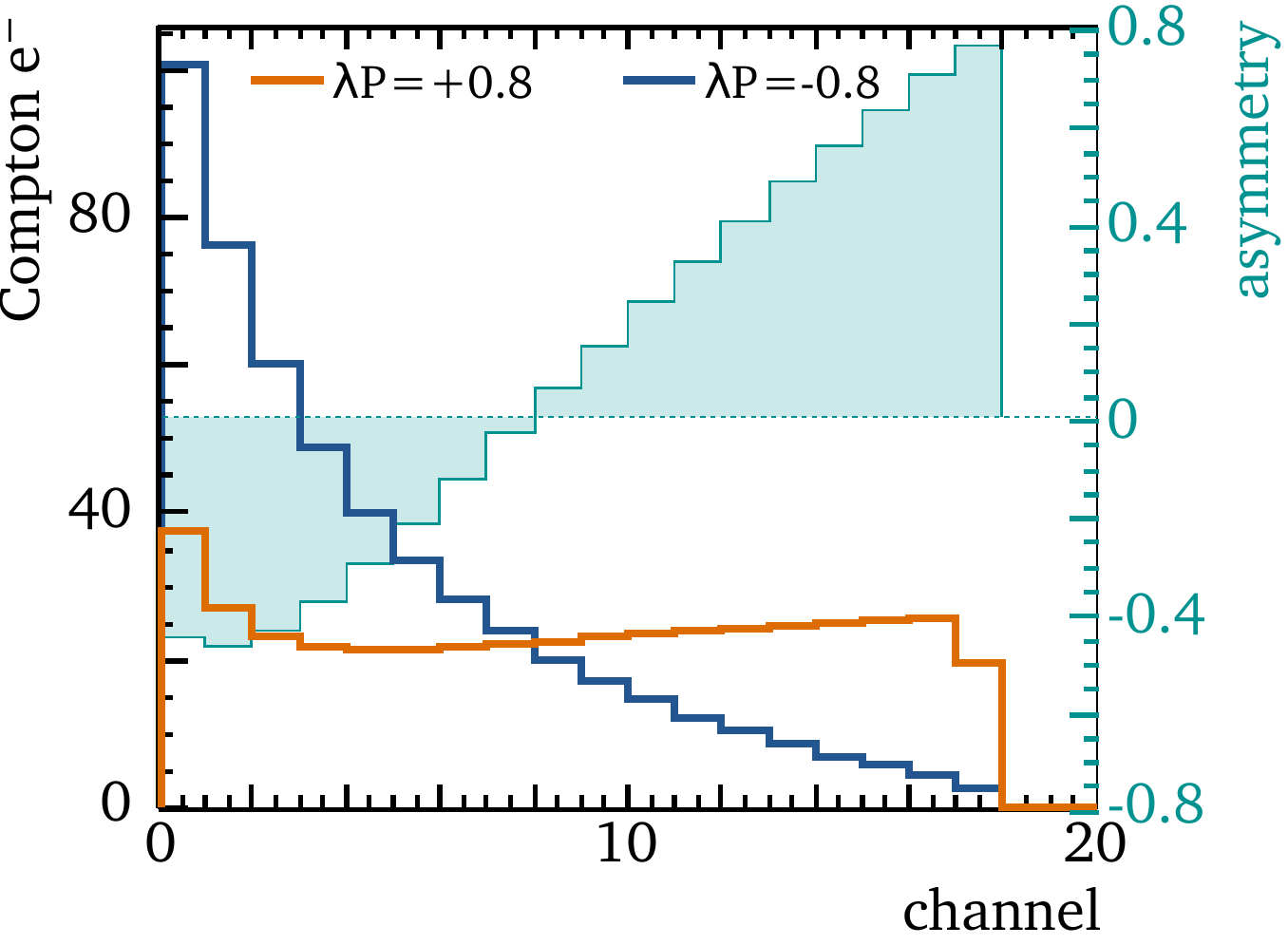}{fig:ncomp20chan}
  \hfill
  \subfloat{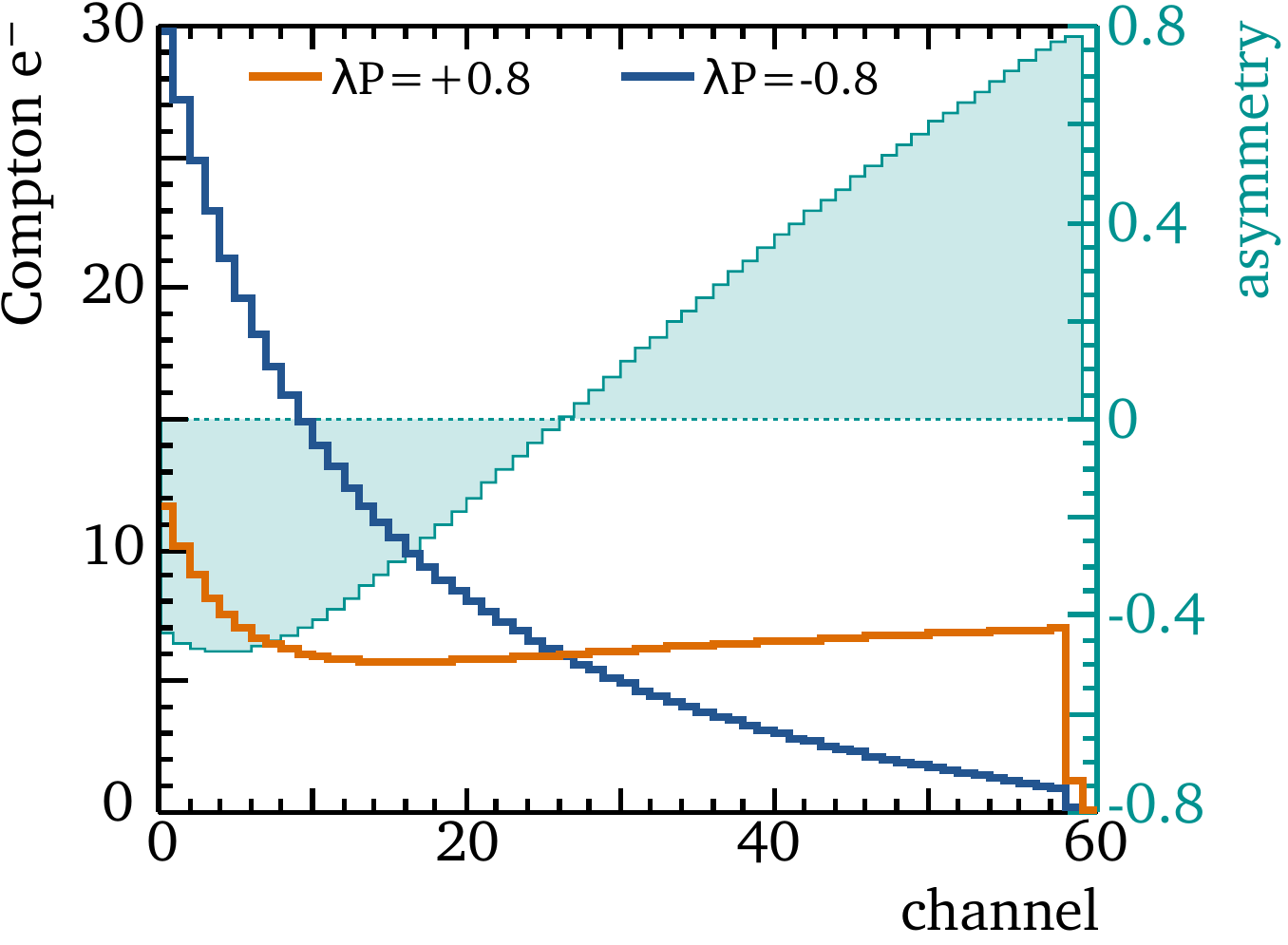}{fig:ncomp60chan}
  \caption{\prosubref{fig:comptonXS}  Differential cross-section in 
  dependence of the energy of the scattered electron for $\lambda\Pol=\pm 80\percent$. 
  \prosubref{fig:lcpol80} Spatial distribution of the Compton scattered 
  electrons at the detector location behind the magnetic chicane for $\lambda\Pol =\pm 80\%$. 
  The filled area indicates the resulting asymmetry (secondary $y$-axis).  
  The bottom plots show the resulting mean number of Compton electrons per detector channel
   for $\lambda\Pol = \pm80\%$ and the corresponding asymmetry  assuming  \prosubref{fig:ncomp20chan}  $20$  channels with a width of \NU{10}{mm} and \prosubref{fig:ncomp60chan}  $60$  channels with a width of \NU{3}{mm}.
    }
  \label{fig:polchannels}
\end{figure}

Several options for detecting the Compton-scattered electrons behind the last magnet of the 
chicane are being considered. The baseline solution is a gas Cherenkov detector consisting 
of $20$ channels, each $1$\,cm wide. The rate asymmetry $\A_i$ with respect to the laser 
helicity is determined in each channel $i$ as
\begin{equation}
\label{eq:asym}
\A_i = \frac{N^R_i - N^L_i}{N^R_i + N^L_i},
\end{equation}
where $N^R_i$ and $N^L_i$ are the count rates in channel $i$ for right- and left-handed laser 
polarisation, respectively. The count rate and asymmetry for each channel for the baseline solution as well
as a detector design with a three times larger channel number are shown in figure~\ref{fig:ncomp20chan} and 
figure~\ref{fig:ncomp60chan}, respectively.

Per-channel measurements of the beam polarisation are then obtained by comparison with the 
predicted analysing power $\AP_i$ of the corresponding channel. The predictions of $\AP_i$ 
are based on the well calculable cross-section for Compton 
scattering~\footnote{With radiative corrections of less than $0.1\%$ currently calculated 
to the order of $\alpha^3$~\cite{Swartz:1997im}.} plus a Monte-Carlo modelling
of the magnetic chicane and the detector geometry and response. The resulting 
per-channel polarisation measurements are then combined in a weighted average, 
where the largest weights $w_i$ are given to the channels with the smallest 
statistical uncertainties, which is equivalent to the channels with the largest 
analysing power~\cite{Gharibyan2001}:
\begin{equation}
\label{eq:weights}
w_i = \frac{1}{(\Pol \cdot \AP_i)^{-2} - 1}
\end{equation}

This measurement principle has already been used successfully at the SLC polarimeter, which was 
the up to now most precise Compton polarimeter, reaching $\deltaP = 0.5\%$~\cite{Abe:2000dq}. 
The gas Cherenkov approach is robust against high data rates, a harsh radiation environment, 
and possible backgrounds. The basic detection mechanism is intrinsically proportional to the 
number of electrons passing a single channel simultaneously, since the amount of emitted 
Cherenkov radiation does not depend on the energy of the particles once they are relativistic. 
The main source of non-linear behaviour are thus the photodetectors.  Table~\ref{tab:polsys} 
shows the anticipated uncertainty budget for the ILC Compton polarimeters assuming a Cherenkov 
detector, and compares it to the uncertainties achieved at the SLC polarimeter.

\begin{table}[hbt]
   \centering
   \begin{tabular}{llll}
      source of uncertainty & & \multicolumn{2}{c}{\deltaP}  \\
      \hline
                                             & & SLC achieved      & ILC goals \\
      \cline{2-4}
      laser polarisation               & & $0.1\percent$  & $0.1\percent$               \\
      detector alignment               & & $0.4\percent$$\phantom{0.15\%}$  & $0.15\percent-0.2\percent$ \quad        \\
      detector linearity               & & $0.2\percent$  & $0.1\percent$               \\
      electronic noise and beam jitter & & $0.2\percent$  & $0.05\percent$              \\
      \hline                                              
      Total                            & & $0.5\percent$  & $0.25\percent$              \\
   \end{tabular}
   \caption{Uncertainty goals for the polari\-sation measurement at the ILC~\cite{spintracking}. For comparison, the systematic uncertainties determined for the SLC polarimeter are also given~\cite{Abe:2000dq}. }
   \label{tab:polsys}
\end{table}

The most challenging improvement has to be achieved with respect to the detector linearity.  
In order to reach the goal of not more than $0.1\%$ contribution from this source to $\deltaP$, 
non-linearities have to be
monitored and corrected for at the level of a few permille~\cite{Eyser2007}. A corresponding 
calibration system has been developed recently~\cite{diss:vormwald}, matching a prototype of 
a gas Cherenkov detector which has been successfully operated in testbeam. With this prototype, 
the alignment requirement has nearly been reached and electronic noise was found to be at a 
negligible level~\cite{testbox_paper}. The influence of beam jitter will also be much smaller 
in the ILC case, since the whole accelerator has been designed to limit the transverse jitter 
to smaller than $10\%$ of the beam sizes~\cite{tdr} in order to achieve the design luminosities. 
Thus, it seems feasible to reach the envisaged precision goal with the baseline design.

The effect of non-linearities would be eliminated completely if the Compton-scattered electrons 
could be detected individually, e.g.\ by a Silicon pixel detector, which would also increase 
significantly the possibilities to control the alignment. Here however, R\&D is still required 
to reach a proof-of-principle level, where the main concern is the high local data rate, 
followed by radiation issues. Employing pixel detectors would enable in addition the measurement 
of transverse beam polarisation~\cite{Etai_LCnote}.

An alternative Cherenkov detector concept which promises a significant reduction of the effects 
of non-linearities and misalignments is the subject of this paper. It is organised as follows: 
In section~\ref{sec:apply2pol}, the concept of polarimetry with Compton electron counting is 
introduced and its benefits are demonstrated using the example of the upstream polarimeter 
of the International Linear Collider. Section~\ref{sec:design} discusses the design of a 
suitable detector matching the requirements and introduces a corresponding prototype. 
In section~\ref{sec:testbeam}, we present results from operating this prototype 
in testbeam before concluding in section~\ref{sec:conclusion}.

%% file: 02_concept.tex
\section{Polarimetry with Compton electron counting}
\label{sec:apply2pol}

The detector concept discussed in this paper aims to resolve individual Compton electron peaks 
in the detected and digitised spectrum of the Cherenkov light. This would allow to build a 
\qqtext{self-calibrating} detector: by
determining the distance between these peaks, the response to single Compton electrons
including the amount of detected Cherenkov light as well as the gain of the photodetector could be
monitored in-situ.  Even better, the shape of such a spectrum could be exploited to directly determine the
number of Compton electrons in this channel for the calculation of the polarisation
asymmetry, without the need to calibrate the photomultiplier gain, its quantum efficiency and geometrical
acceptance for the produced Cherenkov light. It should be noted that this goes beyond the single-peak
resolution offered for low count-rates by Silicon photomultipliers (SiPMs), where each peak corresponds to a fixed number of detected
photons. Thus, the single peaks of a SiPM do not give direct information on the Compton spectrum, but would need to be unfolded for the statistical and wavelength-dependent effects of the Cherenkov light production, propagation and detection. When operated at the higher count-rates required to resolve individual Compton
electron peaks, SiPMs exhibit a significantly higher intrinsic non-linearity and temperature dependence compared to conventional photomultiplier tubes. Therefore, we do not consider SiPMs further in this concept
study.

\begin{figure}[htb]
  \centering
  \includegraphics[width=0.8\linewidth]{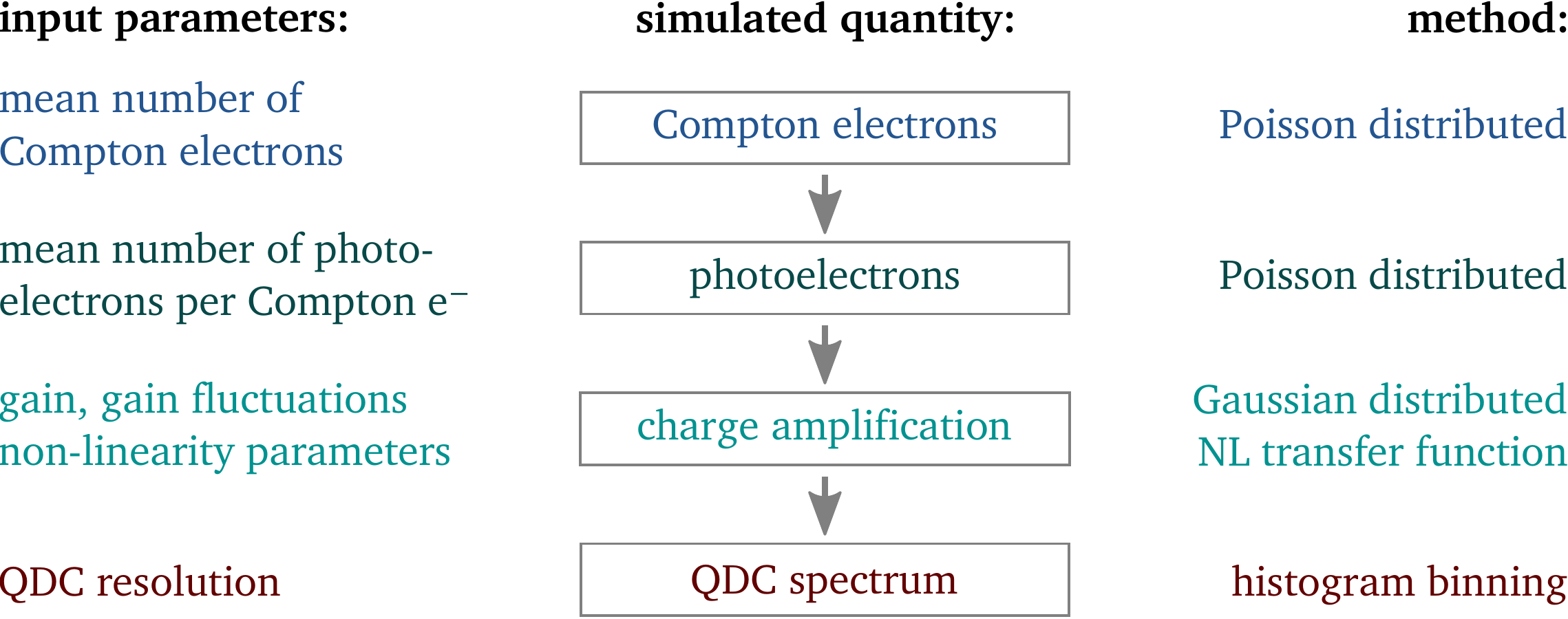}  
  \caption{Schematic overview of the Monte Carlo simulation predicting the digitised
  detector response, with the input parameters for each simulation step in the left column
  and the method to implement the corresponding statistical effects in the right column.}
  \label{fig:MCflow}
\end{figure}

A dedicated Monte-Carlo simulation has been developed to predict the digitised 
photomultiplier response for a given average number of Compton electrons
per detector channel. As illustrated in figure~\ref{fig:MCflow}, the simulation 
starts by determining the actual number of Compton electrons per channel based on
a Poissonian distribution around the mean value. For each Compton electron, the actual 
number of photoelectrons created and emitted from the photocathode of a photomultiplier 
is determined, again assuming Poissonian statistics\footnote{This is equivalent
 to a Poissonian (or for large numbers Gaussian) distribution of 
produced Cherenkov photons and binomial statistics for the detection
probability.}. The charge amplification process is then modelled including Gaussian
gain fluctuations (in the following also referred to as \qqtext{noise}) and, optionally,
a non-linear transfer function. Finally, the resulting charge is digitised with a
fixed resolution. Any potential imperfections of the digitisation step are implicitly
included in the non-linearities of the amplification step, since they have been found
to be tiny for our actual digitiser~\cite{diss:helebrant}. Further noise from the read-out 
electronics is assumed to be negligible in the scope of the detector design studies presented 
here, based on the testbeam experience with the gas Cherenkov prototype~\cite{testbox_paper}.

\subsection{The case for single-peak resolution}

In this section, we will first derive the conditions required to resolve individual peaks
in the digitised photomultiplier spectrum. We will then show how the single-peak resolution
can be exploited for a reliable estimate of the mean number of Compton electrons per channel,
even if the detector behaviour is not perfectly linear.

\subsubsection{Requirements for single-peak resolution}

\begin{figure}[htb]
  \centering
  \subfloat{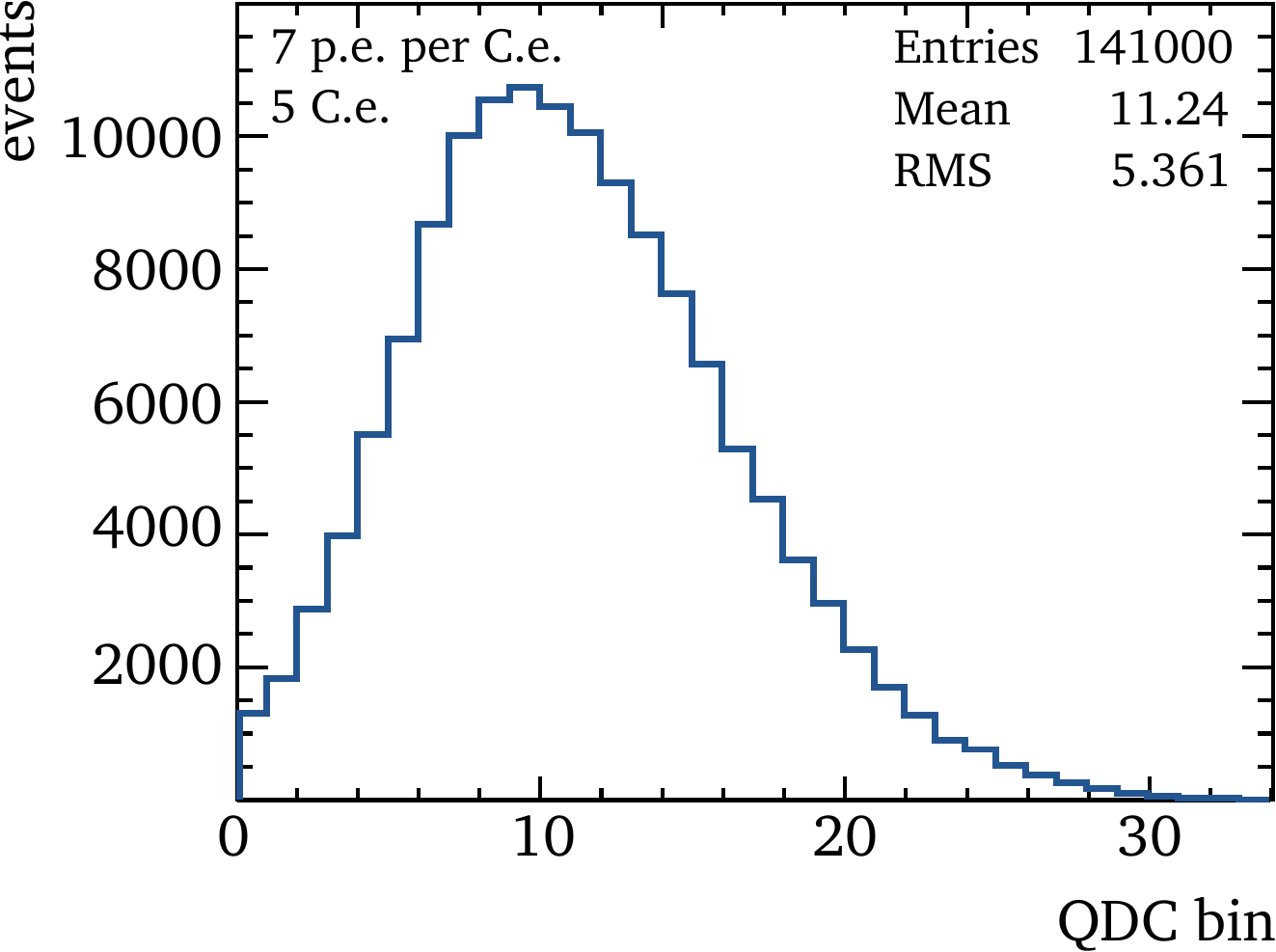}{fig:hspectrum5-7}
  \hfill
  \subfloat{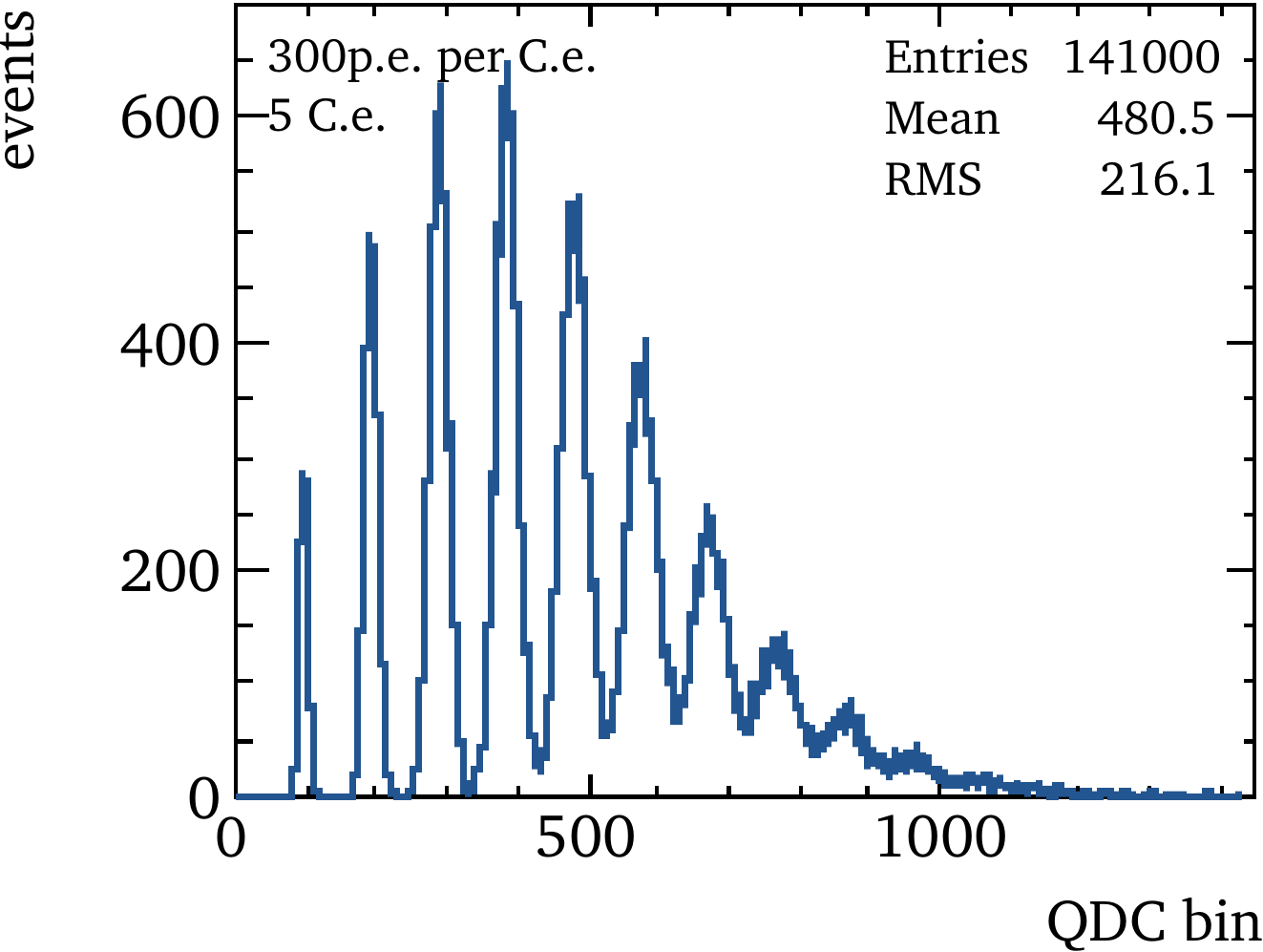}{fig:hspectrum5-300}  
  \caption{Simulated spectra for 5 Compton electrons in the detector with
    \prosubref{fig:hspectrum5-7} 7 and \prosubref{fig:hspectrum5-300} 300 detected photons
    per Compton electron. The simulation assumes that the photodetector signal is
    amplified with a gain of $4\ex{5}$ with a fixed noise level of $1\percent$, followed
    by digitisation with a QDC with a resolution of \NU{200}{fC}.
    }
  \label{fig:hspectrum5}
\end{figure}

Two factors are important to resolve individual Compton electron peaks: the average number 
of Compton electrons
$N_{\mathrm{C.e.}}$ per detector channel, and the number of photoelectrons per electron $
p\colonequals {N_{p.e.}}(N_{\mathrm{C.e.}}=1)$. To illustrate this, figure~\ref{fig:hspectrum5} shows
a simulation of digitised spectra for $N_{\mathrm{C.e.}}=5$, with a
yield of $p=7$ and $p=300$, respectively. Here, a gain $g$ of $4\ex{5}$ with
a fixed noise level of $\frac{\Delta g}{g}=1\percent$ followed by charge 
digitisation\footnote{In the following we will use the abbreviation QDC (charge sensitive 
analogue-to-digital converter).} with a resolution of $w_{\text{QDC}}=$\NU{200}{fC} is 
assumed as example.
For the configuration with $p=7$, the Poissonian distribution of actual electrons per bunch 
crossing is smeared into one broad peak. For $p=300$, however, peaks for the individual actual 
numbers of Compton
electrons $i_\mathrm{C.e.}$ can be easily distinguished.

To resolve individual peaks, the separation $s$ between them should be larger than the peak 
width by a factor $k$, which can be phrased as a requirement of
\begin{equation}
  \label{eq:sepreq}
  s > \text{FWHM} = k\cdot\sigma,  
\end{equation}
with $\sigma$ the width of a Gaussian approximation of an individual peak.

The separation $s$ is determined by $p$, the photomultiplier gain $g$ and 
the digitiser resolution $w_{\text{QDC}}$ as illustrated in figure~\ref{fig:labeled-QDC}:
\begin{equation}
  \label{eq:peaksep}
  s = p \cdot \frac{g \cdot e}{w_{\text{QDC}}} = p \cdot Q,
\end{equation}

where $e$ is the electron charge. The peak width is given by a convolution of
the fluctuations in the actual number of photoelectrons per Compton electron around $p$,
the noise $\Delta g/g$ of the photomultiplier, and $Q$ as defined by equation~\ref{eq:peaksep}, as
\begin{equation}
  \label{eq:sigmapeak}
\sigma = \sqrt{Q^2 \cdot i_{\mathrm{C.e.}} \cdot p  + Q^2  i_{\mathrm{C.e.}} \cdot p \cdot (\frac{\Delta g}{g})^2  + \frac{1}{12}}.
\end{equation}

The first two terms are attributed to the statistics of the photoelectron creation and their amplification, respectively. The term $1/12$ accounts for the digitisation error, which is negligible
for any of the configurations considered here.
To meet the separation requirement given by equation~\ref{eq:sepreq} for at least all peaks up to the mean number of Compton electrons $N_{\mathrm{C.e.}}$, a combination with equations~\ref{eq:sigmapeak},\ref{eq:peaksep} shows that $p$ needs to fulfil 
\begin{equation}
  \label{eqn:quartzmotivCondition}
      p >  \frac{\frac{1}{2}N_{\mathrm{C.e.}} + \sqrt{\frac{1}{4}N_{\mathrm{C.e.}}^2 + \frac{1}{12 \cdot Q^2 }  
                            \left( \frac{1}{k^2} - (N_{\mathrm{C.e.}} \cdot \frac{\Delta g}{g})^2  \right)} }
      {\frac{1}{k^2}   - (N_{\mathrm{C.e.}} \cdot \frac{\Delta g}{g})^2} .
\end{equation}
With a choice of $k=2.35$ for a separation by at least the full width at half maximum, at least $p>30$ is required to 
resolve the case $N_{C.e.}=5$ for the previous example from figure~\ref{fig:hspectrum5}. For the case of $p=300$ as assumed in figure~\ref{fig:hspectrum5-300}, up to $N_{C.e.}=29$ can be resolved. 

Obviously, the modelling of the photomultiplier noise is important for a correct estimation
of the peak separation capability for a given setup. While the example in 
figure~\ref{fig:hspectrum5} assumed a constant noise level of $1\%$, a more realistic approach based 
on a statistical modelling of the amplification 
process~\cite{diss:vauth} leads to a dependency of $\frac{\Delta g}{g}$ on the number of 
photoelectrons which are produced at the cathode. This is taken into account in
the simulation studies presented below and results in changes of about $10\%$ in the value 
of $p$ required to resolve a certain $N_{C.e.}$. Thus equation~\ref{eqn:quartzmotivCondition} 
with a constant noise level remains a useful approximation.

\begin{figure}[tb]
  \centering
  \subfloat[0.423\linewidth]{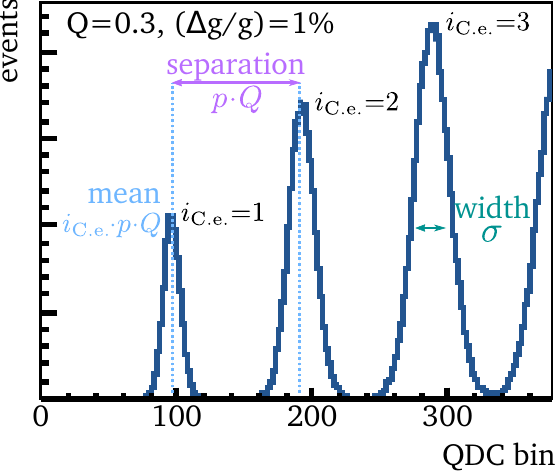}{fig:labeled-QDC}
  \hfill
  \subfloat[0.477\linewidth]{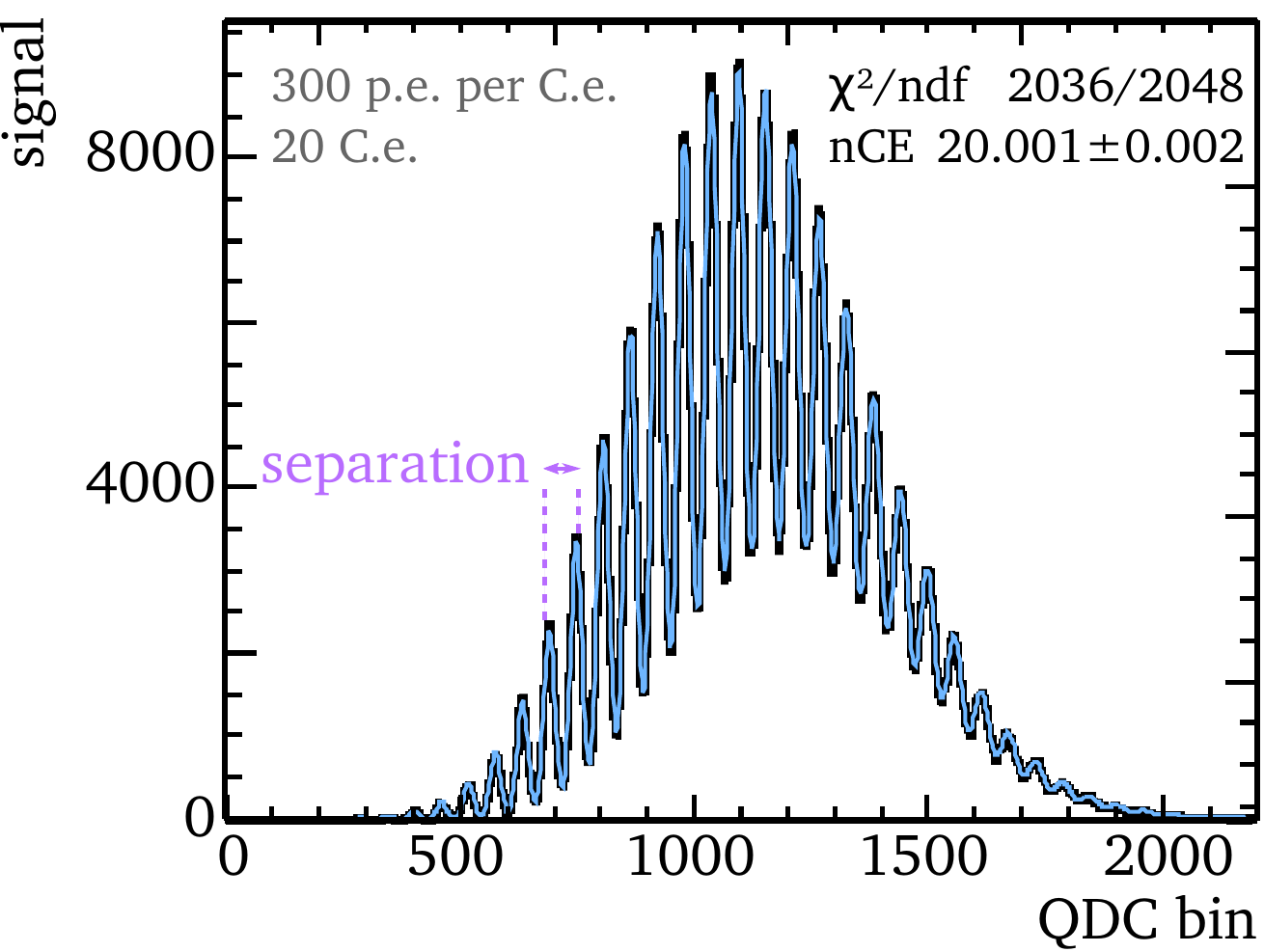}{fig:fit20}
  \caption{\prosubref{fig:labeled-QDC} Illustration of peak position, width and
    separation, for the first three peaks in a simulated QDC spectrum. 
    \prosubref{fig:fit20} QDC spectrum for $N_{\mathrm{C.e.}}=20$  and $p=300$ 
    (black histogram) fitted according to equation~\ref{eq:cefit} (blue curve).}
  \label{fig:hspectrumfit}
\end{figure}

\subsubsection{Reconstruction of $N_{\mathrm{C.e.}}$}

The Monte-Carlo simulation described above has been employed to obtain QDC spectra for
various conditions. As example, the histogram in figure~\ref{fig:fit20} shows the predicted
QDC spectrum for $N_{\mathrm{C.e.}}=20$ assuming $p=300$, $g=2.4\ex{5}$ with detailed modelling 
of the noise, perfect linearity and $w_{\text{QDC}}=\NU{200}{fC}$. There are two possibilities
to reconstruct the average number of Compton electrons from this spectrum: either by taking
the mean (or maximum) of the whole distribution as in the case with no single-peak resolution,
or by extracting $N_{\mathrm{C.e.}}$ from a fit to the detailed shape of the spectrum.

A suitable fit function is the sum of Gaussians with mean ${q}_{2\cdot i_\mathrm{peak}}$ and 
width ${q}_{2\cdot i_\mathrm{peak} +1}$, where each Gaussian describes one of $n_\mathrm{peaks}$
peaks in the spectrum.  Since the number of electrons in a channel is expected to follow a
Poissonian distribution, the height of the peaks is set to the expectation for
$i_\mathrm{peak}$ electrons from a Poissonian distribution of $N_{\mathrm{C.e.}}$
electrons and scaled to the number of events $N_{events}$ in the spectrum:
\begin{equation}
f(x_{QDC}) = N_{events} \cdot  \sum_{i_\mathrm{peak}=0}^{n_\mathrm{peaks}} 
\mathrm{Pois}(i_\mathrm{peak}, N_{\mathrm{C.e.}}) \cdot 
\mathrm{Gaus}(x_{QDC}, {q}_{2\cdot i_\mathrm{peak}}, {q}_{2\cdot i_\mathrm{peak} +1}).
\label{eq:cefit}
\end{equation}
The free parameters of the fit are the mean and width of all $n_\mathrm{peaks}$ peaks and
the central value $N_{\mathrm{C.e.}}$ of the Poissonian. While studying the distance
between the mean values of the individual peaks gives access to the gain linearity, the
central value $N_{\mathrm{C.e.}}$ of the Poissonian is the parameter of interest for the
polarisation measurement. An example for a fit of equation~\ref{eq:cefit} to a spectrum
simulated with $N_{\mathrm{C.e.}}=20$ is shown in figure~\ref{fig:fit20} for $10^7$ accumulated
individual measurements.

\begin{figure}[tb]
  \centering
  \subfloat{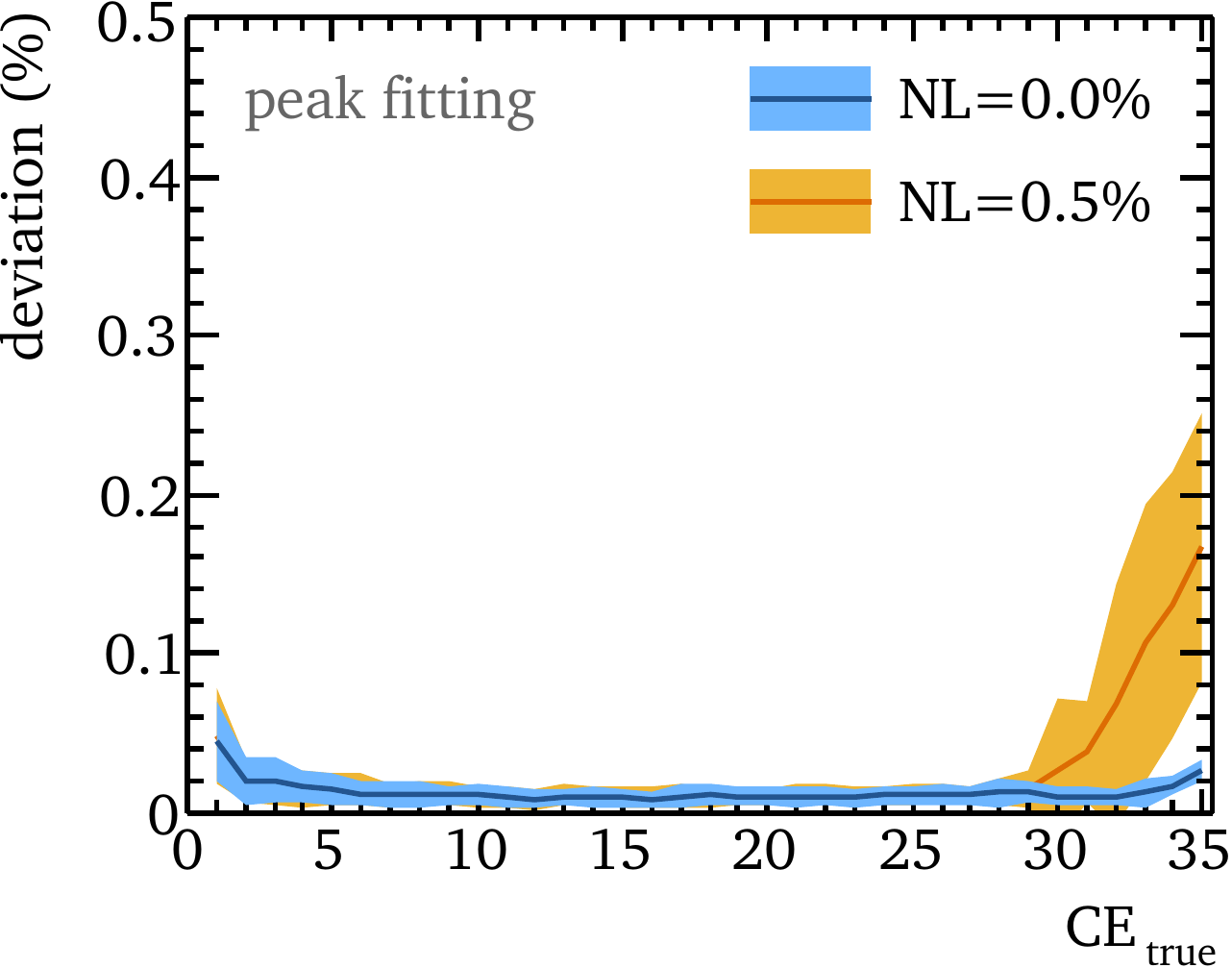}{fig:fit300NL}  
  \hfill
  \subfloat{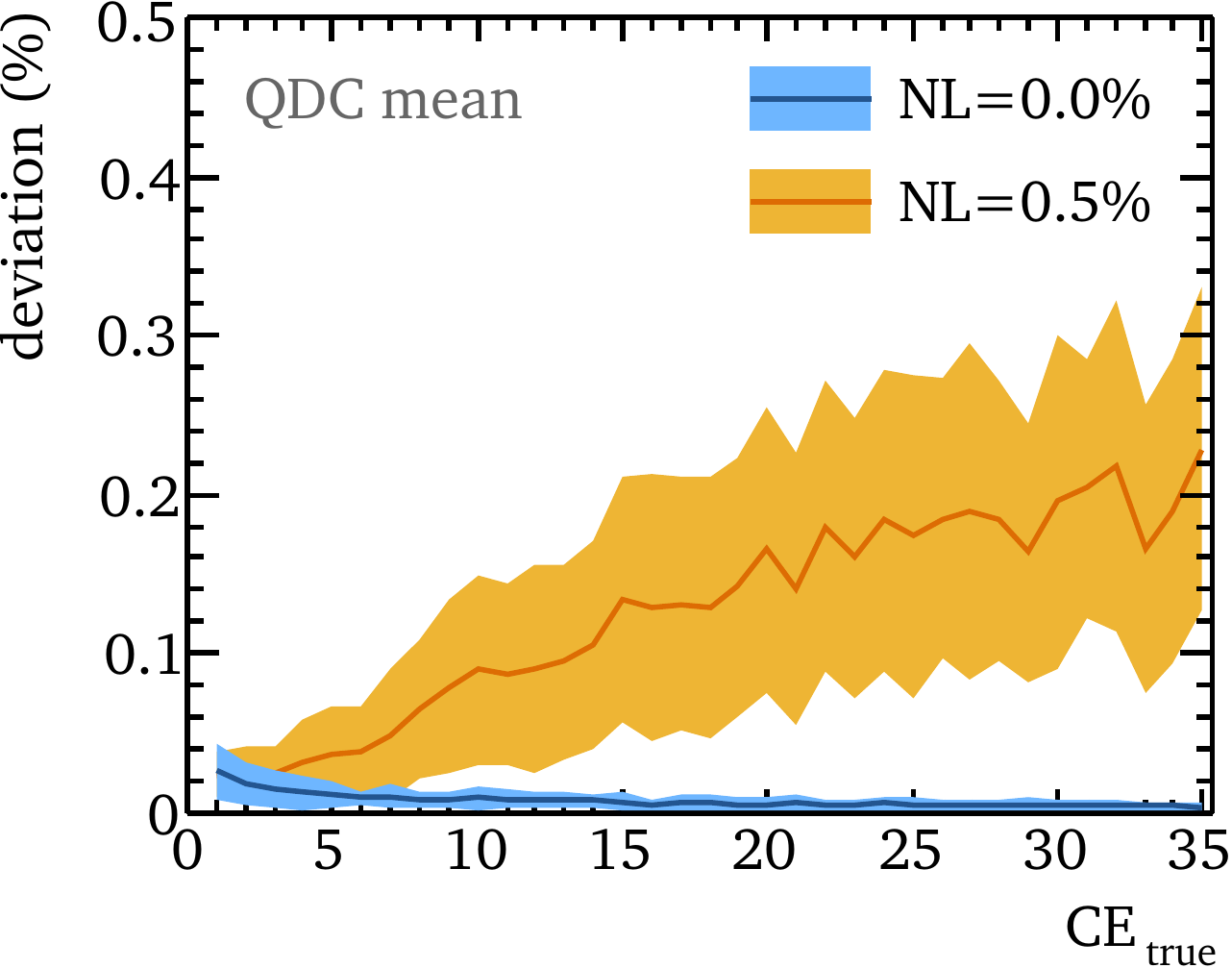}{fig:QDC300NL} 
  \caption{Relative  deviation of the measured $N_{\mathrm{C.e.}}$ from 
    its true value when \prosubref{fig:fit300NL} exploiting single peak resolution and 
    \prosubref{fig:QDC300NL}  when relying  on the mean of the QDC spectrum, respectively.
    The central lines are the mean 
    deviations for $50$ simulation runs per value of $N_{\mathrm{C.e.}}$ with $p=300$, 
    the filled bands represent the RMS around the central value.
    In case of a perfectly linear detector response, both methods perform
    equally well (blue). When a photodetector 
    non-linearity of 0.5\% is simulated (using non-linearity functions with different randomly selected
    parameters for each run), the single-peak based reconstruction
    shows a superior performance (orange).}
  \label{fig:ceFit300}
\end{figure}

The number of parameters to be fitted grows with the number of electrons per detector
channel. For $N_{\mathrm{C.e.}}=20$, peaks up to  $i_\mathrm{C.e.}{\approx} 40$ need to be described 
in the QDC spectrum
and consequently a fit function with ${\approx} 80$ parameters is required. For such a
large number of free parameters, a careful choice of start values for the fit is
essential. Since the distance between neighbouring peaks is constant for a perfectly linear 
photomultiplier and changes only gradually for non-linear gain, an initial estimate for $s$
can be obtained from the discrete Fourier transform of the spectrum. 
From the separation, start values for the Gaussian parameters can be determined by calculating 
the expected mean and width. Since $s$ corresponds to the scale factor between 
initial Compton electrons and QDC bins, a start value for $N_{\mathrm{C.e.}
}$ can be obtained by fitting a Poissonian $P(x')$ with $x'=\frac{x_{QDC}}{s}$ to the 
spectrum.

Figure~\ref{fig:fit300NL} shows the relative deviation of the fit result for $N_{\mathrm{C.e.}}$ 
from its true value for up to $N_{\mathrm{C.e.}}= 35$ in case of a perfectly linear response (blue)
and in the presence of a photodetector non-lin\-e\-arity of $0.5\percent$.
 
Up to ${\sim}25$ Compton electrons, the added non-linearity has little effect. For larger
number of electrons, the description of the spectrum by the fit start values does not model 
the spectrum shape sufficiently well anymore and the fit results starts to degrade.
For comparison, figure~\ref{fig:QDC300NL} shows how much a photodetector non-linearity of 
$0.5\percent$ affects the conventional method using the mean of the QDC spectrum, which is the only 
method for the calculation of the polarisation asymmetry when no single-peak resolution is possible.

\subsection{Application to polarisation measurements at the ILC}\label{sec:ceFit}
  
The results in figure~\ref{fig:ceFit300} where based on accumulating $10^7$ individual measurements.
With the baseline beam parameters of \NU{1312}{bunches} per bunch train at
 \NU{5}{Hz} bunch crossing rate, \NU{3.936\ex{6}}{measurements} could be collected in 
 \NU{10}{minutes}, providing a large enough data sample so that the fit performance
is not limited by statistics.  Since the non-linearity is not expected to change rapidly, 
the actual
polarisation measurement can be done on much smaller datasets, with a determination of the
peak positions from the data taken in the previous \NU{10}{minutes} or, in an offline
analysis, with a moving average over data taken before and after the individual short-time
dataset.

To evaluate the benefits of the fit procedure described above for polarisation measurements
at the ILC, a detector array of 60 channels with a width of \NU{3}{mm} and \NU{0.33}{mm} 
inter-channel spacing was simulated. 
The fast Linear Collider Polarimeter Simulation \texttt{LCPolMC} is
used~\cite{spintracking, Eyser2007} to generate Compton events and subsequently track the
Compton electrons through the polarimeter chicane to the detector.  The
beam and laser parameters used for the simulation are chosen according to the expectations
at the polarimeter locations for the ILC TDR beam parameters~\cite{tdr}. 
The resulting number of Compton electrons per channel is displayed in figure~\ref{fig:ncomp60chan}. 
It is below $N_{\mathrm{C.e.}}=30$  for all channels,  and consequently the requirement for single-peak 
resolution given by equation~\ref{eqn:quartzmotivCondition} is fulfilled.

\subsubsection{Robustness against non-linear detector response}

Based on figure~\ref{fig:ncomp60chan}, the same simulation chain as before is applied to
all channels for both laser helicities and the average number of Compton electrons per
channel is extracted both from single-peak fitting and from the mean of the QDC spectrum. 
The polarisation is then reconstructed as described in section~\ref{sec:intro}.

\begin{figure}[tb]
  \centering
  \subfloat{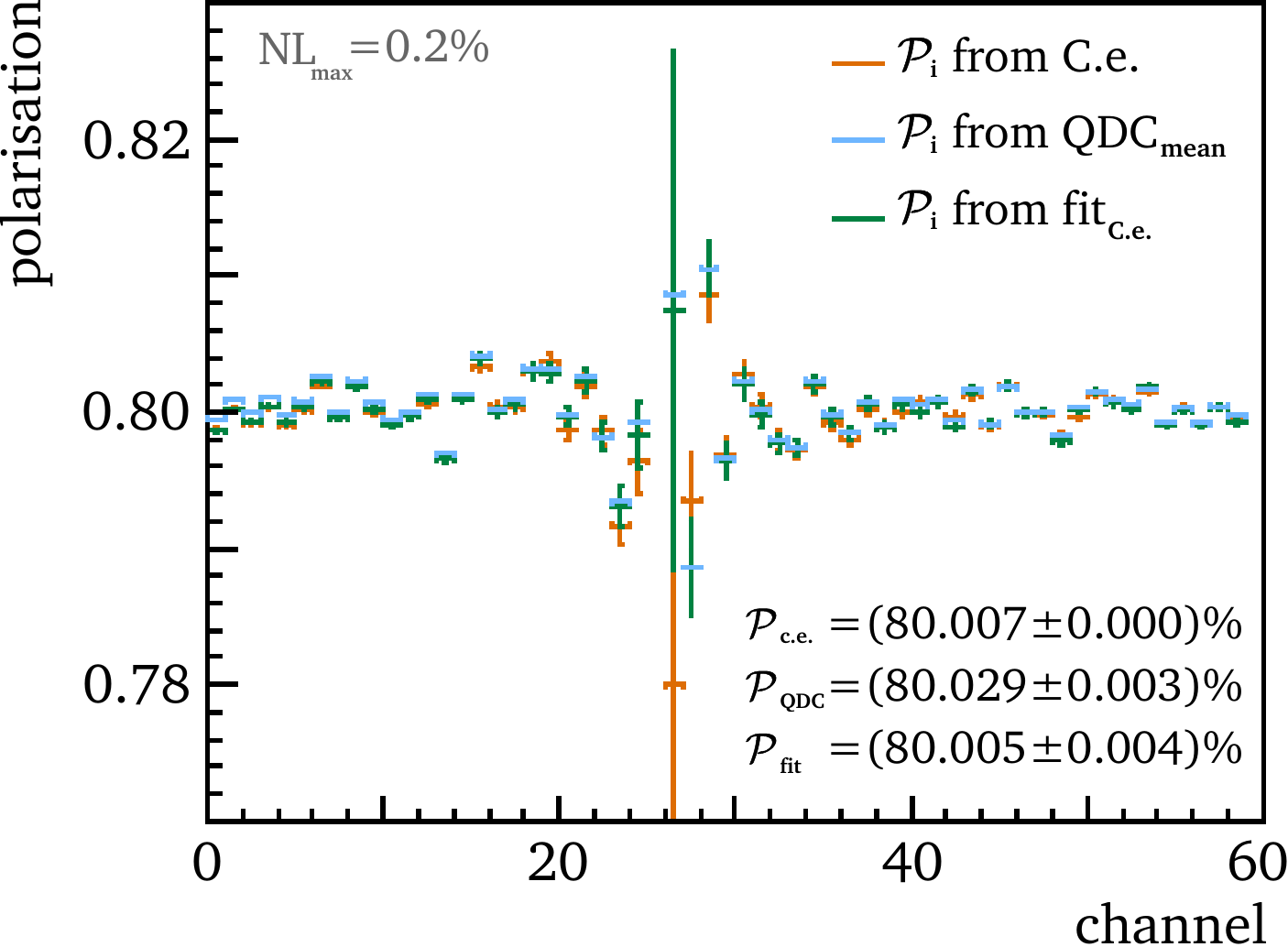}{fig:mypol80NL002}
  \hfill
  \subfloat{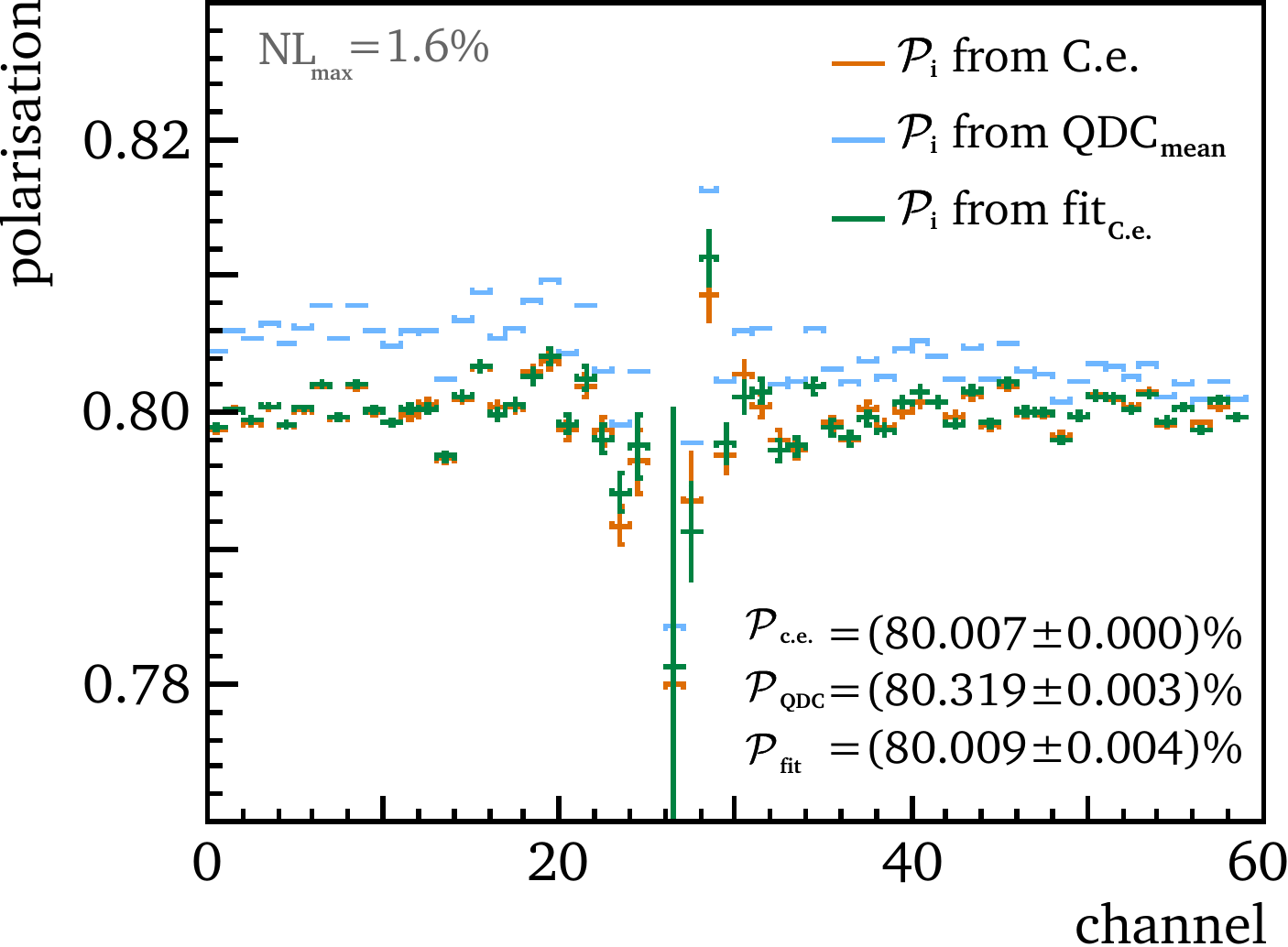}{fig:mypol80NL016}
  \caption{Polarisation calculated for each detector channel $i$ for a simulation of 80\%
    polarisation with  \prosubref{fig:mypol80NL002} 0.2\%  and \prosubref{fig:mypol80NL016} 
    1.6\% photodetector non-linearity, respectively. 
    The orange markers show the polarisation
    $\mathcal{P}_\mathrm{C.e.,i}$ calculated from the primary Compton electrons, the blue
    markers the calculation $\mathcal{P}_\mathrm{QDC,i}$ from the mean of the QDC
    spectrum, and the green markers the results $\mathcal{P}_\mathrm{fit,i}$ for the
    number of Compton electrons determined from the fit to the QDC spectra. 
    The fluctuations around channel 30 are due to the zero-crossing of the 
    asymmetry (c.f.\ figure~\protect\ref{fig:ncomp60chan}).}
  \label{fig:mypol80}
\end{figure}

Figure~\ref{fig:mypol80NL002} compares the polarisation measurements obtained via the two 
reconstruction methods in case of a rather small non-linearity of $0.2\%$ with the ideal 
result obtained directly from the true Compton electron spectrum. While the results from the 
three different methods exhibit no striking difference in this case, a
larger non-linearity of $1.6\percent$ as shown in figure~\ref{fig:mypol80NL016} distorts the 
QDC mean and therefore the measured polarisation significantly, with a relative deviation of 
0.39\percent, while the polarisation calculated from the fit results appears to be 
unaffected, with a deviation of 0.01\percent.

\begin{figure}[tb]
  \centering
  \subfloat{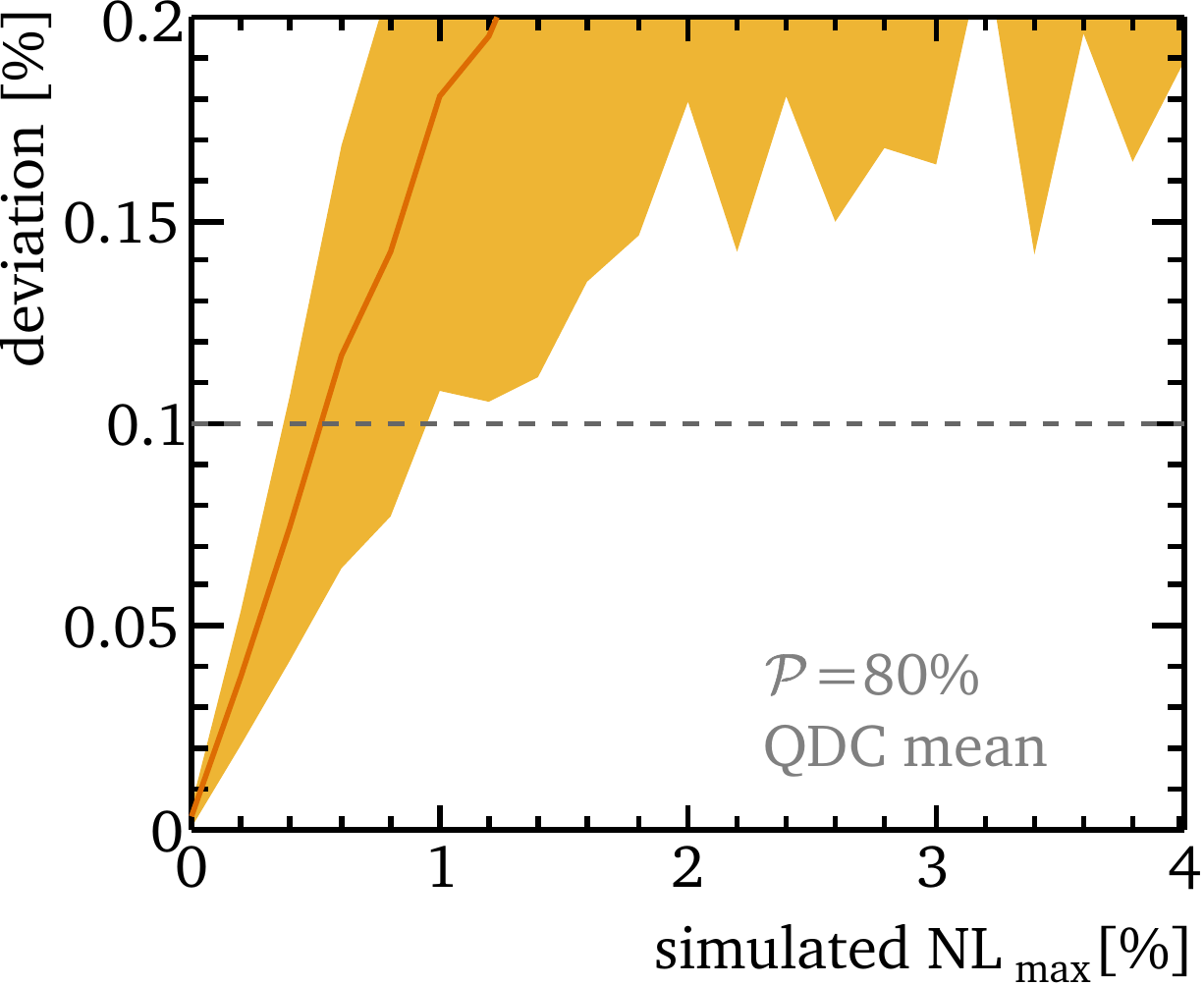}{fig:polfit80QDC}
  \hfill
  \subfloat{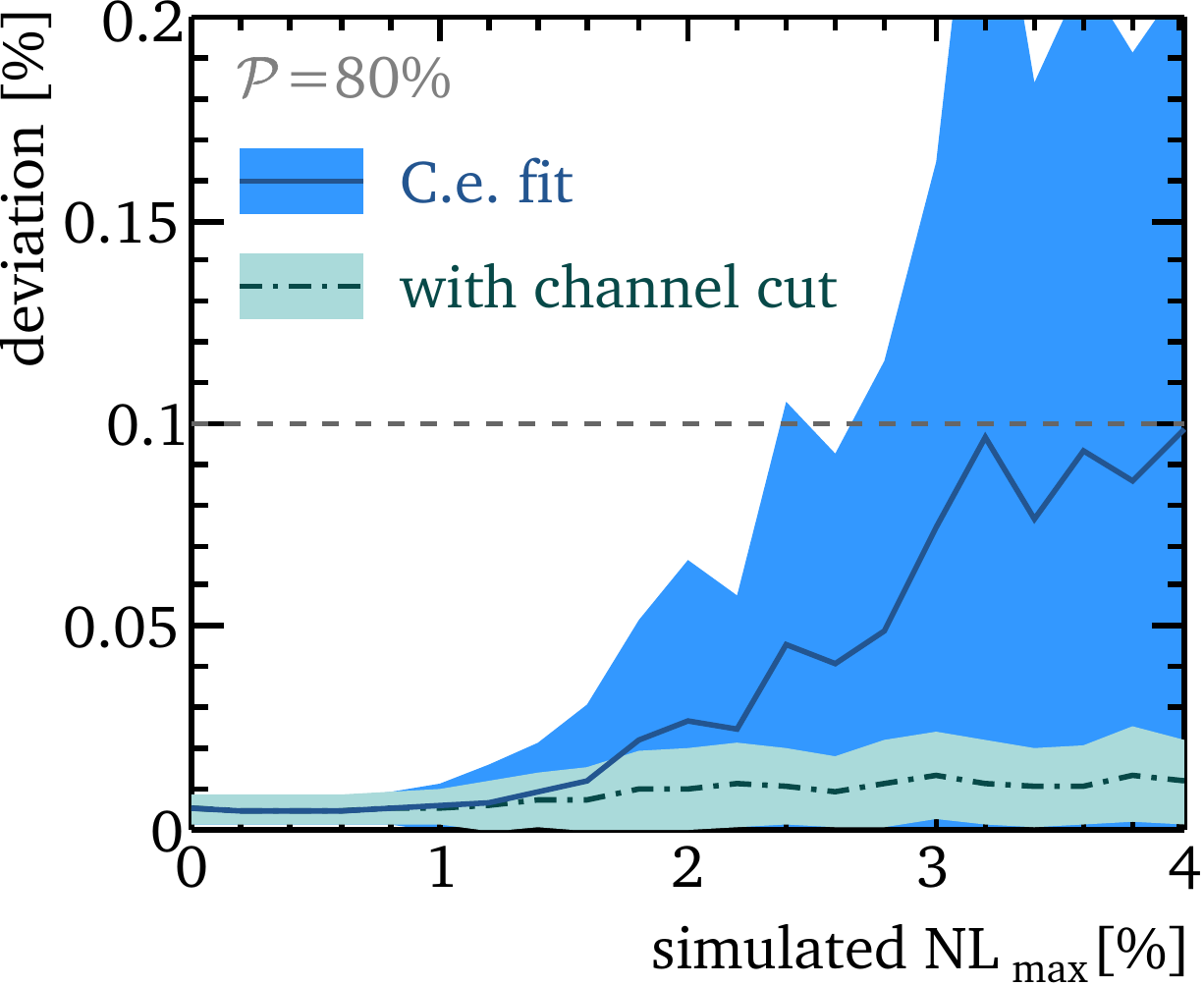}{fig:polfit80Fit}
  \caption{Deviation between the result from the polarisation $\mathcal{P}_\mathrm{C.e.}$
    calculated from the primary Compton electrons and \prosubref{fig:polfit80QDC} the
    polarisation $\mathcal{P}_\mathrm{QDC}$ determined from the mean of the QDC spectrum,
    and \prosubref{fig:polfit80Fit} the polarisation $\mathcal{P}_\mathrm{fit}$ calculated
    with the number of Compton electrons determined from the fit of the QDC spectra, using 
    all detector channels (blue, solid line) or only those channels $i$ whose  polarisations
    $\mathcal{P}_\mathrm{fit,i}$ agree within $\pm 1\% \cdot
    \mathcal{P}_\mathrm{fit}$ with the polarisation calculated from all
    channels (green, dash-dotted line).}
  \label{fig:polfit80}
\end{figure}

For a more systematic survey of the impact of non-linearities, the same procedure as for
figure~\ref{fig:mypol80NL002} was repeated for non-linearities from $0\percent$ to $4\percent$ 
in $0.2\percent$ steps. For each step, random parameters for 100 different functions 
for the photomultiplier non-linearity were picked and used for the spectrum generation. 
 The mean and RMS of the deviation between the polarisation determined
from the mean of the generated QDC with respect to the result for the primary electron
calculation is shown in figure~\ref{fig:polfit80QDC}. The error budget for the detector
linearity contribution to the polarisation measurement is $0.1\percent$ {(see 
table~\ref{tab:polsys})}. The method using the mean of the QDC surpasses the allocated limit for
non-linearities $\gtrsim 0.4\percent$. The results for the polarisation calculation from
the fitted number of Compton electrons is shown in figure~\ref{fig:polfit80Fit}. Fitting
the QDC spectra instead of using the mean can compensate the photodetector non-linearity
well enough to stay within the error budget for non-linearities up to $\sim 2.2\percent$.

This can be further improved by introducing a second step in the calculation of the
polarisation to eliminate the contribution from outlier channels with poor fit performance.
A rough first approach to demonstrate this is a recalculation of the polarisation using only 
the channels which agree within $\pm1\percent \cdot \mathcal{P}_\mathrm{fit}$ with the 
polarisation calculated from all channels. The fraction of channels consequently used in 
the calculation is larger than 90\percent.
Using less channels will decrease the statistical precision slightly. 
However, the benefits of this approach in dealing with larger detector
non-linearities, compared to the use of all channels, are clearly visible in figure~\ref{fig:polfit80}, 
where the deviation
between the polarisation thus obtained with respect to the result for the primary Compton
electrons is shown in green. Even for photodetector non-linearities of $4\percent$, this deviation
is below $0.03\percent$, i.e.\ well within the error budget of~$0.1\percent$, which would
be a contribution towards reducing the overall systematic uncertainty on the polarisation
measurement. For the method using the QDC mean, such an approach would not lead to such 
improvements, since all channels are systematically affected by the non-linearity, as one 
can see when looking at the example in figure~\ref{fig:mypol80NL016}.

\subsubsection{Detector alignment}
Next to non-linearities, the detector alignment with respect to the Compton electron fan is the
other large contribution to systematic uncertainties on the polarisation measurements that
should be taken into account in the design of the polarimeter detectors, c.f.\ 
table~\ref{tab:polsys}. The largest contribution originates from the alignment in the plane
of the magnetic chicane, since it directly affects the analysing power predicted for each channel.

\begin{figure}[tb]
  \centering
  \subfloat{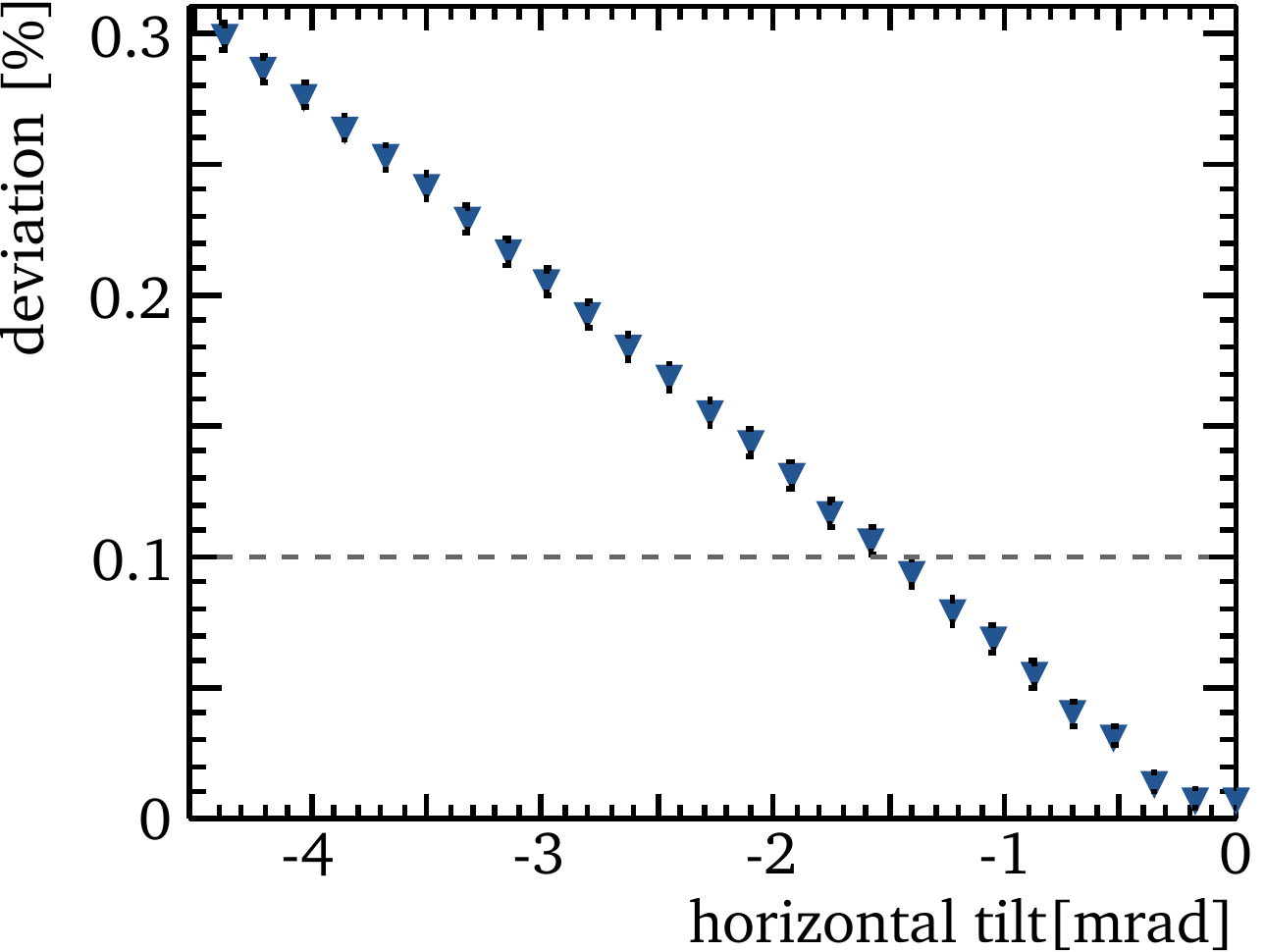}{fig:lcpolRotDev}
  \hfill
  \subfloat{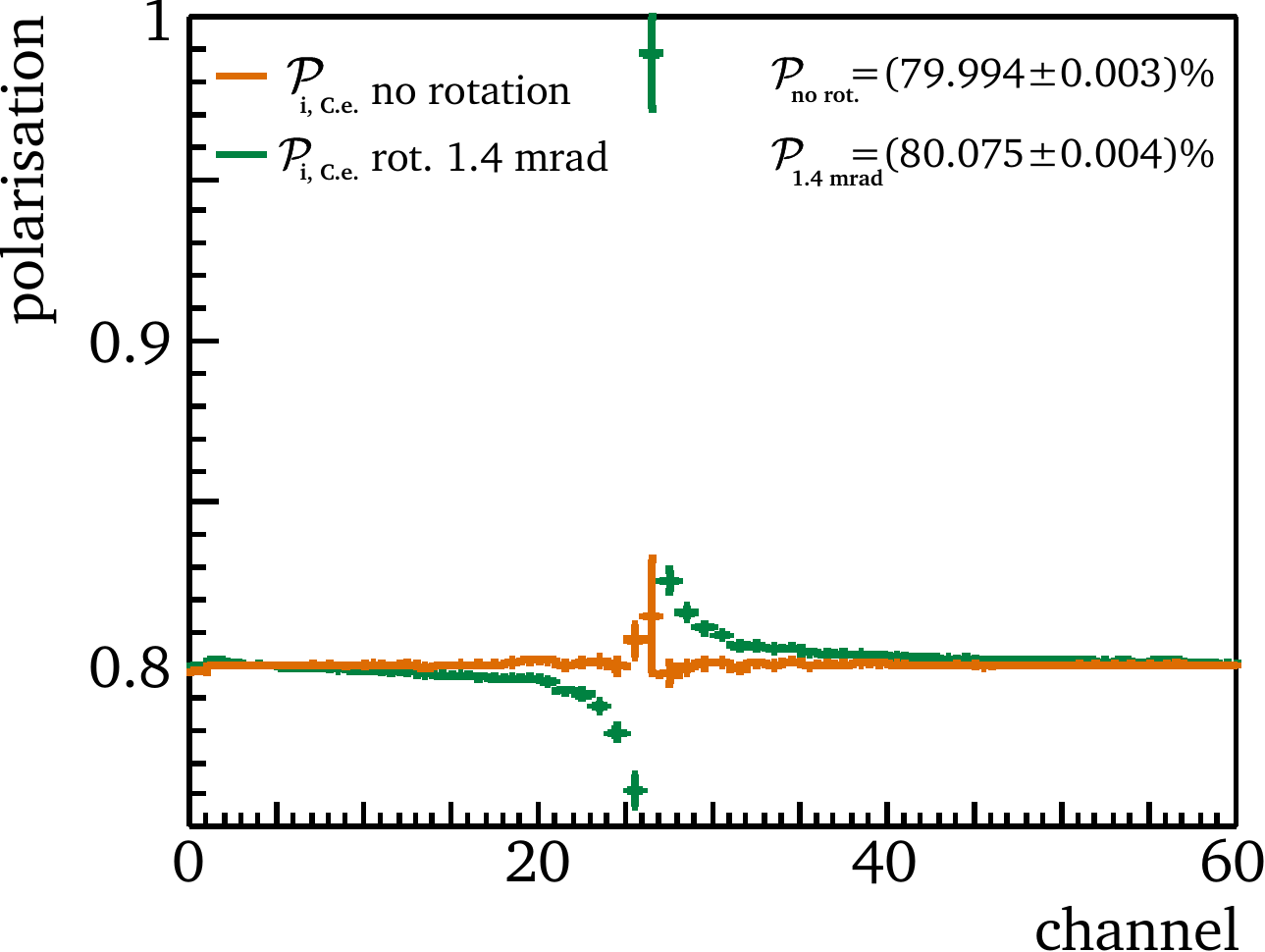}{fig:polfit80Rot}
  \caption{\prosubref{fig:lcpolRotDev} Deviation of the polarisation obtainable from the primary Compton electrons
 in dependence of detector rotation in the horizontal plane for a  simulation of 80\percent polarisation, and 
   \prosubref{fig:polfit80Rot} the polarisation calculated for  each detector channel $i$
 without detector misalignments and in case of a \NU{1.4}{\mathrm{mrad}} tilt in the horizontal plane.}
  \label{fig:lcpolRot}
\end{figure}

Rotations of the detector in the deflection plane of the chicane (i.e.\ in the horizontal plane in case of 
the upstream polarimeter) affect the path of the electrons through the channels.
The amount of Cherenkov light per electron and its propagation to the photodetector could be altered, 
thus changing $p$.
This effect has to be found negligible for all simulated angles up to $5^{\circ}$.
More importantly, such horizontal tilts can introduce cross-talk if a Compton electron
passes through more than one channel. Figure~\ref{fig:lcpolRotDev} shows the impact on the polarisation: 
tilts of the detector in the horizontal plane result in deviations of $0.07\percent/\text{mrad}$. 
For the ILC polarimeters, a tilt alignment of   \NU{1}{\text{mrad}} or better is considered feasible, 
thus the corresponding systematic uncertainty is within the allocated error budget. Horizontal rotations of 
the detector have the strongest impact on the channels with the smallest asymmetries near the zero-crossing 
of the asymmetry, as can be seen in figure~\ref{fig:polfit80Rot}. This characteristic pattern can be resolved 
with high granularity as in case of the 60 channels considered here. Consequently, the systematic uncertainties 
could be further reduced by identifying the presence of misalignments and correcting for them or excluding the 
most strongly affected channels.

Misalignments of the detector position in the horizontal direction can 
be controlled by monitoring the fraction of Compton electrons in the detector 
channel at the Compton edge. 
Figure~\ref{fig:lcpolOffPercent} shows how the fraction
of Compton electrons in the detector channel containing the Compton edge 
over the total number of detected electrons changes as a function of a horizontal 
offset of the detector.
\begin{equation}
  \label{eq:frac60}
  y=\frac{N_\mathrm{C.e}(i_\mathrm{edge})}{\sum\limits_{i=1}^{N_{\mathrm{chan}}} N_\mathrm{C.e}(i)}.
\end{equation}
Averaging over the measurements with both laser helicities eliminates any dependence on the
polarisation, as can be seen in figure~\ref{fig:lcpolOffPercent}. 
A fit of a linear function to this ratio
yields a slope $m$ of \NU{-0.343}{\frac{\percent}{mm}}. 
It should be noted that the expected number of Compton electrons hitting the front face of each
detector channel for a given offset is independent from intrinsic detector 
uncertainties and thus the precision 
of $m$ is only limited by the knowledge of the field of the chicane magnets, which is expected
to be much better than the detector-beamline alignment.

\begin{figure}[tb]
  \centering
  \subfloat{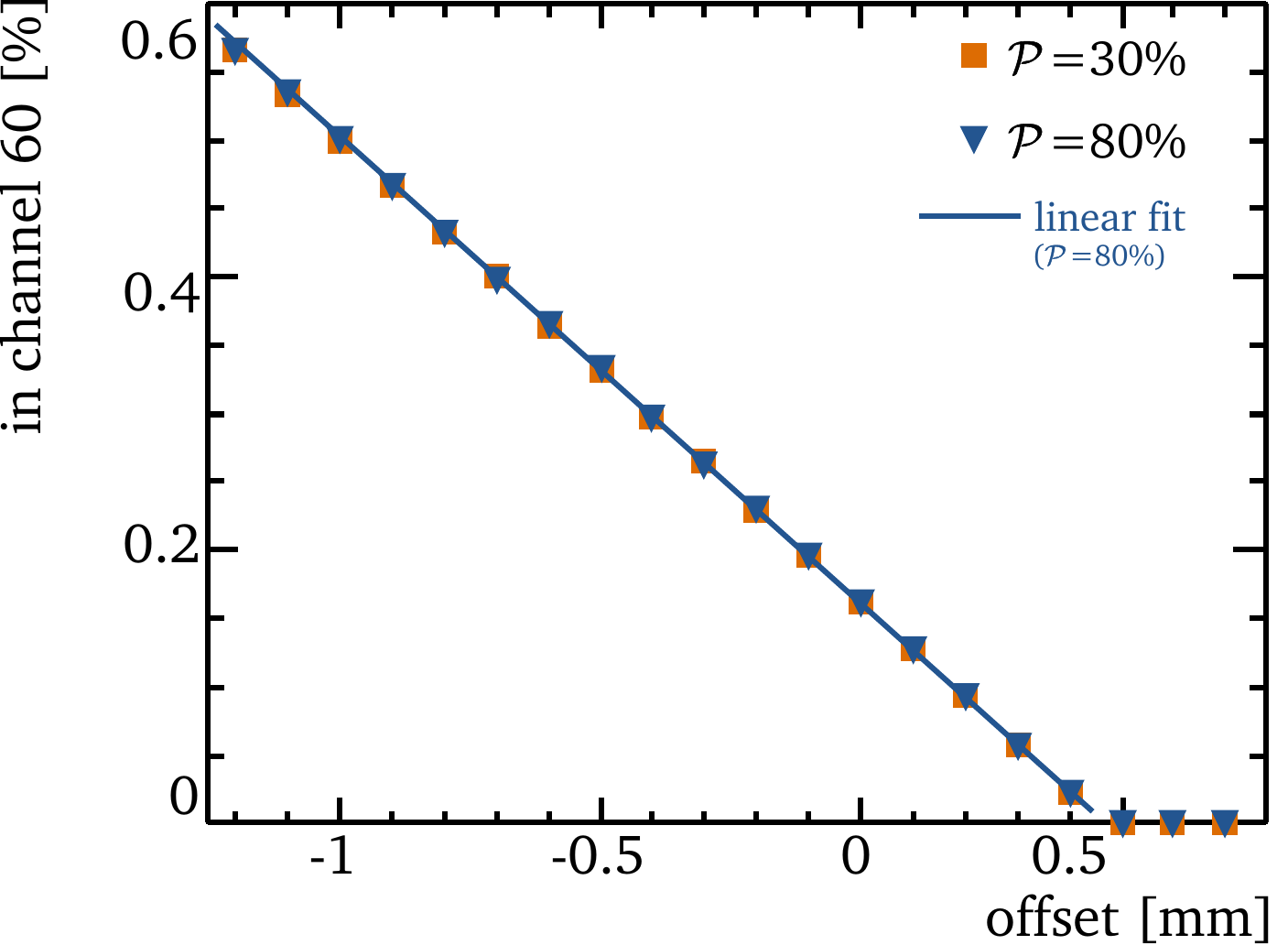}{fig:lcpolOffPercent}
  \hfill
  \subfloat{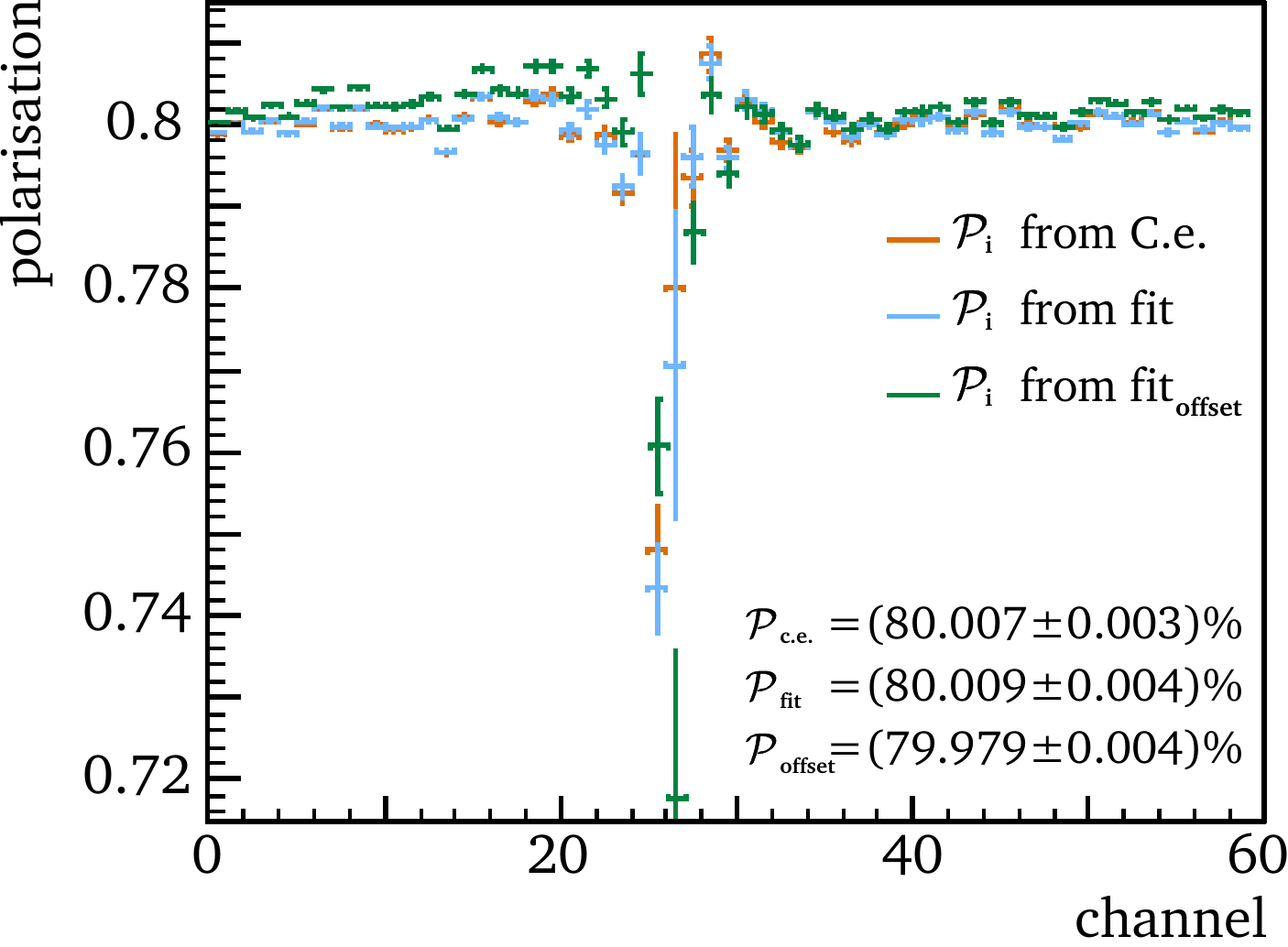}{fig:polfit80Off}
  \caption{\prosubref{fig:lcpolOffPercent} 
  Fraction of Compton electrons expected in the detector channel containing the
    Compton edge as a function of a horizontal offset for $\Pol = 30\percent$ and 
    $80\percent$ polarisation, averaged over the
    measurements with both laser helicities. A first order polynomial has been fitted to the 
    $80\percent$ case.
    \prosubref{fig:polfit80Off} Polarisation $\mathcal{P}_\mathrm{offset,i}$ calculated for 
    each detector channel $i$ for a 
    simulation of 80\percent polarisation with a horizontal misalignment of  $\sim\NU{50}{\mu m}$
    (green). For comparison, the blue (orange) markers show the polarisation $\mathcal{P}_\mathrm{fit,i}$ 
    ($\mathcal{P}_\mathrm{C.e.,i}$) obtained without 
    misalignment from the fit to the QDC spectra (the primary Compton electrons).}
  \label{fig:lcpolOff}
\end{figure}

In case of single-peak resolution, $\Delta N_\mathrm{C.e}(i) = 0.1\percent$ is estimated 
for the determination of the Compton 
electron number per channel based on the performance shown in figure~\ref{fig:fit300NL},
which holds even in case of a not perfectly linear detector response~\footnote{The 
same method can also be applied when relying on the mean of QDC spectrum only. In that
case, however, the precision degrades significantly in case of non-linearities, c.f.\ 
figure~\ref{fig:QDC300NL}.}.
By propagating the assumed uncertainty for each channel to the ratio $y$, 
$\Delta y = \NU{7.9\ex{-4}}{\percent}$ 
is obtained, which translates into a horizontal alignment precision of 
$\Delta x = |\Delta y / m | = \NU{2.3}{\mu m}$.

Such an alignment is sufficient to meet the requirements on the 
contributions to the systematic uncertainty of the polarisation measurement from
the detector alignment. Figure~\ref{fig:polfit80Off} shows the impact that of 
horizontal misalignment of $\NU{50}{\mu m}$ in a simulated polarisation measurement.
The deviation of $\Delta \mathcal{P} /  \mathcal{P} =0.03\percent$ 
caused by the offset is well within the allocated error budget. 
Thus fitting the 
single-electron-peak spectra is also a promising approach towards meeting the 
requirements for the detector alignment.

%% file: 03_design.tex
\section{Design of a quartz Cherenkov detector}
\label{sec:design}

For the baseline gas Cherenkov detector with \NU{1}{cm} wide channels,
the mean number of Compton electrons per channel reaches up to 
$N_{\mathrm{C.e.}} \simeq 100$, while the average yield of photoelectrons per Compton 
electron is $p=6.5$~\cite{testbox_paper}. This configuration is far from
fulfilling equation~\ref{eqn:quartzmotivCondition}.
In order to develop a Cherenkov detector suitable to achieve single-peak resolution,
 there are two possibilities: The number of Compton electrons per 
 channel can be reduced by a factor~2-3 by 
 building smaller channels for the polarimeter. In addition, it is mandatory to increase
 the number of photoelectrons per Compton electron substantially. 
As an approach to achieve the latter, the use of quartz as Cherenkov material 
is considered. The higher refractive index compared to Cherenkov gases results 
in a much higher light yield, as described by the Frank-Tamm 
formula~\cite{Tamm1960}. Compared to e.g.\ perfluorobutane gas, the use of 
quartz would translate into an approximately $200$ times higher intensity of 
the emitted Cherenkov light for relativistic particles. However, this will be 
somewhat mitigated by other effects, such as increased absorption as the produced 
light travels to the photodetector. In this section, we discuss a simulation study 
of these effects and derive
a suitable design for a prototype detector.

\subsection{\GF simulation}
\label{sec:g4sim}

\begin{figure}[tb]
  \centering
  \subfloat[0.495\linewidth]{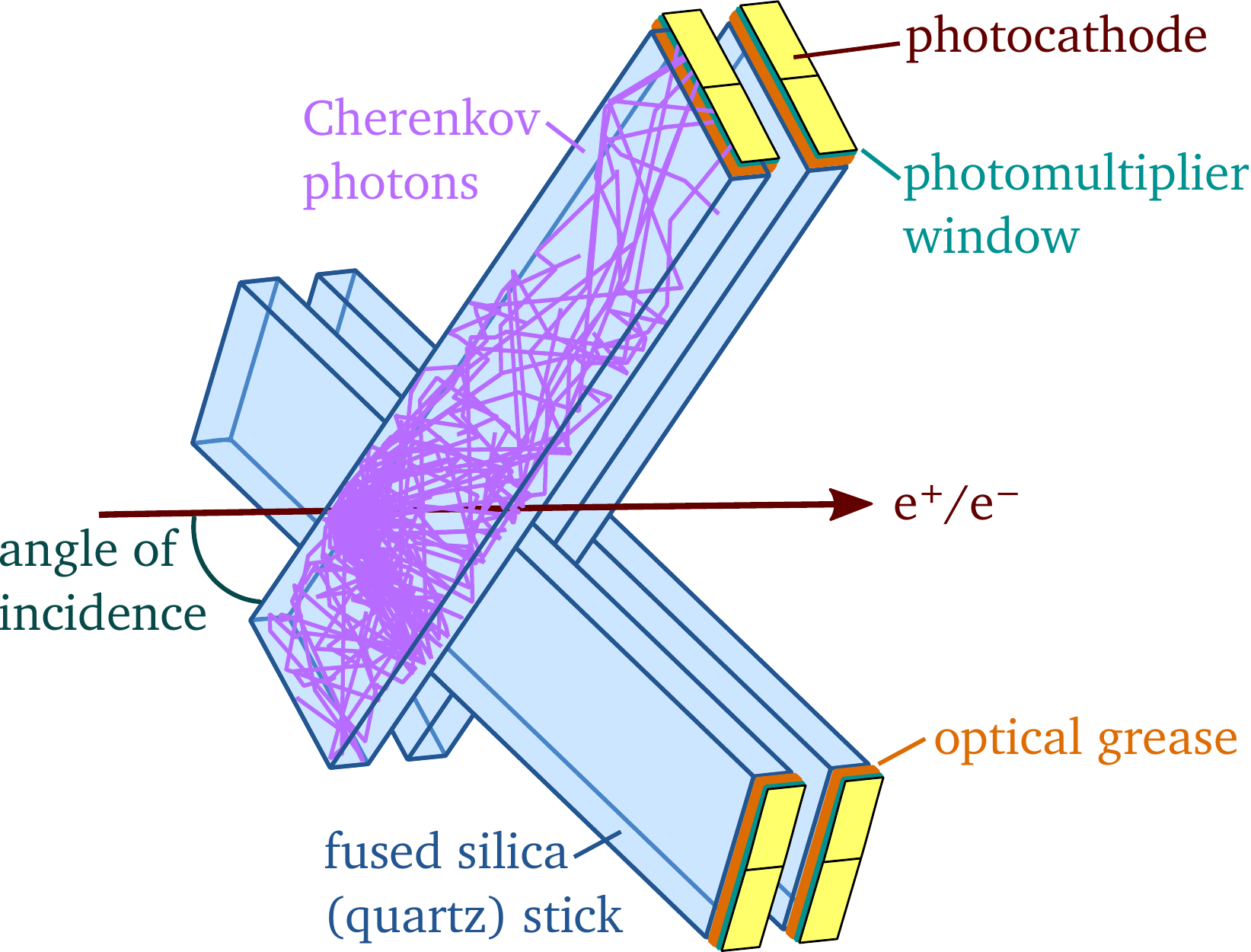}{fig:sim4sticks-rot}
  \hfill
  \subfloat[0.405\linewidth]{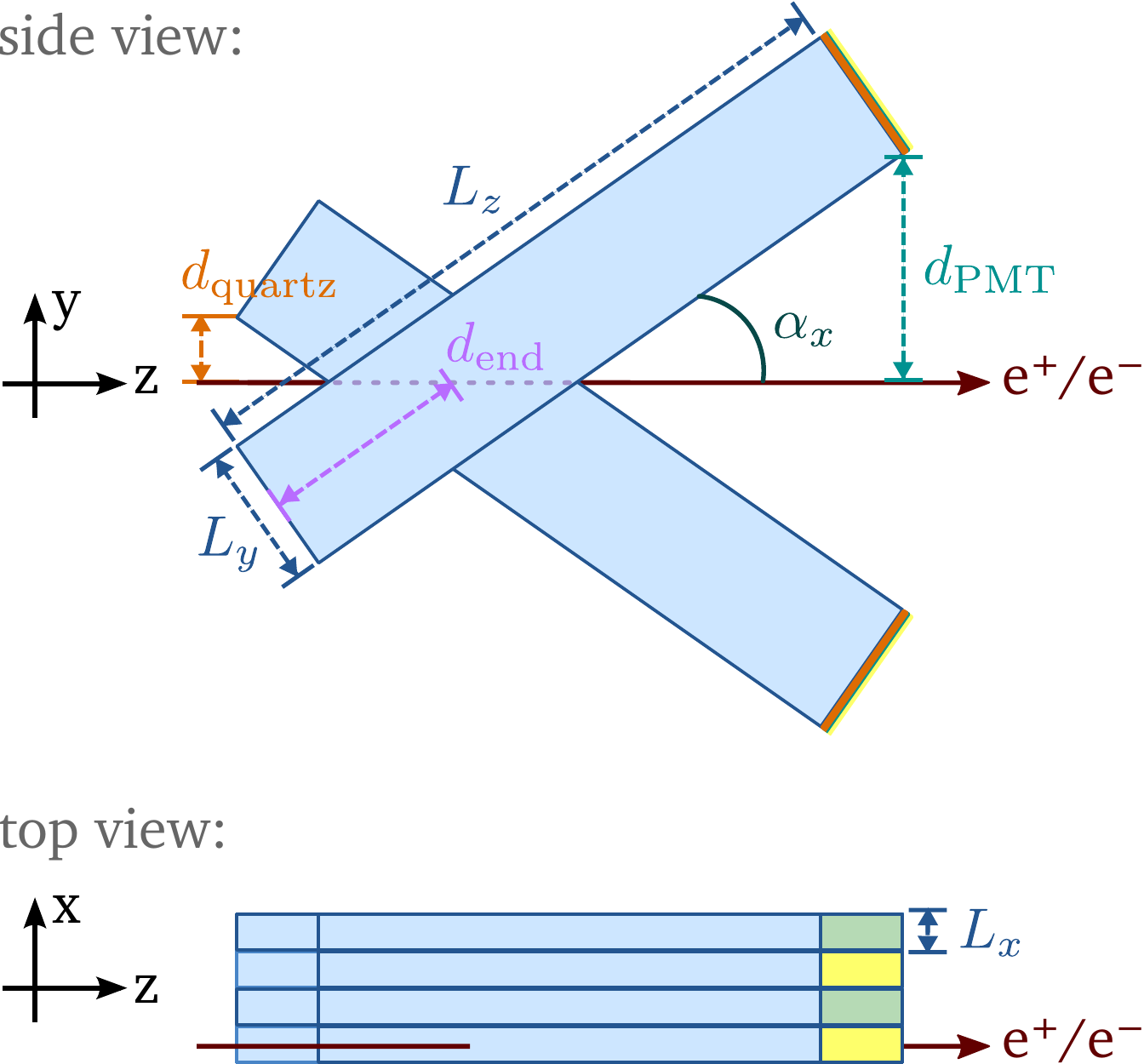}{fig:sim4sticks-TS}
  \caption{Sketch of the simulated geometry. The left side \prosubref{fig:sim4sticks-rot} shows some components of the simulation, in addition demonstrating the added space for electronics due to the rotation of every second quartz block. For a better overview, on the right side \prosubref{fig:sim4sticks-TS} the same geometry is depicted in side- and top-view, along with the geometrical parameters varied in the design studies: the channel dimensions width $L_x$, height $L_y$, length $L_z$; the incidence angle $\alpha_{x}$; as well as the space between quartz and electrons $d_{\mathrm{quartz}}$, between electrons and PMT $d_{\mathrm{PMT}}$, and the distance between the end of the channel and the point where the electron crosses the detector axis $d_\mathrm{end}$.}
  \label{fig:sim4sticks}
\end{figure}

To study the feasibility of a quartz detector for ILC polarimetry and investigate 
different design options, as well as for comparison to data from a prototype 
detector, a detailed simulation of the detector concept based on \GF~\cite{geant4} 
has been developed. 
The simulated geometry is depicted in figure~\ref{fig:sim4sticks}. The main 
elements of the simulated setup are quartz blocks surrounded by a thin layer 
of air and/or aluminium foil, the entrance window and photocathode of a 
photomultiplier tube (PMT) and a layer of optical grease between the 
quartz block and the PMT window. To increase the space available for the 
photodetectors and their readout, every other quartz block is flipped with 
respect to its neighbours.

To produce Cherenkov light, electrons are shot through the quartz. For the Cherenkov photons, all relevant processes are simulated, in particular absorption according to the absorption length of the relevant material, Rayleigh scattering, and boundary processes at the surface between two different media. For the detector material, 
the optical properties of \Spec~\cite{heraeusDandP} have been implemented, and for 
the optical grease the properties of Cargille fused silica matching liquid code 06350 and 50350~\cite{Cargille}. To describe the behaviour of the photons at boundaries between materials, the \textsc{UNIFIED} model~\cite{Moisan96} of \GF is used.

\begin{figure}[tb]
  \centering
     \includegraphics[width=\linewidth]{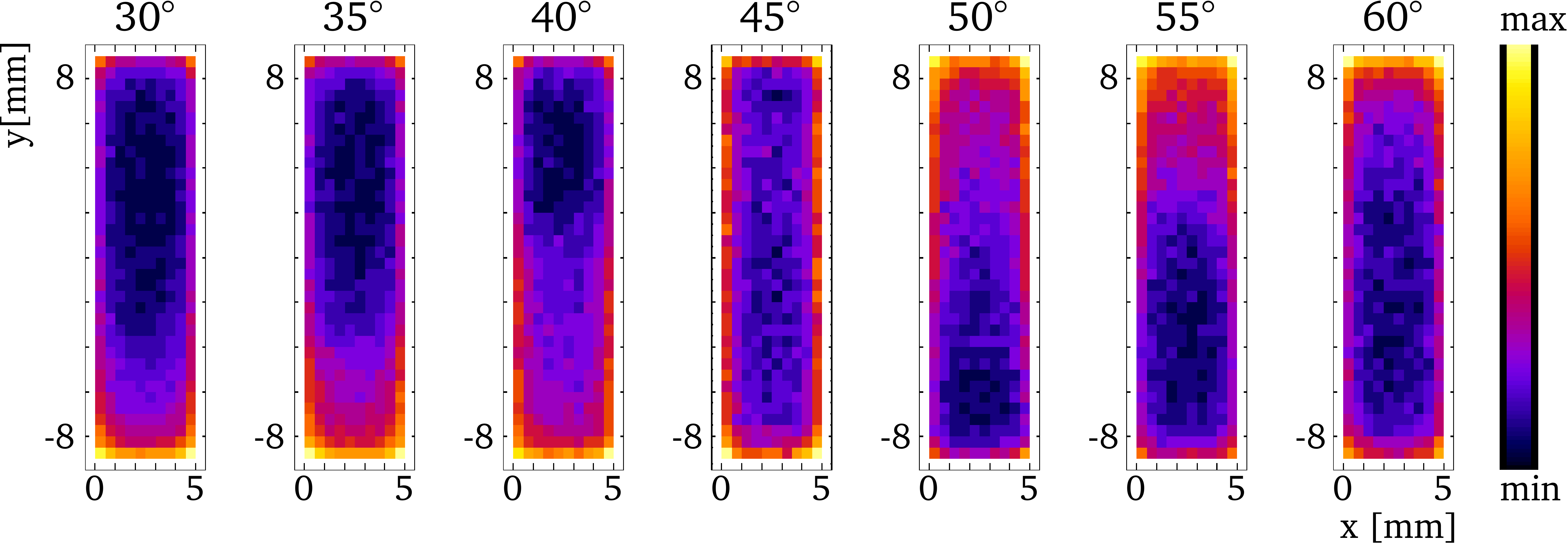}
  \caption{Light distribution on the photocathode for different incidence angles of the electron. The position on the photocathode surface is coloured according to the relative light concentration: the regions with most light for each angle are shown in yellow, with a colour gradient to the regions with the least light in dark blue.}
    \label{fig:AyLightColz}
\end{figure}

Various geometrical properties were varied to study their impact on the light yield: 

\begin{itemize}
\item A larger channel width $L_x$ reduces the number of reflections at the channel 
walls a photon undergoes before detection, which increases the probability to 
arrive at the photodetector.
\item  A larger channel height $L_y$ increases the path length of the electron 
crossing the quartz channel and consequently the amount of Cherenkov light produced.
 Additionally, as in the case of larger width, the number of reflections is reduced.
  Both effects result in an enhanced photon yield.
\item A shorter channel length $L_z$ reduces the distance the photons have to 
travel before detection, which increases the light yield by causing less photons 
to be absorbed inside the quartz.
\item The incidence angle $\alpha_{x}$ of the electron into the quartz channel 
also changes the path length of the electron crossing the quartz channel for 
light production, with a higher light yield for small angles. Another, though 
less pronounced, impact on the number of photons is due to the fact that for 
angles close to the Cherenkov angle, a larger fraction of the photons can reach 
the detector without being reflected on the narrow side faces. 
The spatial distribution of the light on the photodetector surface  is the most 
uniform under such angles, whereas for smaller or larger angles the intensity 
distribution is less homogeneous, as illustrated in figure~\ref{fig:AyLightColz}. 
This effect could possibly aid in the angular alignment of the detector if a 
photodetector with multiple anodes per quartz channel was used.
\end{itemize}

\begin{figure}[tb]
  \centering
  \includegraphics[width=1.\linewidth]{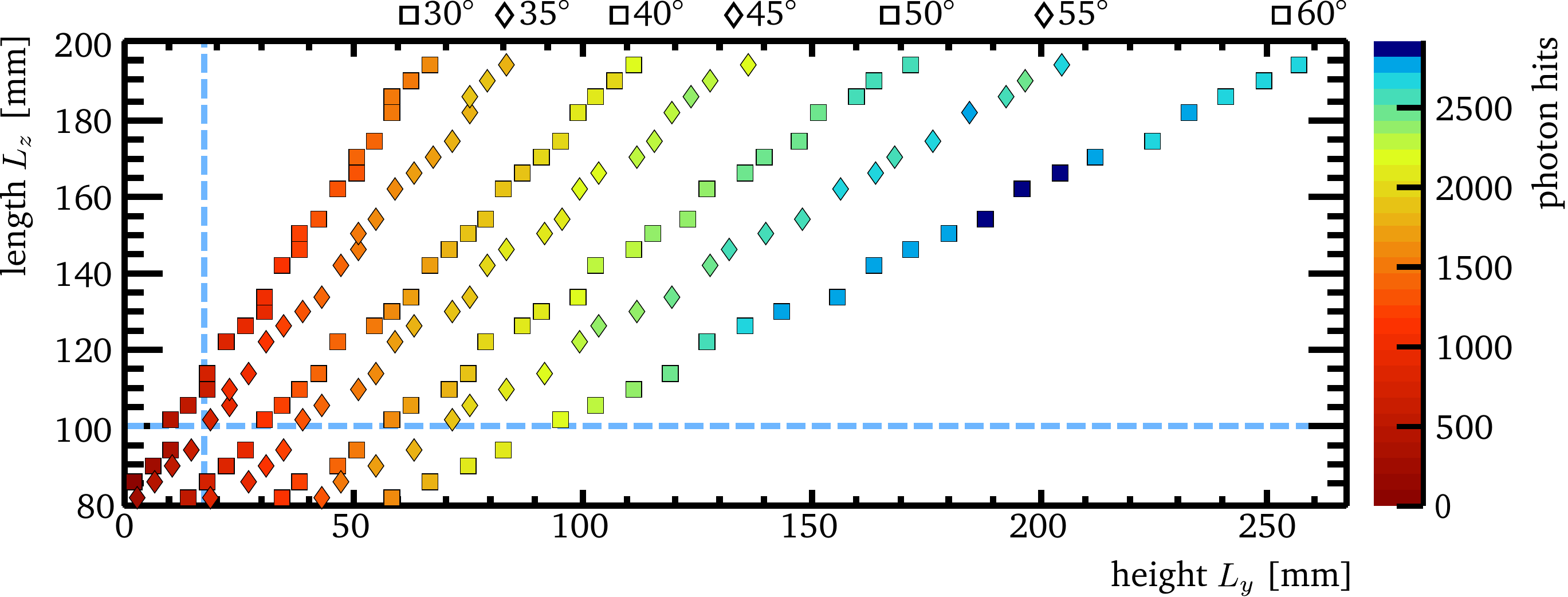}
  \caption{Photon hits for different quartz lengths, heights and angles relative to the electron chosen such that conditions \ref{eq:dist3cmLy} and \ref{eq:dist3cmEnd} are fulfilled.  The dashed blue lines indicate the length and height used for the prototype detector.}
  \label{fig:dist3cm}
\end{figure}

When combining the parameters listed above to chose a suitable detector geometry, some considerations have to be taken into account in addition to the light yield. Small channel width lead to less Compton electrons per detector channel and therefore aid in fulfilling the condition for reaching single peak resolution. Another consideration is that long, narrow channels and large angles move the photodetector further from the Compton electrons to protect them from contact with high energetic particles. To take this into account, the channel height and the electron entrance point were varied along with the length, such that the space between the lower end of the quartz and the electron plane stayed constant at $d_{\mathrm{quartz}}=\NU{10.0}{mm}$ and the shortest distance between the electrons and the photomultiplier was $d_{\mathrm{PMT}}=\NU{30.0}{mm}$.
This translates into a channel height $L_y$ of
\begin{equation}
\label{eq:dist3cmLy}
L_y = \frac{2 \cdot L_{z} \cdot \sin(\alpha_{x}) - d_{\mathrm{quartz}} - d_{\mathrm{PMT}}} {2 \cdot \cos(\alpha_{x})} 
\end{equation}
and a distance $d_\mathrm{end}$ between the end of the detector channel and the electron crossing of the detector axis of 
\begin{equation}
\label{eq:dist3cmEnd}
d_\mathrm{end}=\frac{d_{\mathrm{quartz}} + L_y \cdot \cos(\alpha_{x})}{\sin(\alpha_{x})}
\end{equation}
for each simulated channel length $L_z$ and incidence angle $\alpha_{x}$. The results for channels with a width of $L_x=\NU{5}{mm}$ considering different lengths in the range \NU{50}{mm} -- \NU{300}{mm} and angles in the range 30\degree -- 60\degree are displayed in figure~\ref{fig:dist3cm}. In this range of parameters, the best photon yield of ${\sim} \NU{3000}{photons}$ was achieved for $\alpha_{x}= 60\degree$ and  $L_z=\NU{160.0}{mm}$, which allows a channel height of  $L_y=\NU{197.2}{mm}$ while keeping the chosen  safety margin between incoming electrons and readout electronics. A mechanical setup with such nearly square-shaped channels rather than elongated ones might pose some challenges, e.g.\ covering the readout face of high but narrow with conventional PMTs does not seem reasonable with standard photomultiplier geometries. 
However figure~\ref{fig:dist3cm} shows that there is a large variety of configurations with less extreme aspect ratios which still offer an impressive light yield in the order of ${\sim}1500$ to ${\sim}2000$ photons and which would be more straightforward to implement.

For the construction of the prototype detector (section~\ref{sec:prototype}), smaller dimensions of $L_y=\NU{18.0}{mm}$ and $L_z=\NU{100.0}{mm}$ were chosen, despite the prediction of less (${\sim}600$) photon hits. This choice was partly made to maintain the angular flexibility to scan the full range from 30\degree -- 60\degree to confirm the  incidence angle's effect on the light distributions, but also to match the dimensions of the available photodetectors, since the time and cost for tailor-made photomultipliers were not considered reasonable before a successful demonstration of the concept.

The photon hits given above are the number of photons that reach the photodetector surface. Most common photodetectors have a detection efficiency well below 100\percent. A more realistic estimate of the photon yield which could be achieved will therefore have to take the characteristics of the photodetector into account to evaluate the number of detected photons. Figure~\ref{fig:qQEs} shows the quantum efficiencies of two PMTs with good sensitivity in the UV range.

\begin{figure}[tb]
  \centering
  \subfloat{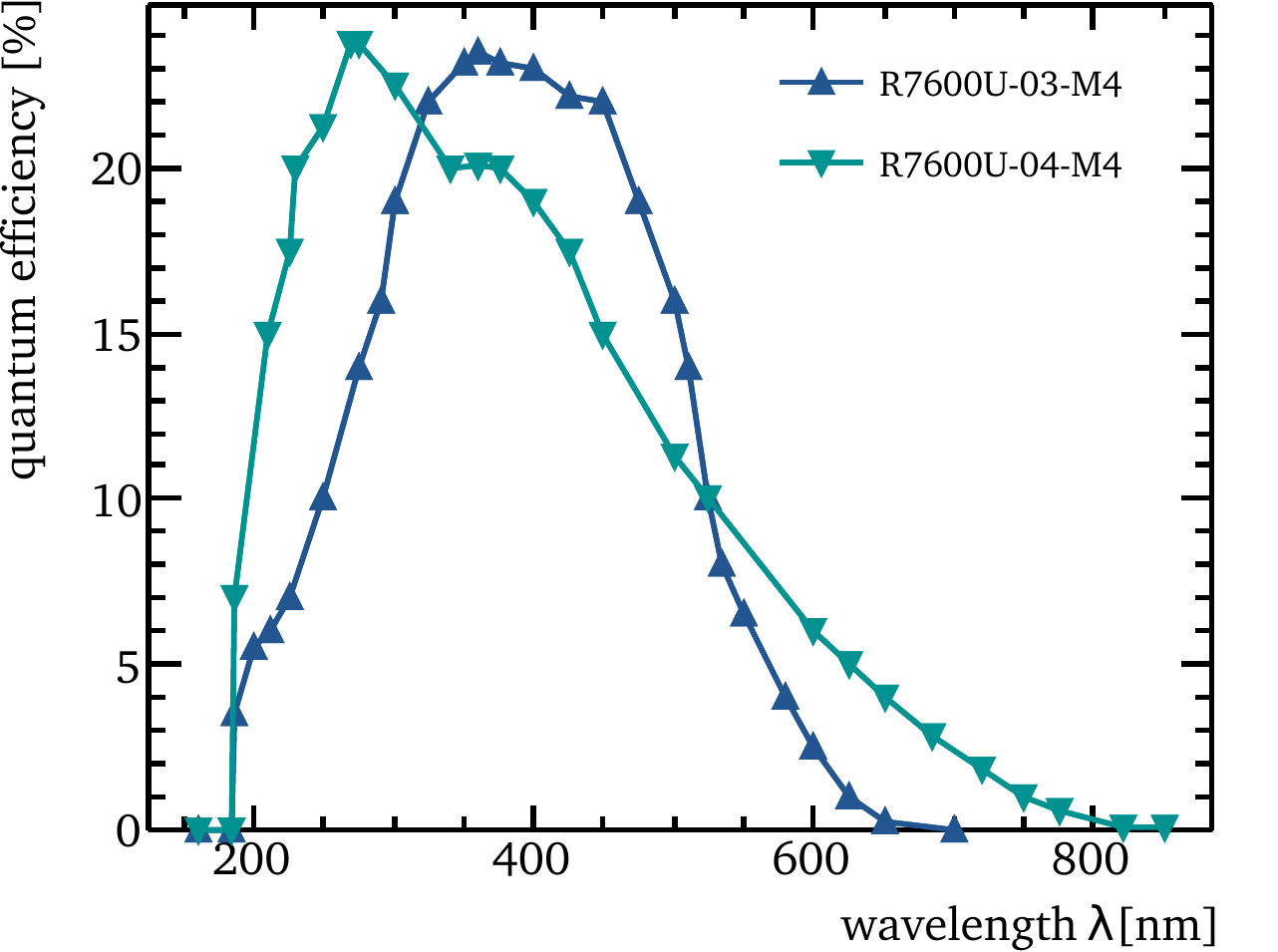}{fig:qQEs}
  \hfill
  \subfloat{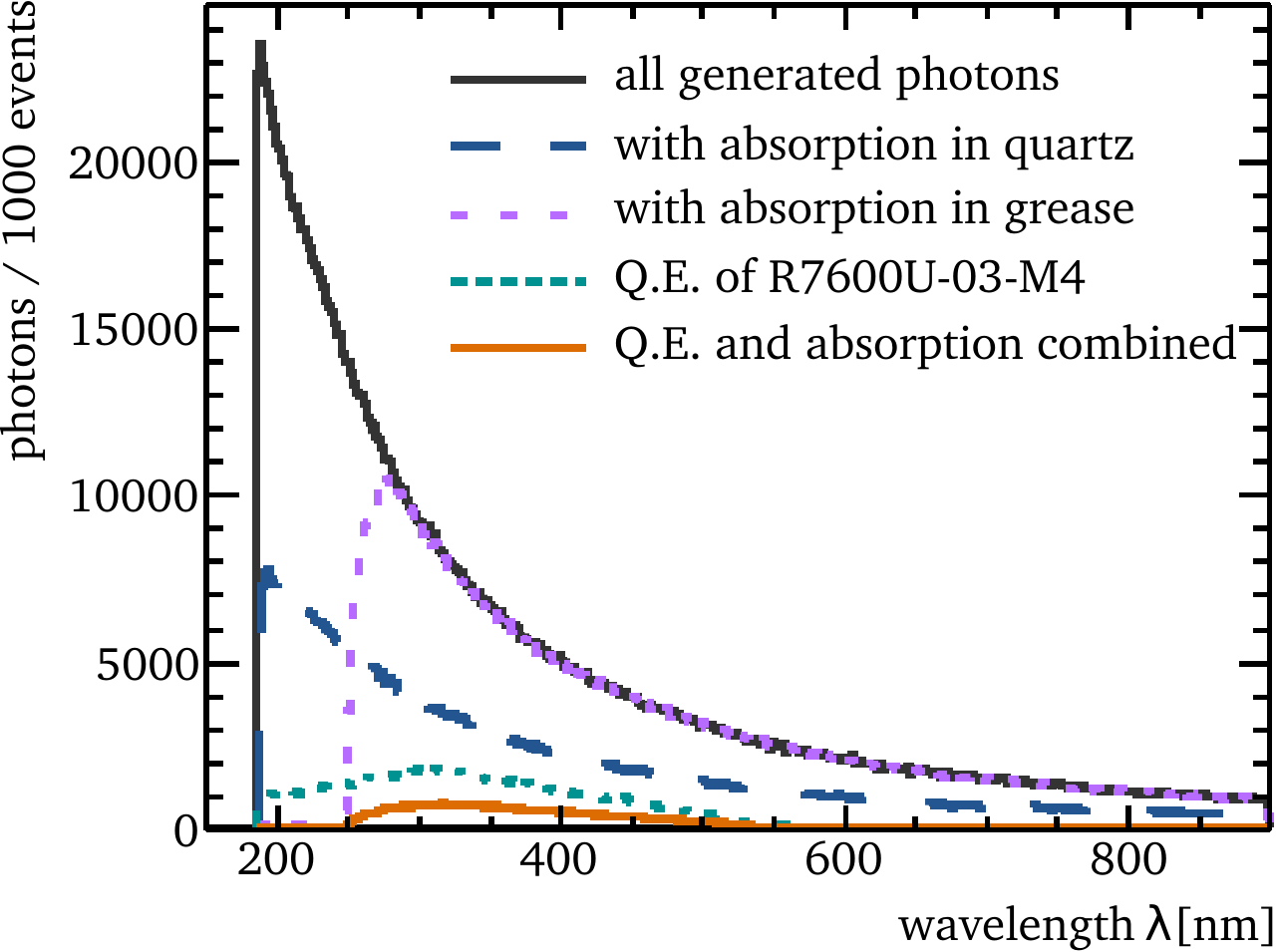}{fig:qTransmission}
  \caption{Effect of photomultiplier choice: the left plot \prosubref{fig:qQEs} shows the quantum efficiencies of two UV sensitive PMTs~\cite{HamamatsuR7600}. On the right \prosubref{fig:qTransmission}, the wavelength spectrum of the Cherenkov photons and the effects of absorption inside the detector and the different quantum efficiencies is shown. }
  \label{fig:qQETrans}
\end{figure}

 The amount of measurable light can be further influenced by the choice of the other materials used in the detector, such as the type of optical grease or the surface finish of the quartz bars.  The impact that the choice of photodetector and materials has on the amount of measurable light is illustrated by figure~\ref{fig:qTransmission}.
Despite the fact that layer of grease is only \NU{1}{mm} thick, the UV component of the Cherenkov light is  severely affected by its short absorption length. This clearly indicates that the choice of a suitable optical grease to couple the photodetectors to the quartz is also of great importance.  The application of another optical grease with lower absorption in the UV range would allow to increase the amount of measurable light for a chosen geometry. To a smaller extent, further improvements could be achieved by polishing the quartz surface. 
Table~\ref{tab:simlightyield} lists the predicted light yield for both the geometry with maximum predicted light yield from figure~\ref{fig:dist3cm} as well as for the prototype  discussed section~\ref{sec:prototype} (with a size of \NU{5}{mm} $\times$ \NU{18}{mm}  $\times$ \NU{100}{mm} and for an incidence angle of $\alpha_{x}= 45\degree$). 
Despite the fact that only $\mathcal{O}(10\percent)$ of the photons are detected with the material choices considered in these first simulation studies, the predicted yield of 60 -- 300 detected  photons per incident electron are promising for achieving a single Compton electron resolution as discussed in section~\ref{sec:apply2pol}. 
For the prototype geometry along with a PMT gain of $g=4\ex{5}$ and a digitiser with a resolution of \NU{200}{fC}, this would correspond to an allowed range of up to $N_{\mathrm{C.e.}}=10$ electrons per channel, while a light yield
of 200 -- 300 as predicted for other geometries would allow $N_{\mathrm{C.e.}}=$28 -- 42 electrons per channel.

\begin{table}[tb]
  \centering
  \begin{tabular}{rrrrr}
\multirow{2}{*}{simulated geometry: \hfill}  & \multicolumn{2}{c}{optimised}       & \multicolumn{2}{c}{prototype}      \\
                                      & ground      &       polished        & ground      &       polished       \\
\cline{1-5}
\multicolumn{1}{r|}{photon hits}      & $3344 \pm 58$ & $3876 \pm 62$ & $576 \pm 24$ & $651 \pm 26$  \\
\multicolumn{1}{r|}{detected photons} & $ 322 \pm 18$ & $ 373 \pm 19$ & $ 57 \pm 7 $ & $ 60 \pm  8$  \\
  \end{tabular}
  \caption{Light yields (with statistical errors) for an optimised geometry and the prototype detector geometry. The values for \qqtext{photon hits} refer to the number of photons reaching the readout at the end of the channel per incident electron, while \qqtext{detected photons} are those remaining after absorption in Cargille 50350 optical grease and application of the quantum efficiency of a Hamamatsu R7600U-03 PMT.}
  \label{tab:simlightyield}
\end{table}

\subsection{Construction of the prototype}
\label{sec:prototype}
The simulation studies indicate that the desired photon yield can be reached. To verify the simulation's predictions, a prototype detector with adjustable incidence angle was built and 
operated at the DESY II testbeam. 

The prototype detector consists of four quartz bars, with every second quartz block flipped with respect to its neighbour, as described in section~\ref{sec:g4sim}. 
The synthetic fused silica brand \Spec~\cite{heraeusDandP} was selected as the most suitable material, based on the high radiation tolerance compared to other natural or synthetic fused silica brands~\cite{Cohen2003}, the low number of optical impurities and the high transmission in the UV range. The dimensions of each quartz block were chosen to be $\NU{5}{mm} \times \NU{18}{mm} \times \NU{100}{mm}$, which is a suitable compromise between a high light yield and off-the-shelf
photodetector geometries. Each quartz block was wrapped in aluminium foil as shown in figure~\ref{fig:pqAlufoil}.

The photodetectors for the prototype detector were chosen according to two main criteria:  high
 quantum efficiency at short wavelength, where the Cherenkov light intensity is the highest, and 
 the dimensions of the sensitive area. Square photomultipliers with an active area of 
 $\NU{18}{mm} \times \NU{18}{mm}$, with a four-anode readout (Hamamatsu \mbox{R7600U-03-M4} 
 and \mbox{R7600U-04-M4}) were employed to read out the two quartz blocks pointing into the same
 direction, with two anodes covering one quartz block as illustrated in figure~\ref{fig:pq1PMT2Sticks}. This layout allows to have the four detector channels directly side-by-side. Figure~\ref{fig:pqPhotoSticksPMT} shows two quartz blocks
 and one of the PMTs compared to a $2$-Euro coin. 

To limit the occurrence of total reflection at the boundary between quartz bar and PMT window, they need to be coupled with optical grease. For the testbeam campaign, Cargille 06350 was used.

The mechanical setup to hold the quartz bars and photomultipliers in place with adjustable angle 
between them is shown in figure~\ref{fig:pqTechDrawing}. The angle between the upper 
and lower channels can be controlled by the movement of a stepping motor with a base step angle of 0.9\degree, which is transformed to 0.5\degree steps by a dedicated gear mechanism contained in the mechanical setup.

\begin{figure}[b]
  \centering
  \subfloat[0.2080\linewidth]{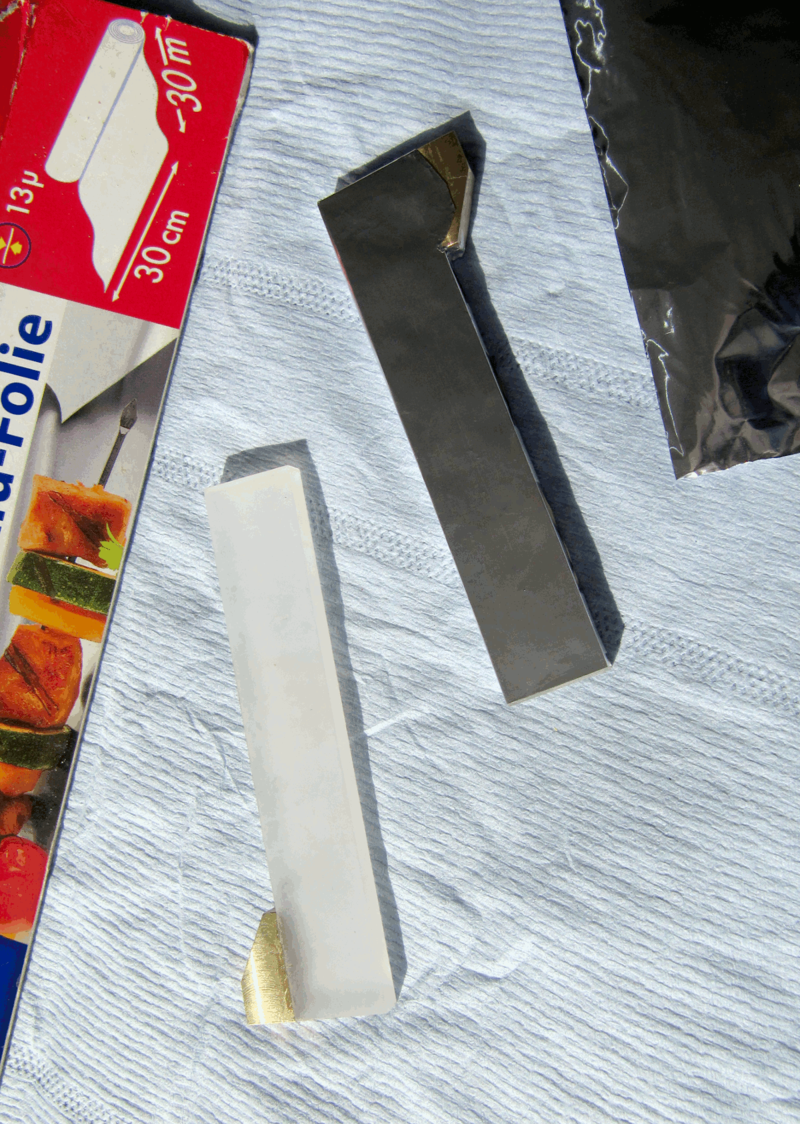}{fig:pqAlufoil}
  \hfill
  \subfloat[0.4035\linewidth]{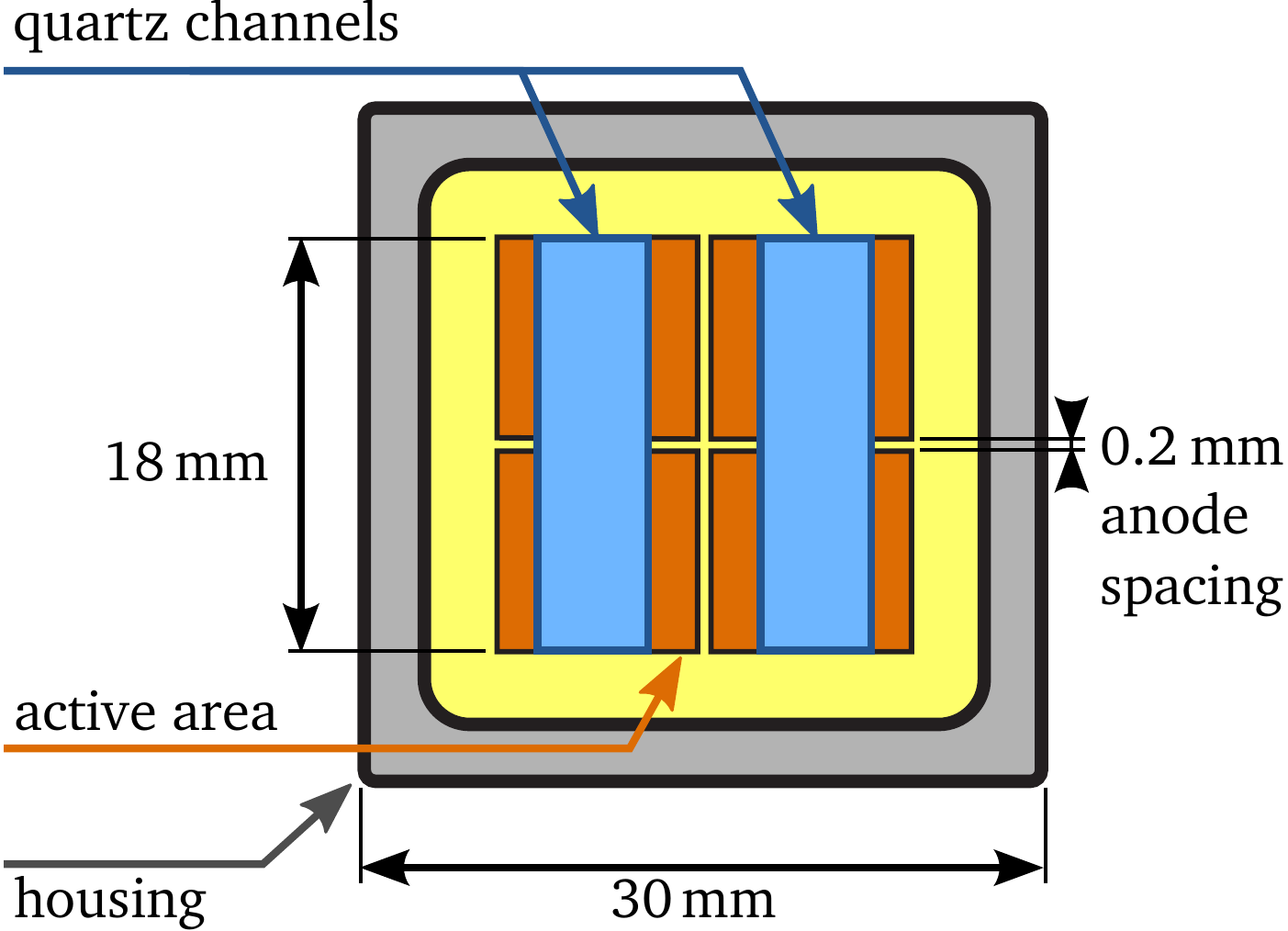}{fig:pq1PMT2Sticks}
  \hfill
  \subfloat[0.2886\linewidth]{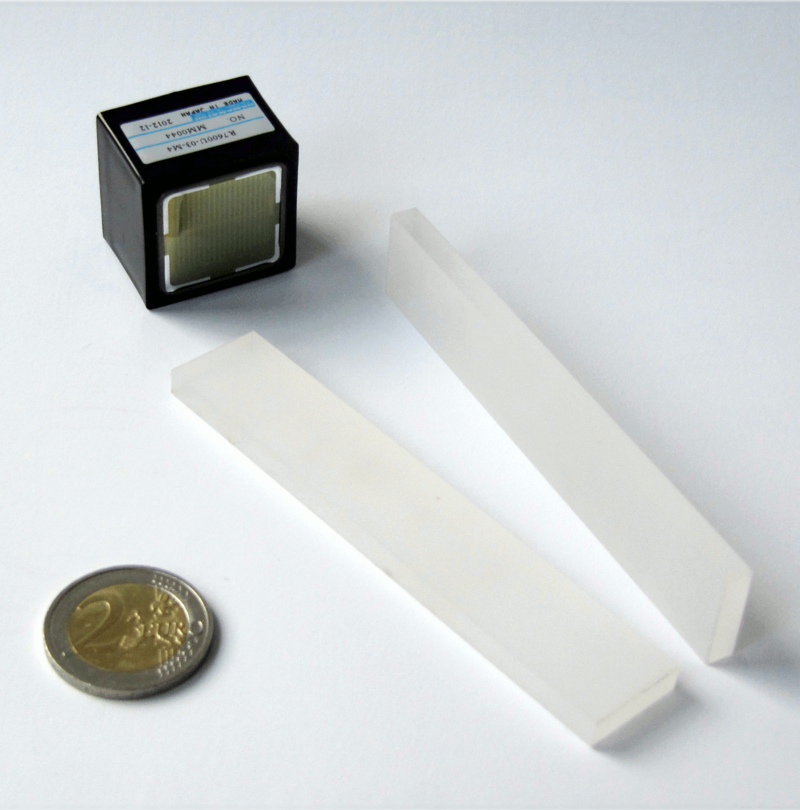}{fig:pqPhotoSticksPMT}
  \caption{
  \prosubref{fig:pqAlufoil} Quartz blocks with the coupling piece used to attach them to the mechanical holder, in the process of being wrapped in aluminium foil.
\prosubref{fig:pq1PMT2Sticks} Sketch to illustrate how one photomultiplier with four readout-anodes can be positioned to provide two measurements for two quartz channels each. 
\prosubref{fig:pqPhotoSticksPMT} Quartz blocks and a photomultiplier with a coin for size comparison.}
  \label{fig:pqQuartzAndHolder}
\end{figure}

%% file: 04_testbeam.tex
\section{Single electron measurements at the DESY testbeam}
\label{sec:testbeam}
The quartz prototype detector was characterised with single electrons at the DESY testbeam~\cite{eudetTB}. 
The electrons are produced by inserting a metal conversion target into a bremsstrahlung beam from the DESY II synchrotron. As a converter target for the generation of the electrons, a copper foil of \NU{5}{mm} thickness was chosen.  With a dipole magnet and a collimator electrons with an energy of \NU{3.75}{GeV} were selected to achieve a high rate for the data taking. 
In combination with an area collimator with aperture of $5 \times \NU{5}{mm}$ located \NU{5}{m} upstream of the prototype detector, this resulted in single electrons reaching the prototype detector with a frequency of ${\approx} \NU{1}{kHz}$ -- \NU{2}{kHz}.

 The quartz detector was placed inside an aluminium box with foam rubber seals to shield it from external light and placed on a base plate movable by two linear precision translation stages, as shown in figure~\ref{fig:tbPhotoTable}.

\begin{figure}[tb]
  \centering
  \subfloat{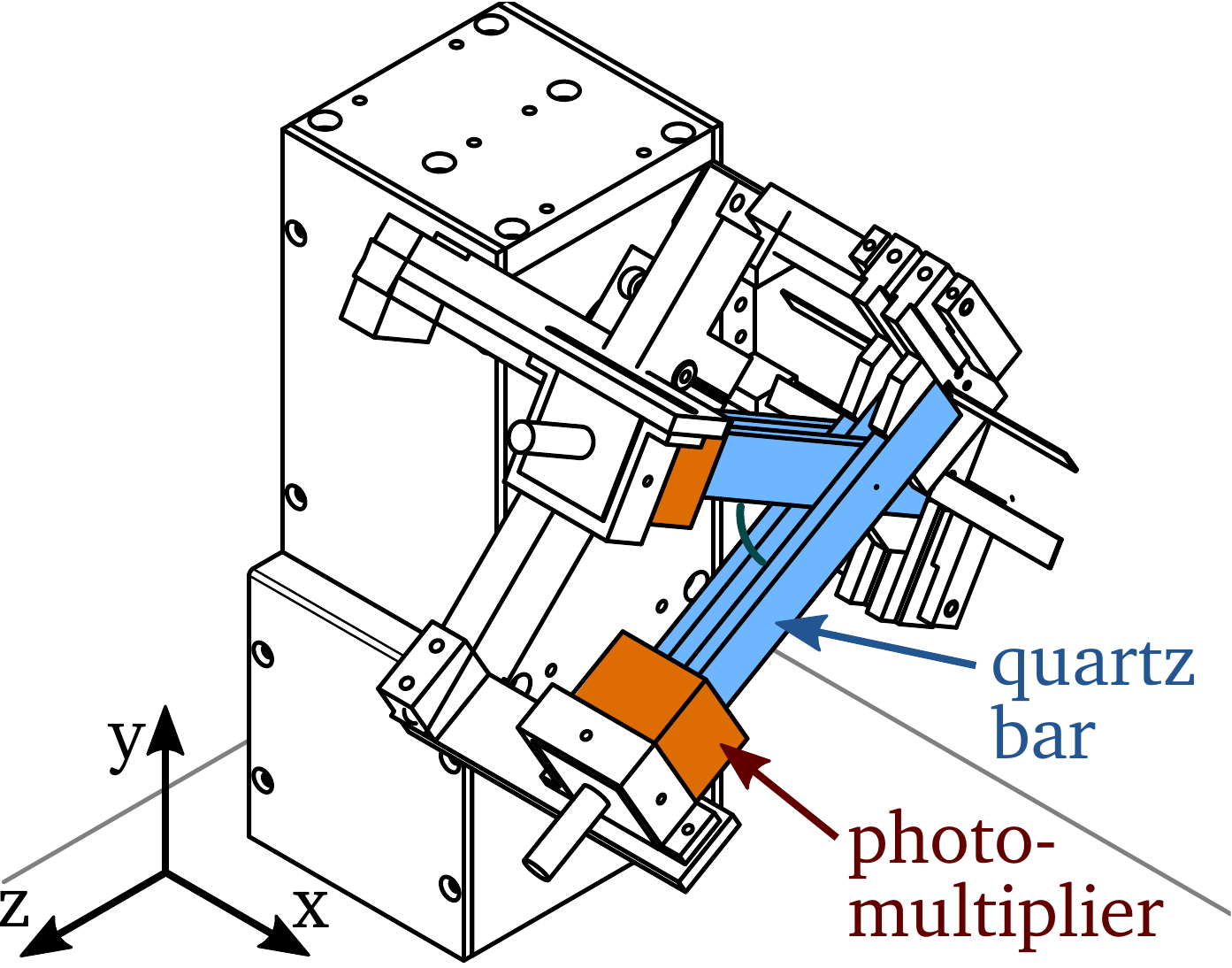}{fig:pqTechDrawing}
  \hfill
  \subfloat{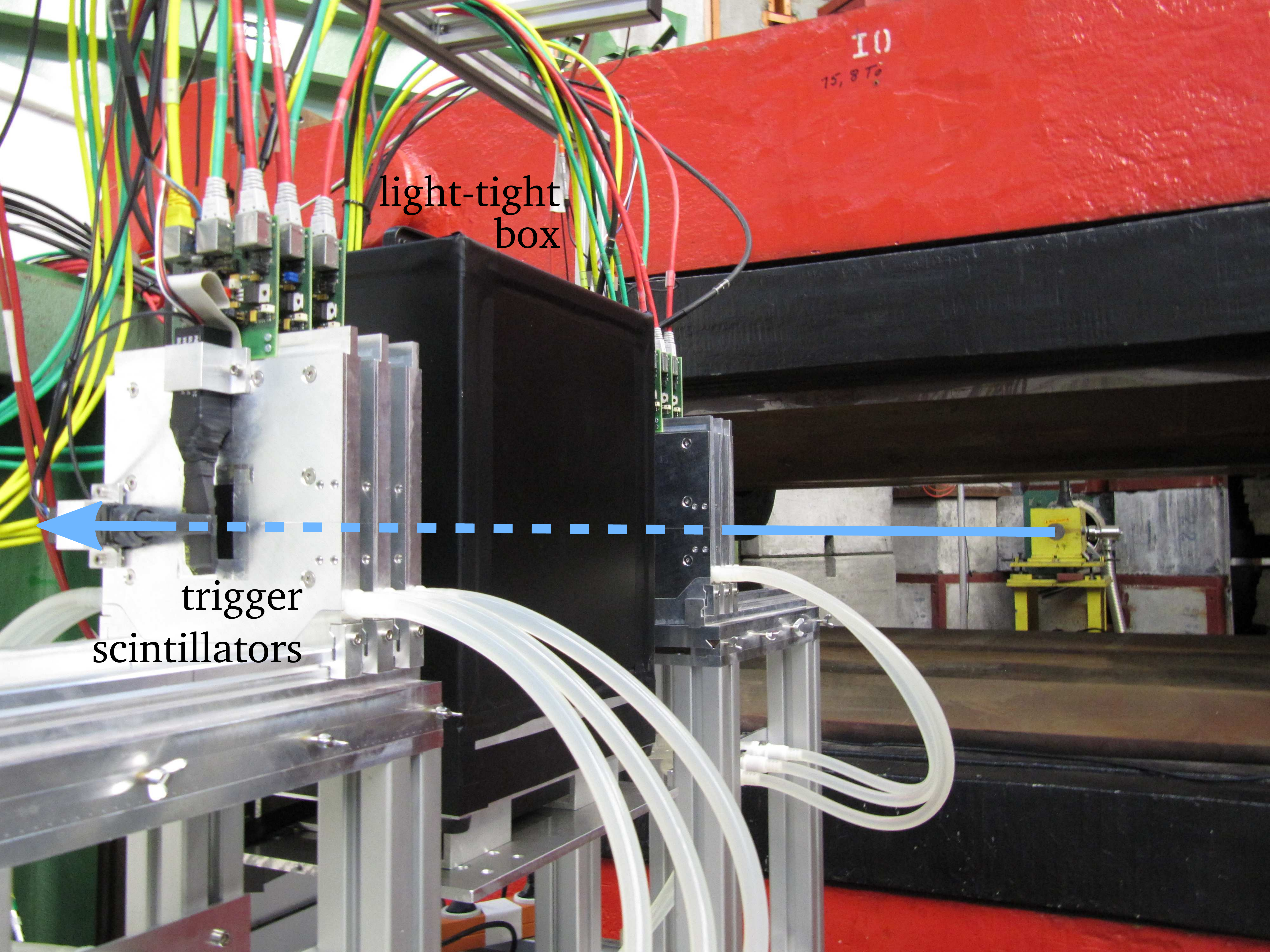}{fig:tbPhotoTable}
  \caption{In the technical drawing \prosubref{fig:pqTechDrawing}, the positioning of the four quartz blocks and two photomultipliers used in the prototype setup are marked. On the photo of the testbeam setup \prosubref{fig:tbPhotoTable} a sketch of the beam path is overlaid on a close-up of the light-tight box with the detector inside on the $x$-$y$-table and two of the trigger scintillators. }
  \label{fig:pqDrawAndPhoto}
\end{figure}

The main components of the setup and the readout chain  during the testbeam campaign are sketched in figure~\ref{fig:tbsetupDrawing}: 
The photomultipliers are operated with a supply voltage of \NU{900}{V}, leading to amplification with a gain of $\sim 1.5\ex{6}$ (c.f.\ section \ref{sec:tblight}). The charge signals of the photomultipliers are read out with a 12-bit charge-sensitive analogue-to-digital converter (QDC) which offers two resolutions, \NU{25}{fC} and \NU{200}{fC}. The digitisation length and cycle are steered by an external gate. To trigger on beam electrons, the generation of the gate signal was started in case of a four-fold coincidence between two trigger scintillators before and after the detector, respectively. A total number of eight QDC channels was used to digitise the signal from the two four-anode photomultipliers. The conventions for channel naming and anode numbering are illustrated in figure~\ref{fig:tbAnodeconfig}. An example for the QDC signal recorded for a PMT anode is shown in \ref{fig:exampleSigDC}.

\begin{figure}[tb]
  \centering
  \subfloat[0.5875\linewidth]{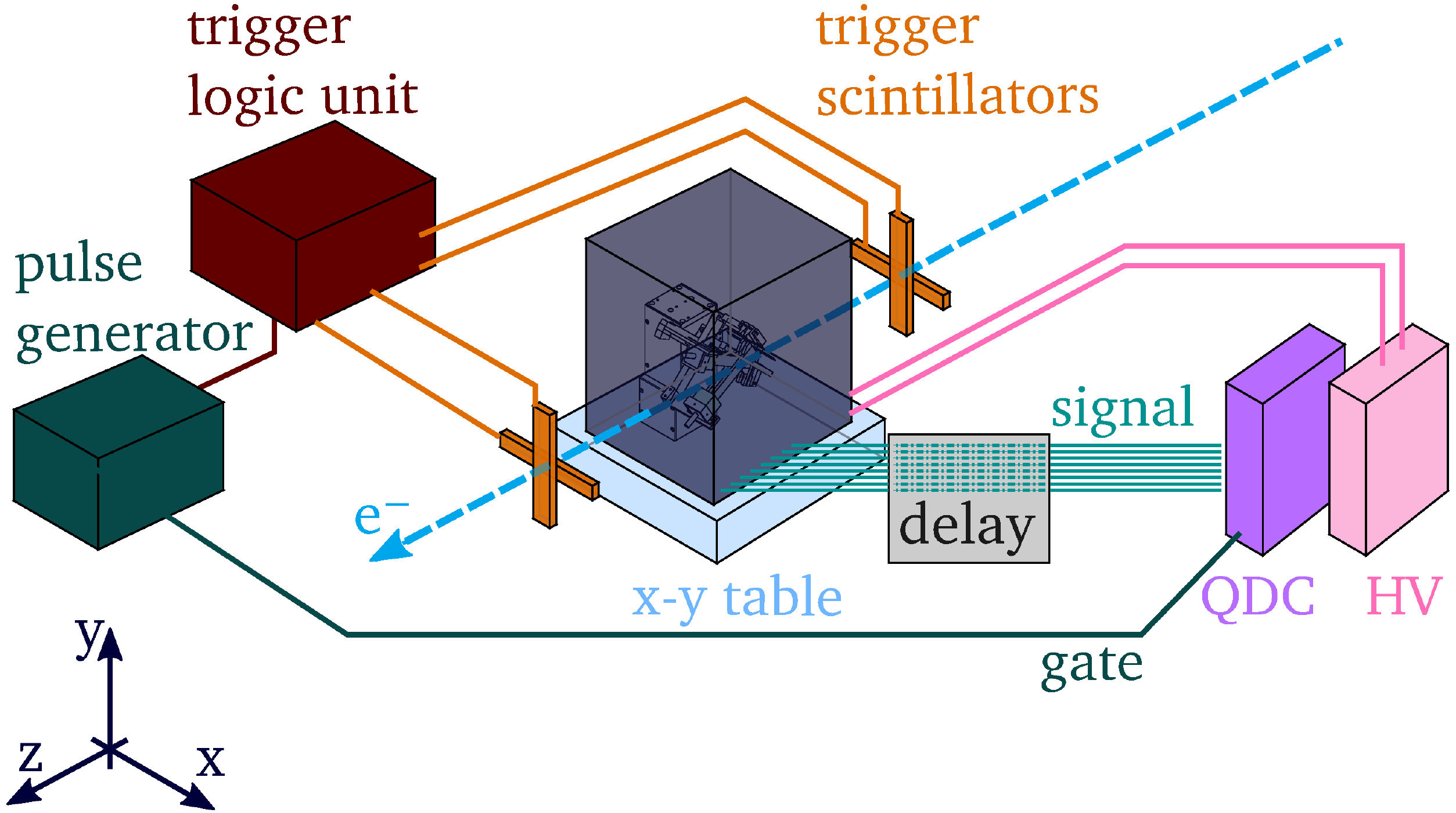}{fig:tbsetupDrawing}
  \hfill
  \subfloat[0.3825\linewidth]{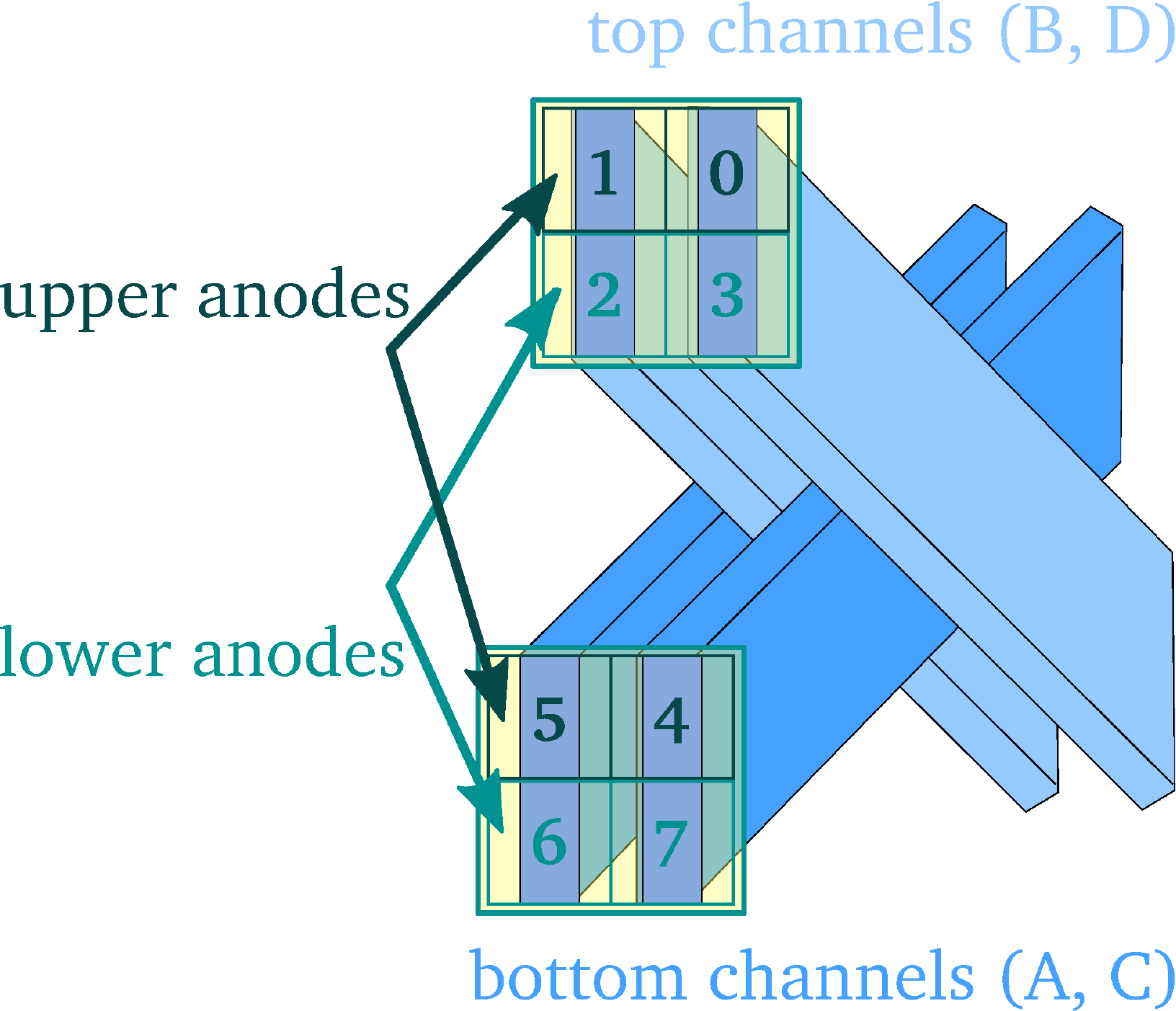}{fig:tbAnodeconfig}
  \caption{\prosubref{fig:tbsetupDrawing}  Experimental setup at the testbeam. The main components of the setup and the path of the readout signal path are sketched. \prosubref{fig:tbAnodeconfig} Channel names and anode numbering used for the description of the testbeam data.}
  \label{fig:tbSetup}
\end{figure}

As a first step the detector response and alignment of the testbeam setup were studied, which consequently allowed to compare the measured data to simulation and evaluate the light yield.

\subsection{Detector response}

The spectrum of each photomultiplier anode contains events in which the data acquisition was triggered, but no electron entered the respective channel. Therefore, in addition to the Cherenkov signal another peak is present at the low edge of the spectrum, caused by the QDC pedestal and broadened by dark current from the photomultipliers.

To exclude events in the analysis of a detector channel where that channel was not hit by the electron beam, a cut threshold was defined based on dedicated dark current measurements (i.e.\ runs without beam) taken over the course of the testbeam period. It was found that applying a threshold of \NU{15}{QDC} bins above the QDC pedestal position eliminates at least 99.7\percent of the events in any dark current spectrum. For the analysis of the data, only events where selected in which both anodes of a channel yield a signal of at least \NU{15}{QDC} bins above the mean of their respective pedestal. The part of the spectrum used in the analysis, compared to the spectrum before any cuts, is depicted in figure~\ref{fig:examplePedSub}.

\begin{figure}[tb]
  \centering
  \subfloat{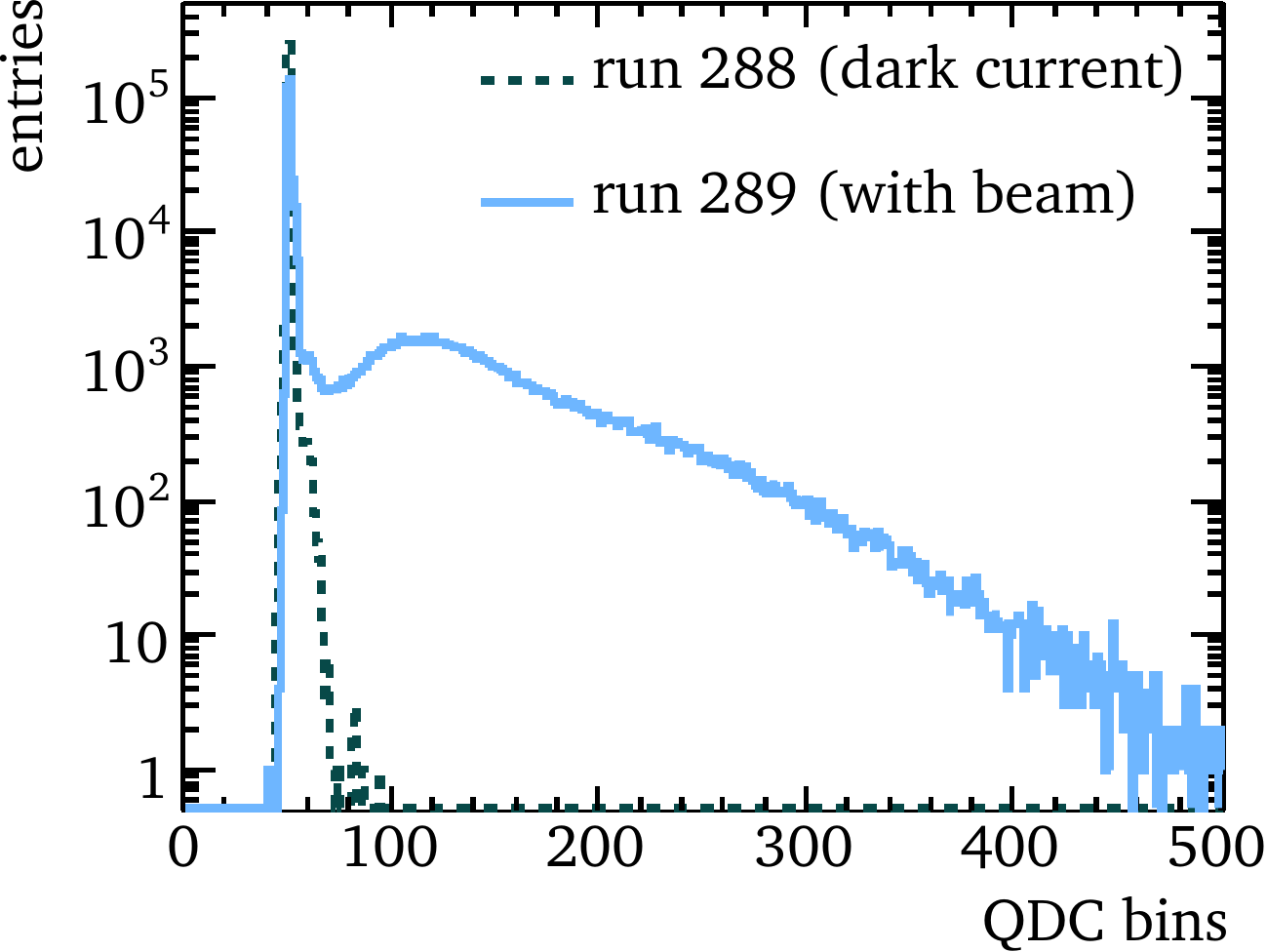}{fig:exampleSigDC}
  \hfill
  \subfloat{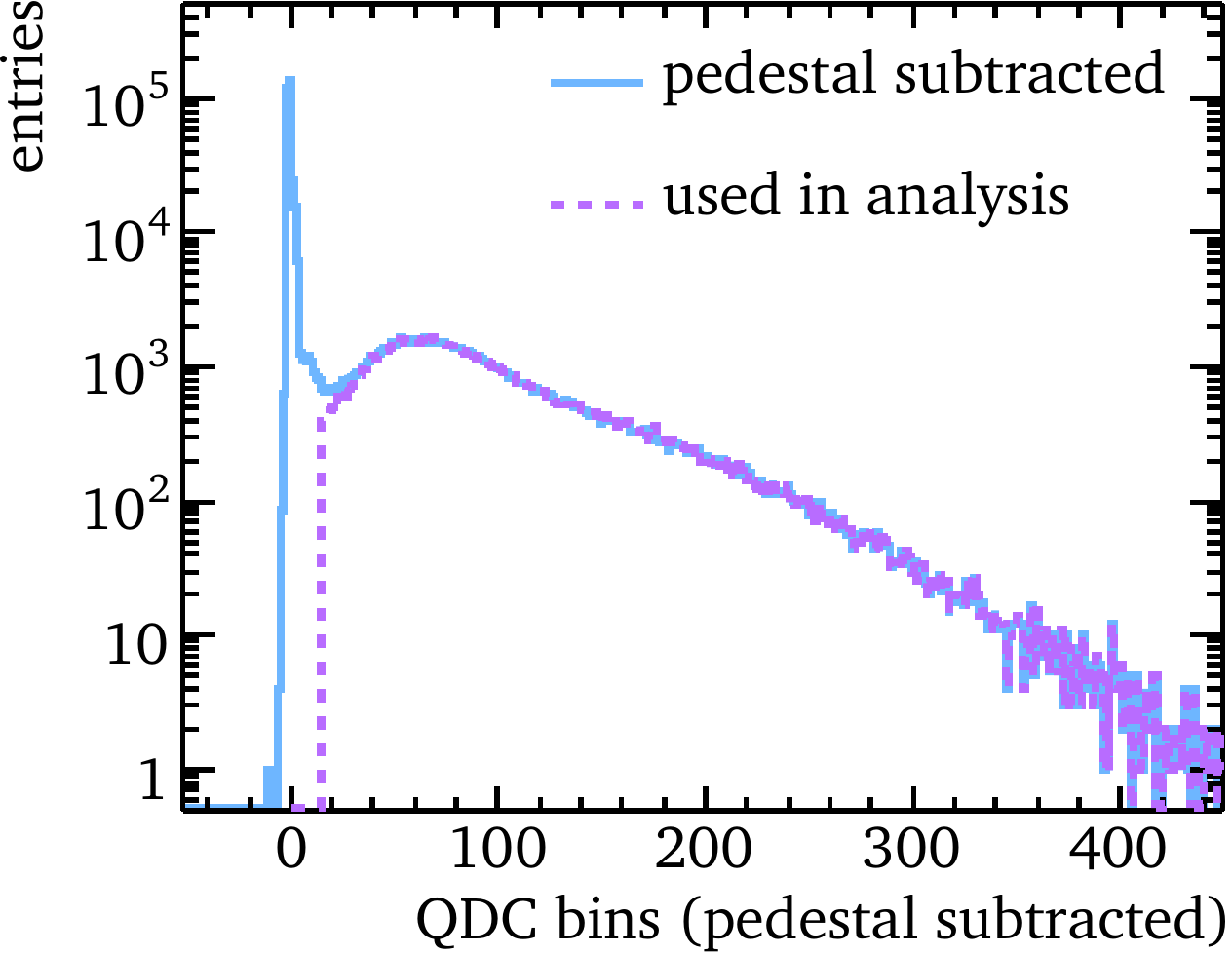}{fig:examplePedSub}
  \caption{Pedestal correction. \prosubref{fig:exampleSigDC} An example for an uncorrected QDC spectrum  is shown in light blue, recorded in anode 5 with the beam centred on channel A and a beam incidence angle of 45\degree. A dark current spectrum for the same anode is drawn in dark green (dashed line). \prosubref{fig:examplePedSub} The same QDC spectrum with shifted $x$-axis to correct for the pedestal contribution. The light blue histogram contains all events, while the violet histogram (dashed line) is made up of only the events used in the analysis, i.e.\ events in which both anodes of a channel see a signal of at least \NU{15}{QDC} bins above the pedestal.}
  \label{fig:PedCor}
\end{figure}

\begin{figure}[tb]
  \centering
  \subfloat{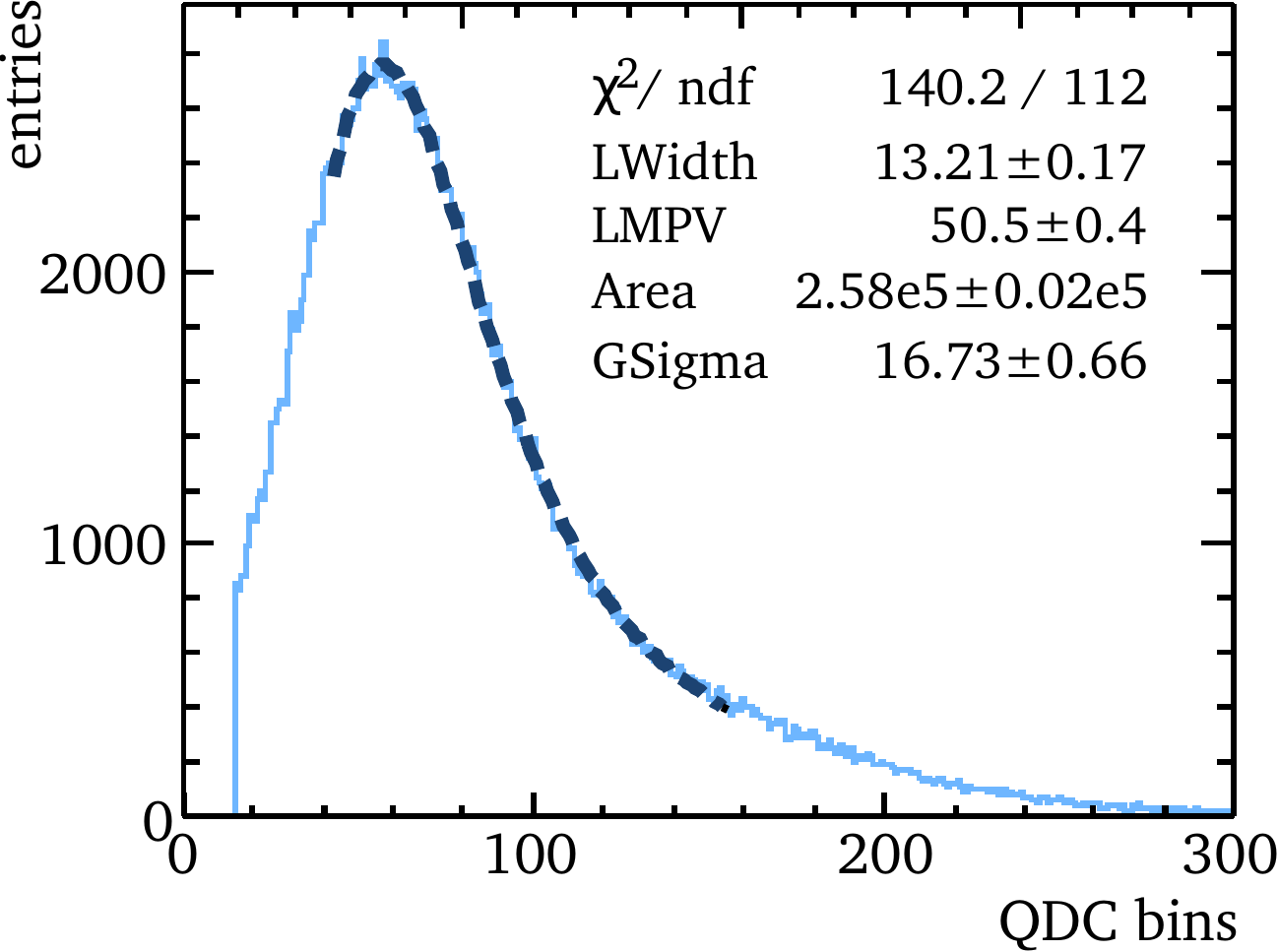}{fig:lanFitData}
  \hfill
  \subfloat{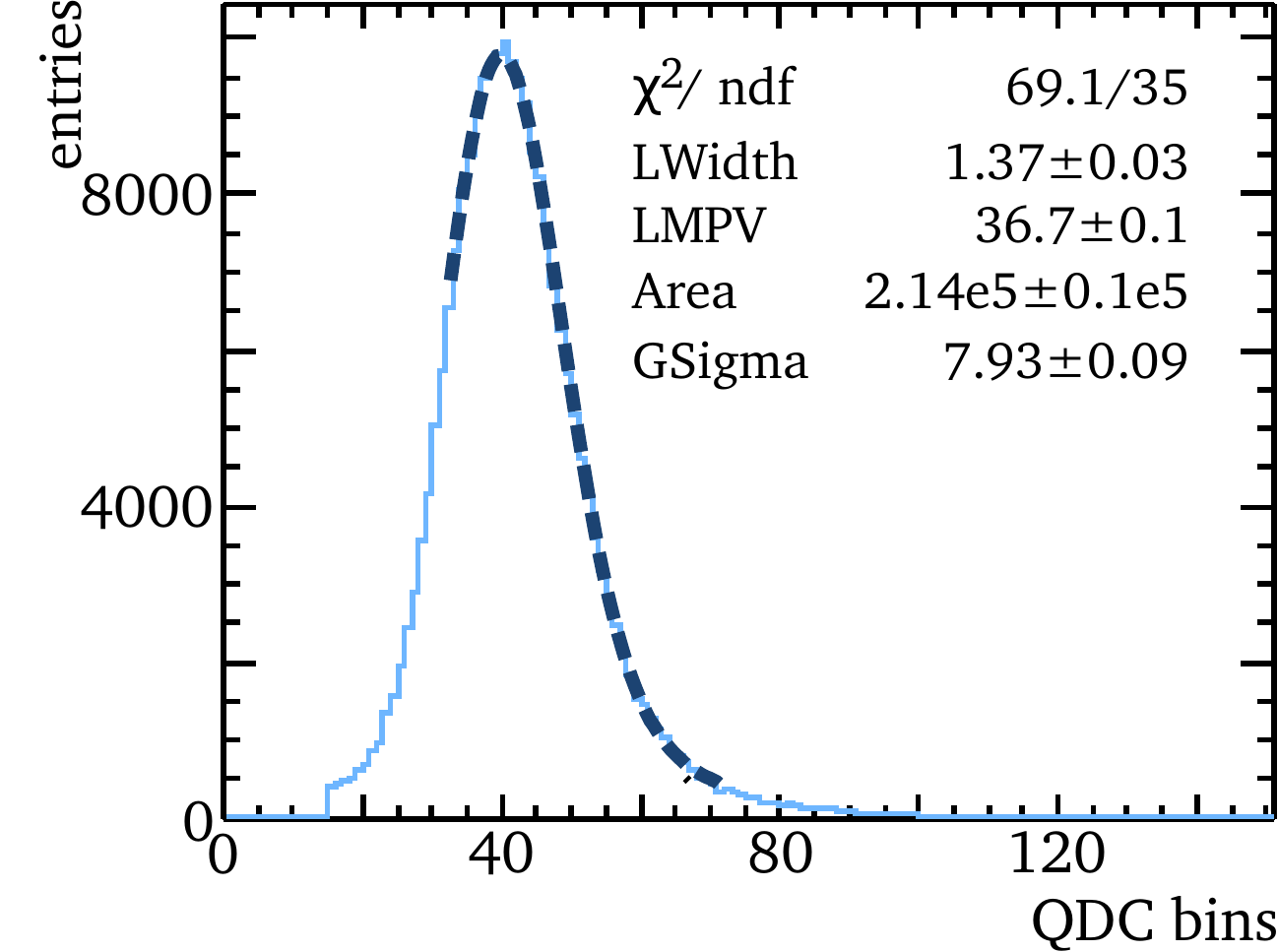}{fig:lanFitSim}
  \caption{Fit of a Landau function convoluted with a Gaussian to the signal for \prosubref{fig:lanFitData} anode 2 in a data taking with the beam centred on channel B, under an incidence angle of 30\degree and \prosubref{fig:lanFitSim} the corresponding simulation.  The cut-off at low QDC bins is due to the pedestal subtraction cut.}
  \label{fig:lanFit}
\end{figure}

Figure~\ref{fig:lanFit} shows the signal of a single anode in both data and simulation. 
To convert the number of detected photons in the simulation to QDC bins, a gain of $1.3\ex{6}$ (c.f.\ section~\ref{sec:tblight}) and the same digitiser resolution as in the data were assumed.\\
In the data, the QDC signal has a noticeable tail to higher charges. The most likely explanation is the presence of secondary electrons in addition to the main beam electron, which would produce some additional light when they hit one of the quartz channels. It is currently not clear where these stem from. The wall thickness of the light-tight box surrounding the detector is only \NU{4}{mm}, which does not account for enough shower electrons to  explain the amount of additional light observed. Other possible contributions along the beam line have not been investigated so far. \\
To account for the tail in the distribution, the QDC signals were fitted with a fit function consisting of a Landau function convoluted with a Gaussian. The free parameters of the function are the width and most probable value (MPV)  of the Landau function, the sigma of the convoluted Gaussian and a normalisation constant for the peak area. Whenever the testbeam data is compared with the simulation, the same fit is applied to the simulation. For the simulation, the width of the Gaussian is dominant, and the relation between MPV and the Gaussian width is consistent with the expectations from the statistical modelling used for the fast simulation introduced in section~\ref{sec:apply2pol}. In the data, the Gaussian width is roughly twice as large as one would expect from noise in the amplification process alone. Since this broadening was not observed in a separate setup in the laboratory, it is attributed to the experimental environment at the testbeam, such as the long signal cables. The width of the Landau part of the fit function required to describe the large tails in the data is comparable to the Gaussian width.

The most probable value of the fitted Landau function was used as a measure for the amount of light in a channel,  while the area parameter is related to the number of times a channel was hit by the beam. To illustrate the stability of the detector response and the fit procedure's outcome, figure~\ref{fig:stability30} shows the fitted most probable value for both anodes of a channel for a number of data taking runs at identical beam position over a period of several hours.

\begin{figure}[tb]
  \centering
  \subfloat{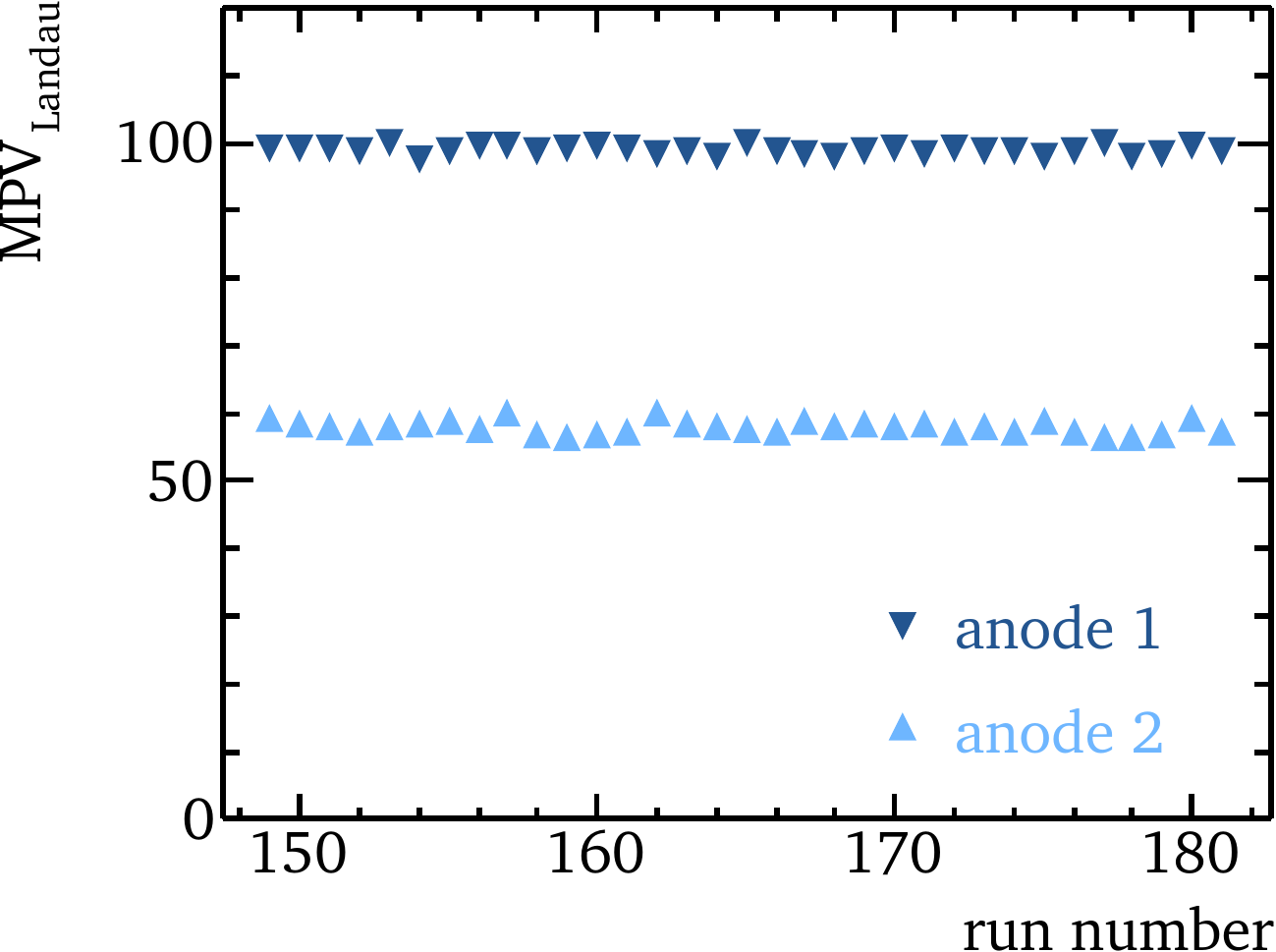}{fig:stability30}
  \hfill
  \subfloat{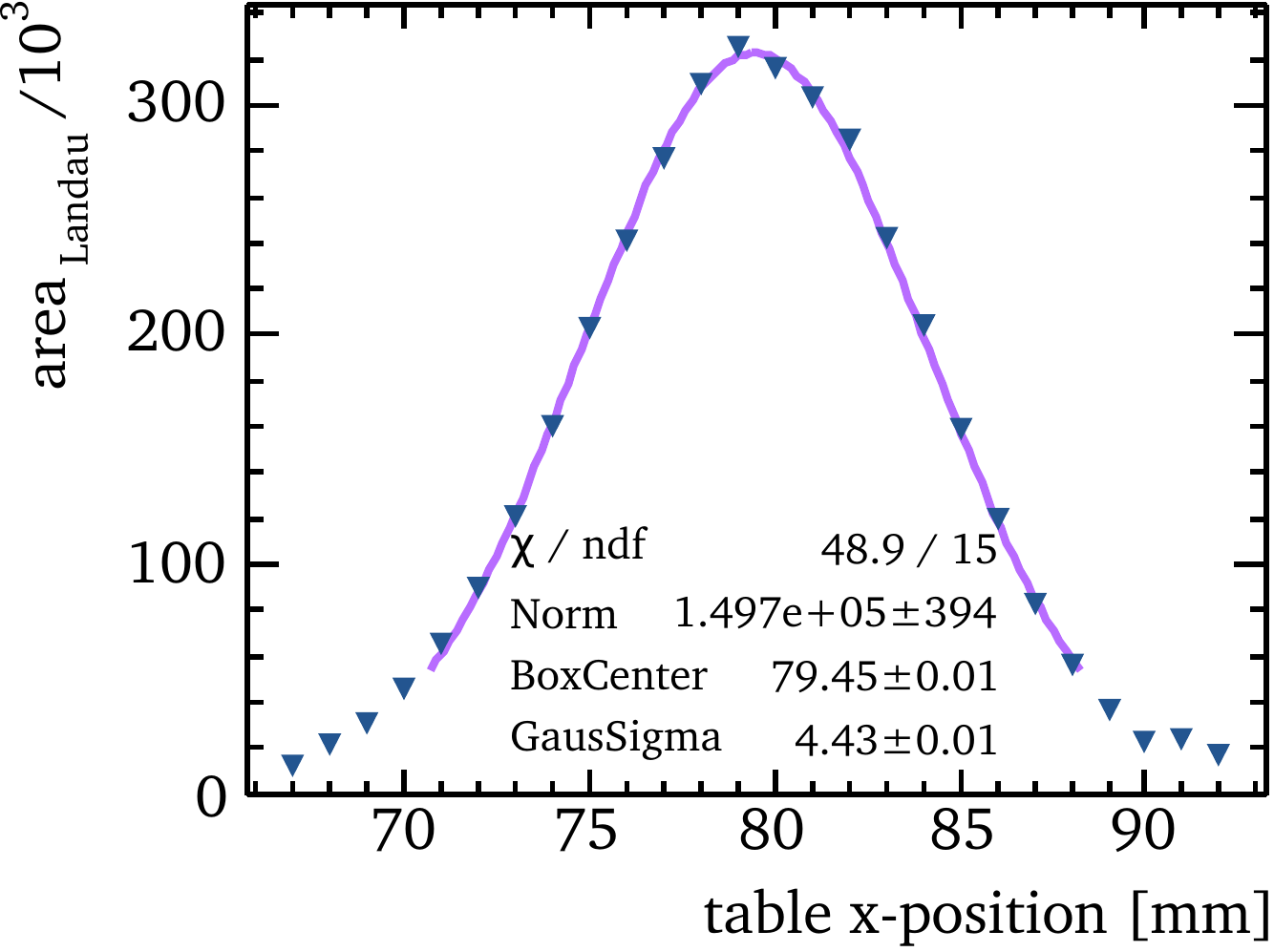}{fig:tbXfit}
  \caption{\prosubref{fig:stability30} Fit results anodes 1 and 2 for 32 for consecutive measurement runs taken over a period of 11 hours  (beam on channel B, under an angle of 30\degree). The observed light yield stays constant.  \prosubref{fig:tbXfit} Landau area of anode 1 observed at different $x$-positions of the detector, fitted with a box function convoluted with a Gaussian (as in equation~\ref{eq:boxfit}).}
  \label{fig:stabilityAndX}
\end{figure}

\subsection{Horizontal alignment and beam profile}
To determine how well the prototype detector can be aligned from data alone, the detector position was scanned by varying the position of the movable table in $x$-direction in \NU{1}{mm} steps. Studying the observed number of events in each detector channel for the different $x$-positions allows to determine the position of the channel with respect to the beam, as well as derive information on the size and shape of the beam spot. For a pointlike beam and perfect alignment, the channel response would be expected to have a \qqtext{box-like} shape, whereas a finite beam-size and tilts around the vertical axis will smear out the edges of this step response function.
To determine the channel centres, the measurements for each anode were fitted with a box function convoluted with a Gaussian. The  box function is 1 for all $x$ within $\pm \NU{2.5}{mm}$ of its central value, to describe the prototype channel width of \NU{5}{mm}, and the Gaussian is used to account for the smearing. The convolution is implemented as sum within four sigma of the Gaussian, i.e.\
\begin{equation}
f(x) = p_0 \cdot \sum_{\tau=p_1-4\cdot p_2}^{p_1+4\cdot p_2}  \text{box}(\tau, p_1) \cdot   \frac{1}{{p_2 \sqrt {2\pi } }}e^{{{ - \left( {x - \tau } \right)^2 } \mathord{\left/ {\vphantom {{ - \left( {x - \tau] } \right)^2 } {2\cdot p_2 ^2 }}} \right. \kern-\nulldelimiterspace} {2\cdot p_2 ^2 }}},
\label{eq:boxfit}
\end{equation}
where the free parameters are a normalisation factor~$p_0$, the centre position  $p_1$ of the channel box function, and the sigma  $p_2$ of the~Gaussian. As an example, the fit to one of the anodes is shown in figure~\ref{fig:tbXfit}.
The channel centre positions derived from these fits are listed in table~\ref{tab:channelcenters}. The fitted centre positions to the data from the upper anodes of each channel all agree within \NU{0.05}{mm} with the fitted positions for the respective lower anodes, indicating that the detector position can be determined within a precision of the same order, which meets the requirements for the horizontal alignment precision.

\begin{table}[tb]
   \begin{tabular}{c|cc|cc}
   channel & anode & channel centre & anode & channel centre \\
   \hline
   A & 5 & \NU{ (74.12 \pm 0.01) }{mm}  & 6 & \NU{ (74.07 \pm 0.01) }{mm}  \\
   B & 1 & \NU{ (79.45 \pm 0.01) }{mm}  & 2 & \NU{ (79.47 \pm 0.01) }{mm}  \\
   C & 4 & \NU{ (85.54 \pm 0.01) }{mm}  & 7 & \NU{ (85.57 \pm 0.01) }{mm}  \\
   D & 0 & \NU{ (91.10 \pm 0.03) }{mm}  & 3 & \NU{ (91.15 \pm 0.02) }{mm}  \\
   \end{tabular}
\caption{Channel positions fitted to $x$-scan data, with statistical errors from the fit. The central values for the two anodes per channel agree within \NU{0.05}{mm} or better with each other. The positions are given in the coordinate system of the movable table's control software. }
  \label{tab:channelcenters}
  \end{table}

In the fit of equation~\ref{eq:boxfit} to the $x$-scan data of all eight anodes, the sigma of the convoluted Gaussians were found to be $\sigma=\NU{4.5 \pm 0.1}{mm}$. By simulating $x$-scans with different beam spots and comparing them to the measurement, it could be verified that a beam profile with this extension would cause the smearing observed in the testbeam data. Therefore, all comparisons of the testbeam data to simulations were performed for simulations with a beam profile of $\sigma=\NU{4.5}{mm}$.

\subsection{Vertical alignment}
At the beginning of the testbeam campaign, the position at which the detector was centred with respect to the beam in vertical direction was determined by eye to be at a table position of $y=\NU{76}{mm}$ (in the coordinate system of the steering software for the movable table). To check this estimate, data was taken at different $y$-positions over a range of \NU{40}{mm}, with the beam centred between channel C and D. Once the beam moves far enough down from the vertical centre that it enters channel D through the end face rather than the narrow side face, the path length for light production and thus also the amount of detected light  will decrease rapidly. The same is true for channel C when the beam entrance point is moved upwards high enough. This effect was used to determine the centre in $y$-direction by comparing the data to a simulated $y$-scan. Two parameters need to be adjusted in this comparison: the first is a scaling constant to account for the difference in light yield in data and simulation; the second is an offset in the $y$-position, i.e.\ how far the $y$-axis of the simulation (with the detector centre at position  \NU{0}{mm}) has to be shifted to match the data (with $y$-positions given in the coordinate system of the table software). 

A separate fit was done for both channel C and D. In these fits, both anodes of the respective channel were compared to the simulated $y$-scan, allowing individual scaling factors for each anode in addition to the common offset in $y$-direction. Figure~\ref{fig:yscanData} shows the data for one channel together with the scaled and shifted simulated $y$-scans. The fit to the data of channel C returns a vertical centre $y$-position of $\NU{72.5}{mm} \pm \NU{0.1}{mm}$, the fit to channel D a $y$-position $\NU{80.0}{mm} \pm \NU{0.2}{mm}$. The poor agreement is most likely due to large beam spot size of~$\sigma=\NU{4.5}{mm}$. Therefore the distinct edges  of the channel edges, which would be the most discriminating factor in the alignment, are washed out. This effect is illustrated in figure~\ref{fig:yscanSIM}, where simulated $y$-scans for both $\sigma=\NU{0.1}{mm}$ and $\sigma=\NU{4.5}{mm}$ are shown. As another possible explanation a misalignment in the $x$-$z$-plane has been considered: The ground plate on which the detector box was standing was only affixed to the movable table on one side, which could allow the whole setup to be tilted around the $z$-axis. This would induce correlations between the $x$- and $y$-coordinate, which would affect channel C and D differently and could lead to different  vertical centre positions. However, in the simulation small misalignments in $\alpha_z$ ($\alpha_z<5\degree$) were found to have negligible impact on the light yield, and a larger tilt would have been visible by eye. 

The fit results do nevertheless confirm that the first estimate by eye of a centre at $y_{\text{table}}=\NU{76}{mm}$ was accurate within a few millimetres. Under conditions where the vertical spread of the incident particles is smaller, a better alignment precision is expected.\footnote{More details on this can be found in \cite{diss:vauth}.}

\begin{figure}[tb]
  \centering
  \subfloat{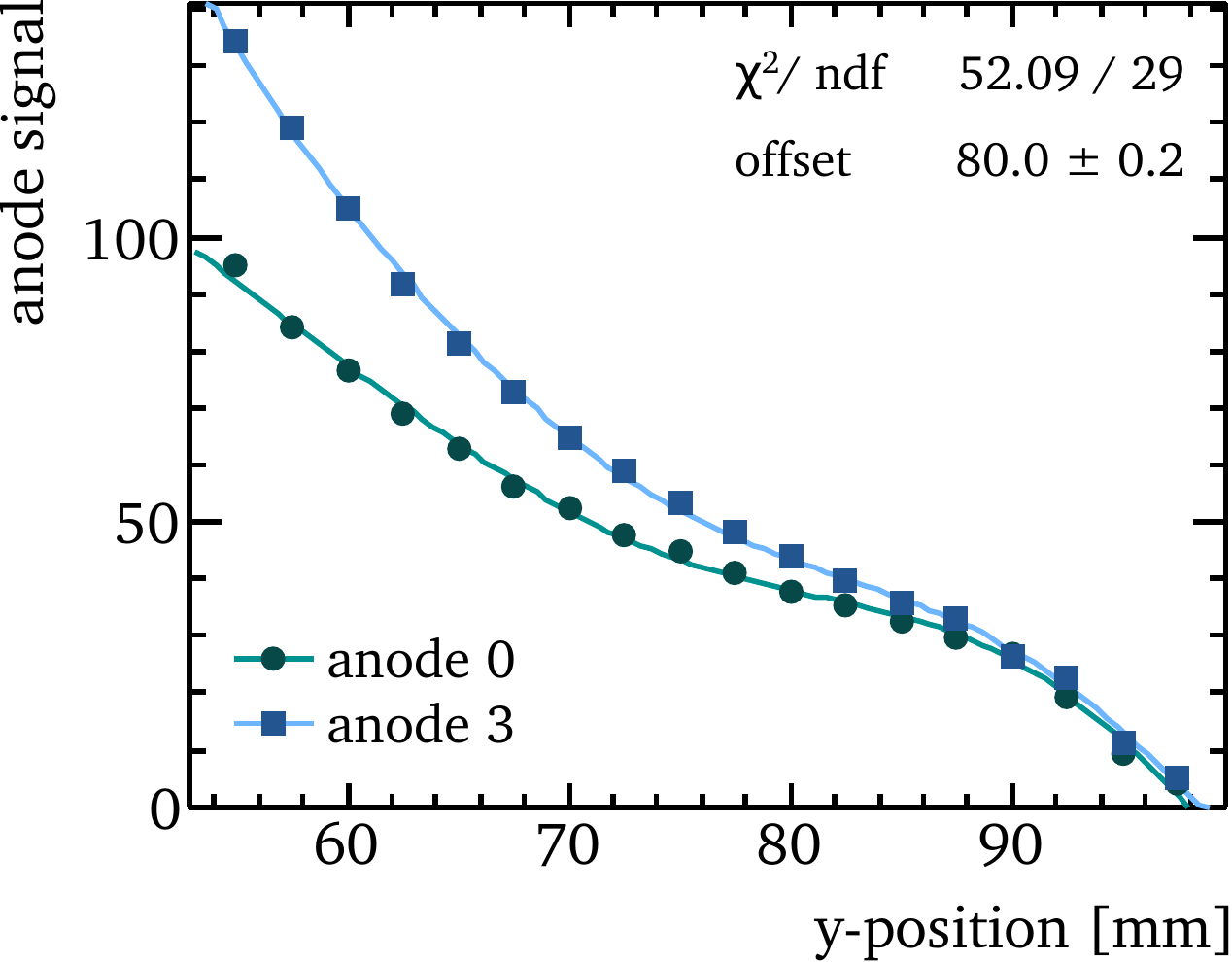}{fig:yscanData}
  \hfill
  \subfloat{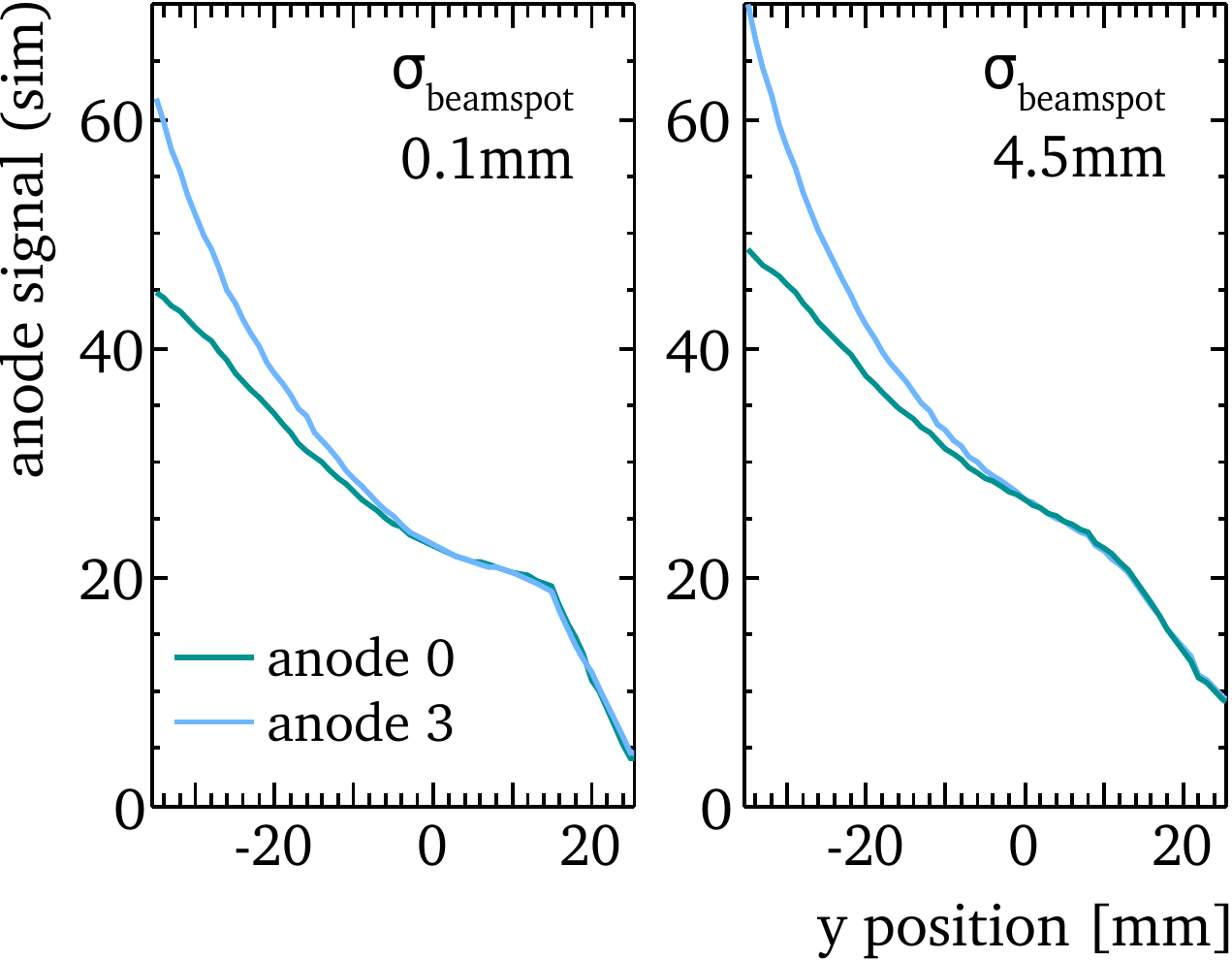}{fig:yscanSIM}
  \caption{Signal at different $y$-positions of the detector \prosubref{fig:yscanData} observed at the testbeam for both anodes of channel D (markers). The solid lines are the fits of a simulated $y$-scan to the signal of both anodes per channel, with a shift of the $y$-axis as a free parameter. \prosubref{fig:yscanSIM} Simulated $y$-scans with beam profile $\sigma_\mathrm{beamspot}=\NU{0.1}{mm}$ (left) and  $\sigma_\mathrm{beamspot}=\NU{4.5}{mm}$ (right).}
  \label{fig:yscan}
\end{figure}

\subsection{Angle scans}

The prototype detector was equipped with a stepping motor to change the angle between the incident electron and the detector channels.  The angle was calculated from the stepping motor's internal step counter. At the edges of the allowed movement range, a mechanical end switch was located, which stopped the motor movement when it was pressed. In some cases the stepping motor experienced difficulties on leaving the switch, causing a disagreement between the angle calculated from the step counter and the actual angle. This becomes evident by comparing the detector response for individual angle scans taken under otherwise unchanged conditions. Figure ~\ref{fig:AyOffset1} shows the signal for both anodes of channel B for two consecutive angle scans, the first one from 30\degree to 60\degree (where the end switch was reached), immediately followed the second angle scan back in the opposite direction. In order to reach an agreement between the detector responses, the angle coordinate of the second scan had to be shifted by 1.15\degree. Three such bi-directional data sets were taken during the testbeam campaign. The angle correction required to obtain the best agreement between opening and closing angle scans for these three data sets were 1.15\degree, 0.69\degree and 0.78\degree, leading to the conclusion that the given angles for the testbeam data are afflicted with an uncertainty of $\mathcal{O}(1\degree)$.

\begin{figure}[tb]
  \centering
  \subfloat{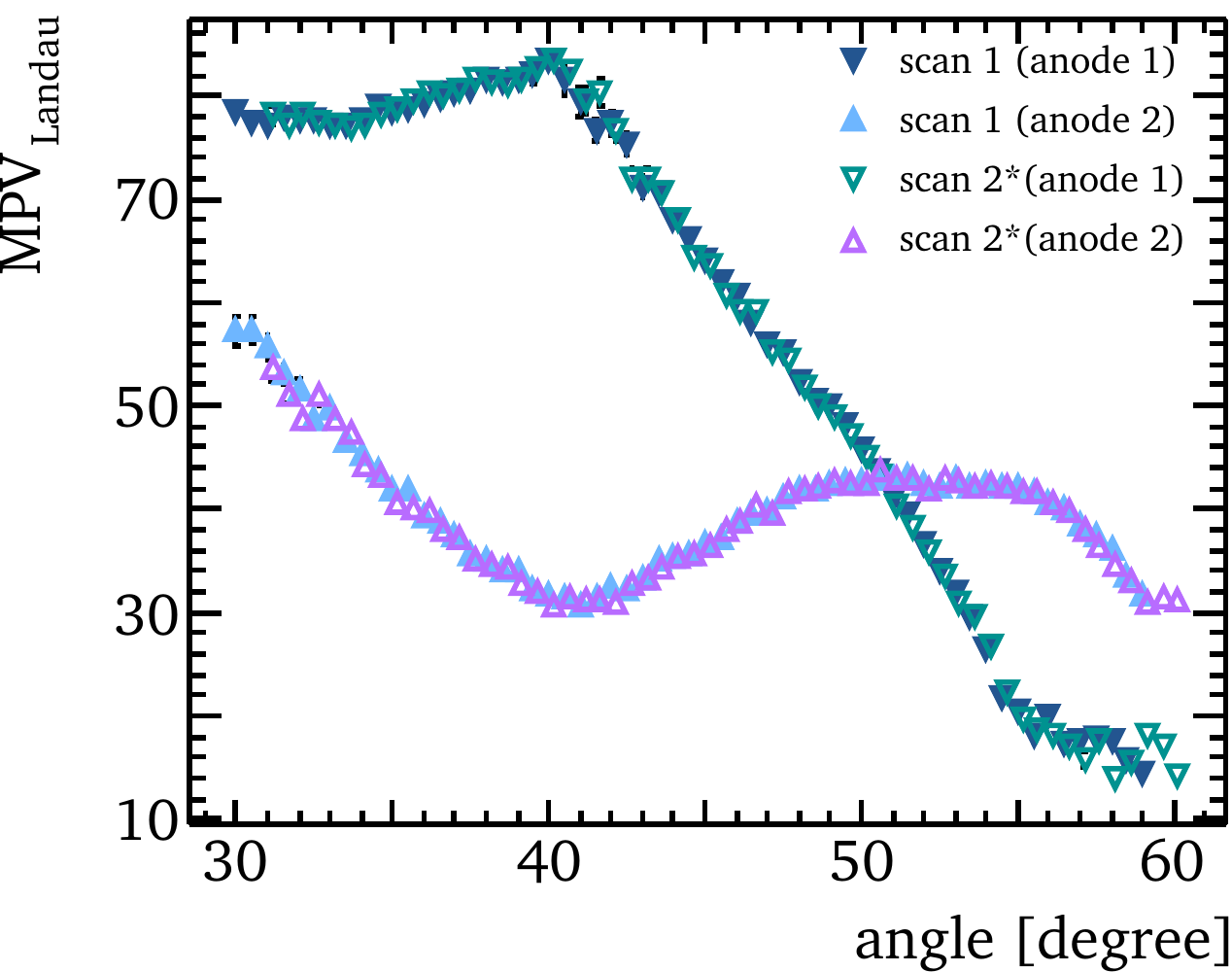}{fig:AyOffset1}
  \hfill
  \subfloat{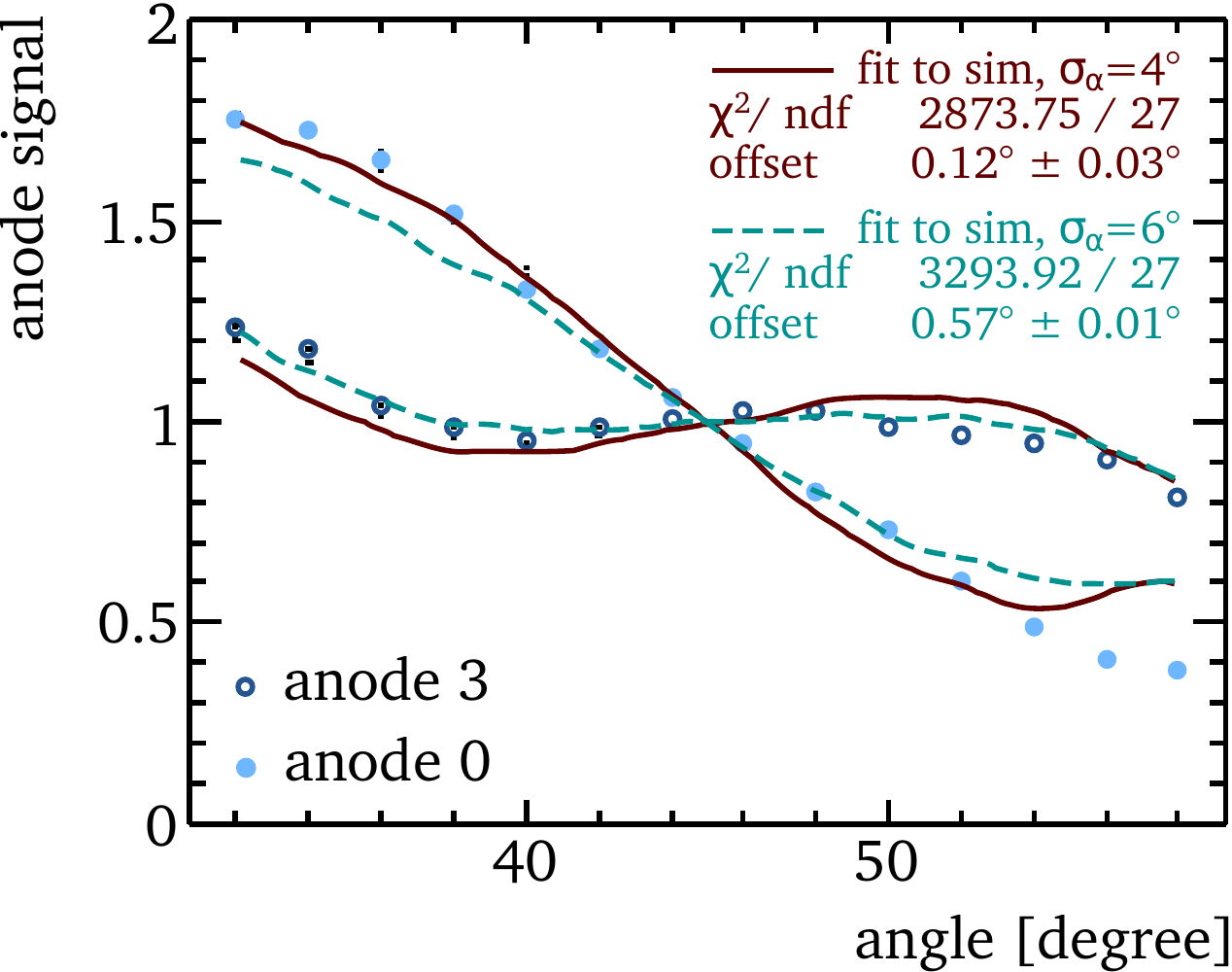}{fig:tbA2Ss46}  
  \caption{Angular dependence of light yield. \prosubref{fig:AyOffset1} Signal for both anodes of channel B for two consecutive angle scans, the first one from  30\degree to 60\degree (solid markers), the second one in the opposite direction (open markers). The second scan is shifted by 1.15\degree. \prosubref{fig:tbA2Ss46} Shape of the angular response function for an angle scan at $y_{\text{table}}=\NU{70}{mm}$, compared to the best-matching simulations at $\sigma_{\alpha}=$4\degree and $\sigma_{\alpha}=$6\degree.}
  \label{fig:tbA2SDandF}
\end{figure}

The data of different angle scans taken during the testbeam campaign were compared to a number of simulated angle scans with different choices for the microfacet distribution parameter  $\sigma_{\alpha}$, describing the surface smoothness of the quartz channels, and for the offset in the vertical direction $\Delta y$, to study the possibility to determine these parameters from the data:
 The surface smoothness of the quartz bars used in the prototype detector was not known. For a more polished surface, i.e.\ a smaller $\sigma_{\alpha}$, more directed reflections occur. Consequently, a change in the incidence angle of the beam is expected to cause larger changes compared to reflections on a rougher surface. The vertical alignment within an uncertainty of several millimetres also had to be taken into account. An offset in the vertical direction moves the central point of the beam crossing by  $\frac{\Delta y}{\sin\alpha}$ and therefore affects the detector response  especially at small angles.

In order to compare the shape of the detector response in the testbeam data to the different simulations independent of the absolute light yield, the response functions for both data and simulation were normalised to their respective signal at 45\degree, allowing for a small offset in angle to account for the observed uncertainty in the angle reading for the data discussed above. The current simulation does not describe the angular response satisfactorily for angles $\gtrsim 55\degree$. To illustrate this, figure~\ref{fig:tbA2Ss46} shows simulated angle scans with both  $\sigma_{\alpha}=4\degree$ and $\sigma_{\alpha}=6\degree$ compared to the data of an angle scan at $y_{\text{table}}=\NU{70}{mm}$.  The differences in the shape of the response function hint that additional factors must be present which are not well modelled in the current simulation.\\
No definite conclusions could be drawn for the vertical offset, since no single value was found to describe the data of several angle scans significantly better than any other. For any given vertical offset, the best agreement was found for $\sigma_{\alpha}=5\degree \pm 1\degree$. This was the case for all angle scans which were compared to the simulation, leading to the conclusion that a surface smoothness in this range would be a good description of the quartz bars used in the prototype detector.

\subsection{Light yield}
\label{sec:tblight}

In order to compare the light yield observed at the testbeam to the predictions from the simulation, the gain of each photomultiplier anode has to be known. According to the \qqtext{typical values} from the datasheets, the gain for the R7600U-03 photomultipliers should be twice as high as for the R7600U-04 photomultipliers. In the data however, no discernible difference between the signal of these two photomultiplier types used at the testbeam was found. Therefore, rather than using the datasheet values, a separate test setup was used to determine the actual gain values of the PMTs used with the prototype: inside a light-tight box, a small amount of UV light was produced with an LED and further reduced by a filter mounted in front of the photocathode. A more detailed description of the LED test setup can be found in~\cite{diss:vormwald}.
The filter strength and LED voltage were chosen such that ${\approx} \NU{95}{\percent}$ of the time only the dark current signal was observed.Assuming that the photons reaching the photomultiplier are Poissonian distributed, and that the quantum efficiency can be approximated by a binomial distribution with 20\percent probability for detection, this means that in the ${\approx} 90\percent$ of the remaining 5\percent of the events, in which a signal was observed, this signal stems from a single photon. After subtracting the pedestal, the remaining signal was fitted with a Gaussian to determine the gain. The fitted mean provides the mean number of electrons measured in the QDC, which corresponds to the gain in the single photon~case. Data was taken at PMT supply voltages from \NU{700}{V} to \NU{900}{V}. The determined gain followed a logarithmic behaviour as expected. As an example, the fitted gain for all four anodes of one of the R7600U-04 photomultipliers is shown in figure~\ref{fig:gainscanZB4785}.

\begin{figure}[tb]
  \centering
  \subfloat{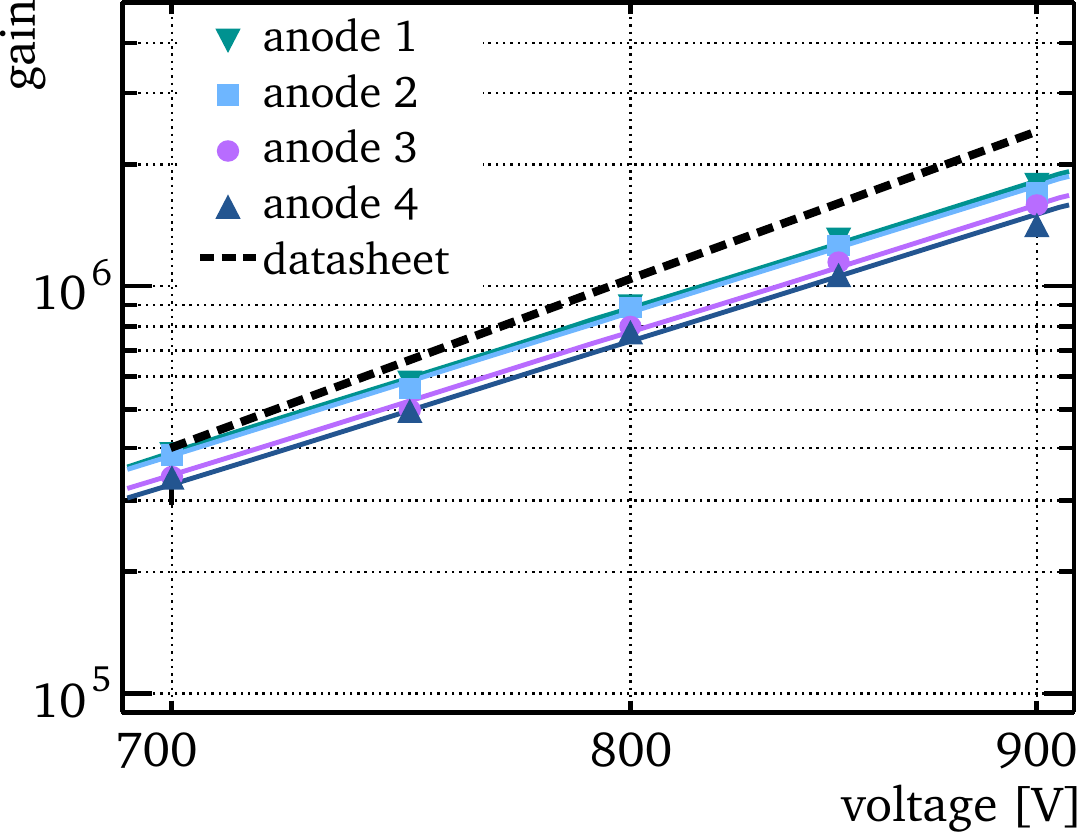}{fig:gainscanZB4785}
  \hfill
  \subfloat{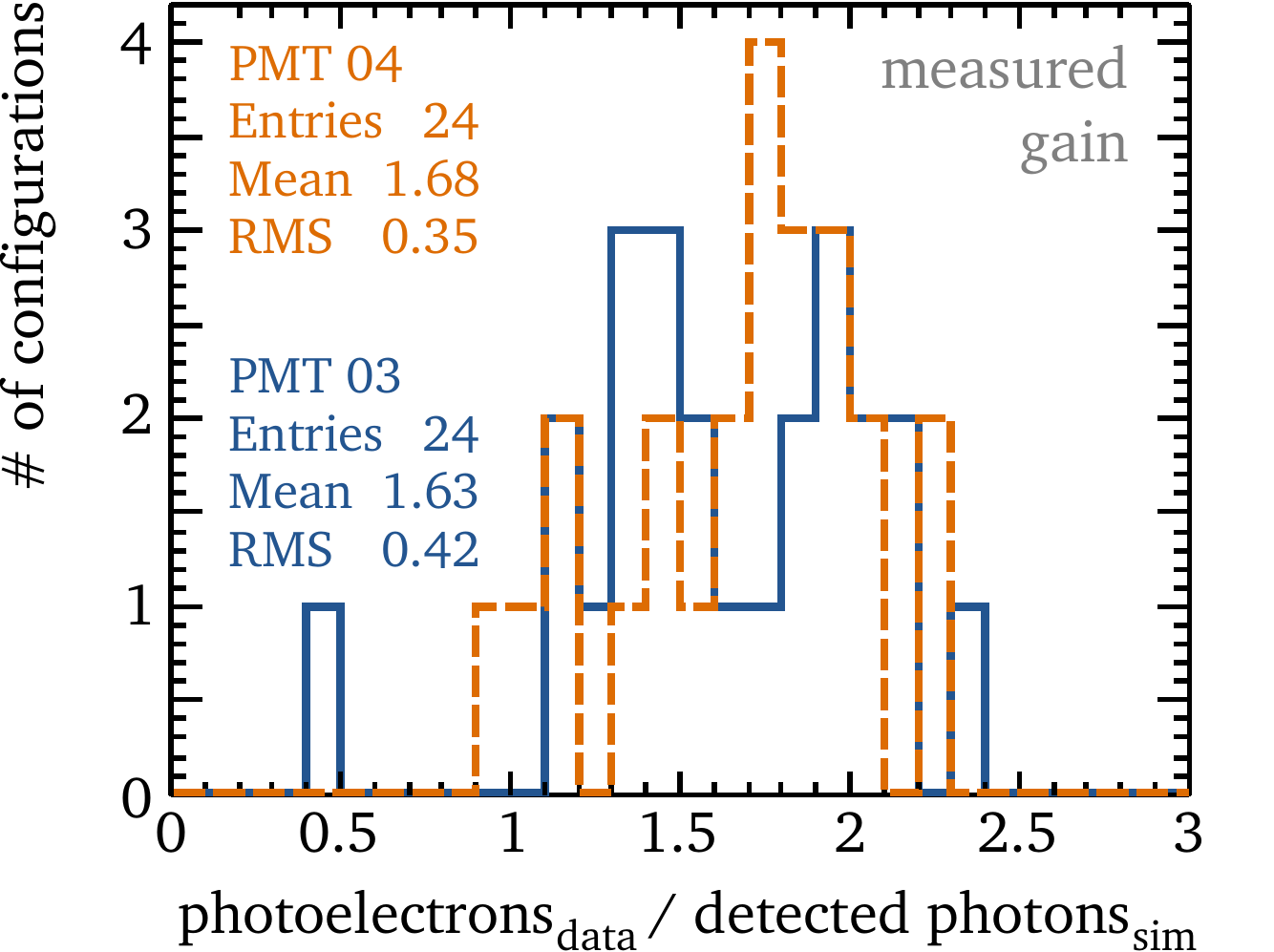}{fig:lightratioHisto}
  \caption{\prosubref{fig:gainscanZB4785} shows the fitted gain of one of the R7600U04 photomultipliers for different supply voltages, with the dashed line indicating the typical values provided in the datasheet. \prosubref{fig:lightratioHisto} Ratio between the light yield calculated from the data (using the measured gain) and the number of detected photons predicted by the simulation, for all anodes of both photomultipliers, for measurements at three different incident angles.}
  \label{fig:lightyield}
\end{figure}

Once the gain of the photomultipliers is known, the measured QDC signals can be compared to the simulation's predictions for the number of detected photons. For each detector channel, data taken under three different angles (35\degree, 45\degree, 55\degree) with both photomultiplier types was used for the comparison. For the configuration with the highest light yield, at 35\degree, summing up the signal for both anodes of a channel yields $107 \pm 12$ photoelectrons, while the simulation predicts ${\approx} 70$ photoelectrons for this angle. \\
The ratio between the observed number of detected photons in the data and the predictions from the simulation, i.e.\
$\frac{\#\text{photoelectrons}_\text{data}}{\#\text{detected photons}_\text{simulation}}$,
 for all configurations are shown in figure~\ref{fig:lightratioHisto}.
The alignment of the prototype detector at the testbeam with the precisions identified in the previous sections introduces systematic uncertainties in this ratio. While the error on the horizontal alignment is negligible, a systematic uncertainty on the light yield prediction from the simulation of $\frac{\Delta p}{p}=2\percent$ for the angular alignment with $\Delta \alpha_x = 1\degree$ and  $\frac{\Delta p}{p}=10\percent$ due to the uncertainty of $\mathcal{O}(\NU{5}{mm})$ for the vertical alignment have to be considered.

 On average, the amount of detected light calculated from the data is  higher than expected based on the simulation by a factor of \mbox{$1.7 \pm 0.4_\text{(stat.)} \pm 0.2_\text{(syst.)}$}.
This might be partly an effect of additional light provided by secondary electrons.
 In addition to this, a number of assumptions where made in the simulation, such as the roughness of the quartz surface, and the thickness of the optical grease layer. In regard of this, an agreement better than factor~2 between the data and the light yield in the simulation is considered quite satisfactory, especially since the simulation underestimates the amount of light and therefore provides a conservative estimate of the measurable light.

%% file: 05_conclusions.tex
\section{Conclusions}
\label{sec:conclusion}

Precise polarimetry is crucial for precision measurements at future lepton colliders, such as the International Linear Collider. At the ILC, two Compton polarimeters per beam aim for an accuracy of $\delta \mathcal{P} / \mathcal{P} = 0.25\percent$, limited by the linearity
and the alignment of the baseline gas Cherenkov detectors. 

A novel concept for detecting the Compton-scattered electrons using quartz as Cherenkov medium has been developed. Due to its much larger light yield per incident electron, it allows to resolve individual peaks in the measured Cherenkov light corresponding to different numbers of Compton electrons. This concept can improve the systematic uncertainty of the polarisation measurement substantially. In particular, it allows to control non-linearities in the response of the photodetectors used to detect the Cherenkov light, but also aids in alignment of the detector array.

A four-channel prototype detector has been built and operated in a first testbeam campaign. The detector response to single electrons has been measured and compared to the simulation.  The light yield per electron, which was a factor of \mbox{$1.7 \pm 0.4_\text{(stat.)} \pm 0.2_\text{(syst.)}$} higher than expected from the simulation, was found to be suitable for the construction of a detector capable of resolving individual peaks in the required dynamic range at the upstream polarimeter of the International Linear Collider.
 
The testbeam data allowed a horizontal alignment of the prototype detector on the order of \NU{0.05}{mm}. While the vertical alignment and study of the angular dependence of the detector would benefit from further investigation under better constrained conditions, i.e.\ with a smaller beam spot size and  a more accurate determination of the vertical alignment independent of the data, 
they were sufficient  to allow to determine that the surface roughness of the quartz used for the prototype is best described by a microfacet parameter  $\sigma_{\alpha}=5\degree \pm 1\degree$. The noise level of the electronic setup at the testbeam was not adequate, leading to a signal width factor~2 larger than considered attainable. 

The testbeam results do not change the conclusions from the simulation studies significantly. The differences between simulation and testbeam data in the angle scans are too small to affect the design optimisation. The larger observed light yield leads to an expectation of a possibly higher light yield for all detector geometries, which might allow to use smaller quartz channels. However, the observed light yield should first be verified in future tests of the current prototype with multiple electrons and less noisy electronics to establish that the achieved light yield does indeed allow to resolve individual Compton electrons.

Simulation studies extrapolating the testbeam results to ILC conditions showed that single-peak fitting can compensate photodetector non-linearities up to 4\percent and allows a horizontal alignment to about $\NU{10}{\mu m}$. The impact of tilts in the horizontal plane ($0.07\percent/\text{mrad}$) stays similar as for the baseline gas detector concept. The non-linearity compensation and improved alignment would reduce the effect of the two leading sources of systematic uncertainties on the polarimeter measurements significantly, to a level comparable
to the knowledge of the laser polarisation. This could reduce the total systematic uncertainty
to even $\delta \mathcal{P} / \mathcal{P} = 0.20\percent$, at least in case of the upstream polarimeter. In case of the downstream polarimeter, further studies are needed to investigate
whether the low Cherenkov threshold of quartz poses a problem in the harsh background conditions.

%% file: bib.tex
\bibliographystyle{apsrev}
